\documentclass[twocolumn]{aastex61}

\usepackage{amsmath}
\usepackage{gensymb}

\shorttitle{The Proper Motion of Pyxis}
\shortauthors{FRITZ et al.} 
\begin{document}

 \title{The Proper Motion of Pyxis: the first use of Adaptive Optics in tandem with HST on a faint halo object}
 \correspondingauthor{T.~K.~Fritz}
\email{tk4fw@virginia.edu}

\author[0000-0000-0000-0000]{T.~K.~Fritz} \affil{Department of Astronomy, University of Virginia, Charlottesville, 530 McCormick Road, VA 22904-4325, USA}
\author{S.~T.~Linden} \affil{Department of Astronomy, University of Virginia, Charlottesville, 530 McCormick Road, VA 22904-4325, USA}
\author{P.~Zivick} \affil{Department of Astronomy, University of Virginia, Charlottesville, 530 McCormick Road, VA 22904-4325, USA}
\author{N.~Kallivayalil} \affil{Department of Astronomy, University of Virginia, Charlottesville, 530 McCormick Road, VA 22904-4325, USA}
\author{R.~L.~Beaton} \affil{The Observatories of the Carnegie Institution for Science, 813 Santa Barbara St., Pasadena, CA 91101, USA}
\author{J.~Bovy} \affil{Department of Astronomy \& Astrophysics at University of Toronto,
50 St. George Street M5S 3H4
Toronto, Ontario, Canada}\affil{Center for Computational Astrophysics, 162 5th Ave, New York, New York 10010, USA}\affil{Alfred P. Sloan Fellow}
\author{L.~V.~Sales} \affil{Department of Physics and Astronomy, University of California Riverside, 900 University Avenue, CA 92507, USA}
\author{S.~T.~Sohn} \affil{Space Telescope Science Institute,
3700 San Martin Drive,
Baltimore, MD 21218, USA}
\author{D.~Angell} \affil{Department of Astronomy, University of Virginia, Charlottesville, 530 McCormick Road, VA 22904-4325, USA}
\author{M.~Boylan-Kolchin} \affil{The University of Texas at Austin, Department of Astronomy, 2515 Speedway, Stop C1400,Austin, Texas 78712, USA}
\author{E.~R.~Carrasco} \affil{Gemini Observatory, Southern Operations Center, AURA, Casilla 603, La Serena, Chile}
\author{G.~J.~Damke} \affil{Department of Astronomy, University of Virginia, Charlottesville, 530 McCormick Road, VA 22904-4325, USA}
\author{R.~I.~Davies} \affil{Max Planck Institut f{\"u}r Extraterrestrische Physik, Postfach 1312, D-85741, Garching, Germany}
\author{S.~R.~Majewski} \affil{Department of Astronomy, University of Virginia, Charlottesville, 530 McCormick Road, VA 22904-4325, USA}
\author{B.~Neichel} \affil{LAM - Laboratoire d’Astrophysique de Marseille,
38, rue Frederic Joliot-Curie, 13388 Marseille, France}     
\author{R.~P.~van der Marel} \affil{Space Telescope Science Institute,
3700 San Martin Drive,
Baltimore, MD 21218, USA}

\begin{abstract}
We present a proper motion measurement for the halo globular cluster Pyxis, using HST/ACS
data as the first epoch, and GeMS/GSAOI Adaptive Optics data as the second, separated by a
baseline of $\sim5$ years. This is both the first measurement of the proper motion of Pyxis and the
first calibration and use of Multi-Conjugate Adaptive Optics data to measure an absolute proper
motion for a faint, distant halo object. Consequently, we present our analysis of the Adaptive
Optics data in detail. We obtain a proper motion of 
$\mu_{\alpha}\,\cos(\delta) = $1.09$\pm$0.31 mas yr$^{-1}$ and $\mu_{\delta} =$0.68$\pm$0.29 mas yr$^{-1}$.
From the proper motion and the line-of-sight velocity we find the orbit of
Pyxis is rather eccentric with its apocenter at more than 100 kpc and its pericenter at about 30 kpc.
We also investigate two literature-proposed associations for Pyxis with the recently discovered
ATLAS stream and the Magellanic system. Combining our measurements with dynamical
modeling and cosmological numerical simulations we find it unlikely Pyxis is
associated with either system. We examine other Milky Way satellites for possible association
using the orbit, eccentricity, metallicity, and age as constraints and find no likely matches in
satellites down to the mass of Leo II. We propose that Pyxis probably originated in an unknown
galaxy, which today is fully disrupted. Assuming that Pyxis is bound and not on a first approach, we
derive a 68\% lower limit on the mass of the Milky Way of 0.95$\times10^{12}$ M$_\odot$.
\end{abstract}


\section{Introduction}
\label{sec:intro}

Of the globular clusters of the Milky Way, 
Pyxis \citep{DaCosta_95,Irwin_95} is one of the most distant \citep[$\sim 40$  kpc;][]{Sarajedini_96}. Even though the relatively high line-of-sight extinction of E(B-V)$\approx0.25$ \citep{Dotter_11} adds a large uncertainty on this measurement, Pyxis clearly resides in the halo. 
 Pyxis has a metallicity of [Fe/H]$=-1.45\pm0.1$; \citep{Palma_00,Dotter_11,Saviane_12} 
 and, from comparison to
 theoretical isochrones, is 11.5$\pm$1 Gyr old, roughly 2 Gyr younger than inner Milky Way globular clusters of the same metallicity \citep{Dotter_11,Saviane_12}.
Together, these measurements suggest that Pyxis belongs to the somewhat younger population of halo globular clusters that have likely been accreted by the Milky Way \citep{Zinn_93}.

Indeed, the three-dimensional location of Pyxis is quite suggestive of a complicated origin.  
\citet{Irwin_95} speculate that Pyxis originates from the Large Magellanic Cloud (LMC), based on the proximity of Pyxis to its orbital plane.
More recently, \citet{Koposov_14} discovered the ATLAS stellar stream
 and three globular clusters, including Pyxis, were considered potential progenitors of the stream. 
The proper motions of both NGC\,7006 and M\,15 were inconsistent with the implied orbit of the stream,
 and Pyxis, with no proper motion measurement, became the most likely candidate. 
Both scenarios, an LMC origin and that Pyxis is being tidally stripped, can be tested with a proper motion measurement.

Validation of these scenarios has larger implications than just the origin of Pyxis. 
Detailed observations of stellar streams coupled with numerical simulations can be used to constrain the potential of the Milky Way halo \citep{Koposov_10,Kuepper_15,Bovy_16b} as well as the mass function of subhalos within it \citep{Yoon_11,Erkal_16,Bovy_16}.
Longer streams improve such constraints: currently, the ATLAS stream is rather short (12$^{\circ}$ \citet{Koposov_10}), but if Pyxis is its progenitor system, then it would be one of the longest streams currently traced within the Milky Way halo.
Three-dimensional motions for halo objects, like Pyxis, can also be used to constrain the rotation curve at relatively large distances and thus the mass of the Milky Way.

Absolute proper motions for halo objects have been obtained with ground-based seeing-limited observations with long time baselines \citep[$\geq15$ years; e.g.,][]{Dinescu_99a,Fritz_15}, 
or with the \emph{Hubble Space Telescope} (HST) \citep[e.g.,][]{Piatek_07,Kallivayalil_13},
 which permits measurements over shorter time scales due to its better spatial resolution ($\sim$3-5 years). 
Ground-based adaptive optics have the potential to provide the HST-quality spatial resolution to enable shorter baseline proper motion measurements from the ground.
 Indeed, adaptive optics techniques have been well-established for
 proper motions \citep[e.g.,][]{Gillessen_09} 
using single
   conjugated adaptive optics systems. Now, multi-conjugated adaptive
   optics (MCAO) systems provide a larger field of view and have already been used for relative proper motions \citep{Ortolani_11,Massari_16}, in which the motion of a source is measured relative to another object in the Milky Way halo.
The larger field-of-view of MCAO makes it possible to now also find faint
background galaxies in the same images which can be used to get absolute
motions, like what has been done already with seeing-limited images \citep[e.g.,][]{Dinescu_97} and with HST \citep[e.g.,][]{Kalirai_07,Sohn_13,Pryor_15}. 
The GeMS/GSAOI system in operation at Gemini South
\citep{Rigaut_14,Carrasco_12} is the first AO system that combines the
large sky-coverage of laser guide-stars with the wide
diffraction-limited field-of-view of MCAO \citep{Davies_12} enabling
observations of targets without bright stars which are necessary for a
system without lasers.

In an on-going multi-year Gemini Large Program (LP-GS-2014B-2; PI: Fritz\footnote{A program summary is available at the following URL {\url http://www.gemini.edu/?q=node/12238\#Fritz}}), 
 we are using the GeMS/GSAOI system to measure absolute proper motions for a set of Milky Way halo tracers, including Pyxis.
While systems like GeMS/GSAOI are currently rare, planned instrumentation for 30-meter class facilities include wide-field AO imaging \citep[one example being MICADO, see][for details]{Davies_16} 
and AO-based proper motions from the ground will become more fruitful. 
The development of proper motion analysis techniques that use such instrumentation, thus, are valuable for future efforts. 

The Gemini Large program is still ongoing (we require 3 year baselines) and the second epoch of AO imaging for Pyxis has not yet been obtained. 
We can, however, utilize archival optical HST imaging, taken in 2009, as our first epoch. 
This sets a five year baseline between the first optical observations and our second near-infrared AO imaging.

The paper is organized as follow:
 In Section~\ref{sec:dataset} we present the data used in this study 
 and in Section \ref{sec:catalogs} we discuss our techniques to measure photometry and positions in our datasets.
We describe the methods used in the measurement of the proper motion of Pyxis in Section~\ref{sec:deriv_pm}.  
In Section~\ref{sec:origin} we use the proper motion to constrain its orbit and to explore the possible origin of Pyxis.
We conclude in Section~\ref{sec:summary}.

\section{Imaging Data Set} \label{sec:dataset}

In this section, we describe the imaging used to measure the proper motion of Pyxis. Technical details for our imaging datasets are summarized in Table \ref{tab:datasummary}.

\begin{table*}
\centering
\caption{Summary of Imaging Data} \label{tab:datasummary}
\begin{tabular}{l c c c c c l}
 \hline \hline
Telescope+Camera & Filter  & $\textit{MJD}$ &  ExpTime [s] & N$_{obs}$ & Resolution  &  Notes \\
                 &         &  days          &   [s]        & & $\arcsec$ pixel$^{-1}$ &    \\
 \hline 
HST+ACS/WFC      & $F606W$ & 55115.7        &  517         &  4  & 0.05 &    \\
                 &         & 55115.7        &  50          &  1  & 0.05 &  Used to recover saturated stars \\
                 & $F814W$ & 55115.7        &  557         &  4  & 0.05 &     \\
                 &         & 55115.7        &  55          &  1  & 0.05 &  Used to recover saturated stars   \\
\hline
Gemini-S + GeMS/GSAOI & $K'$ & 57030.3      & 120          & 30  & 0.02 &   Obtained in 5 6-image sets.\\
\hline \hline
\end{tabular}
\end{table*}

\begin{figure*}
\begin{center}
\includegraphics[width=0.947 \columnwidth]{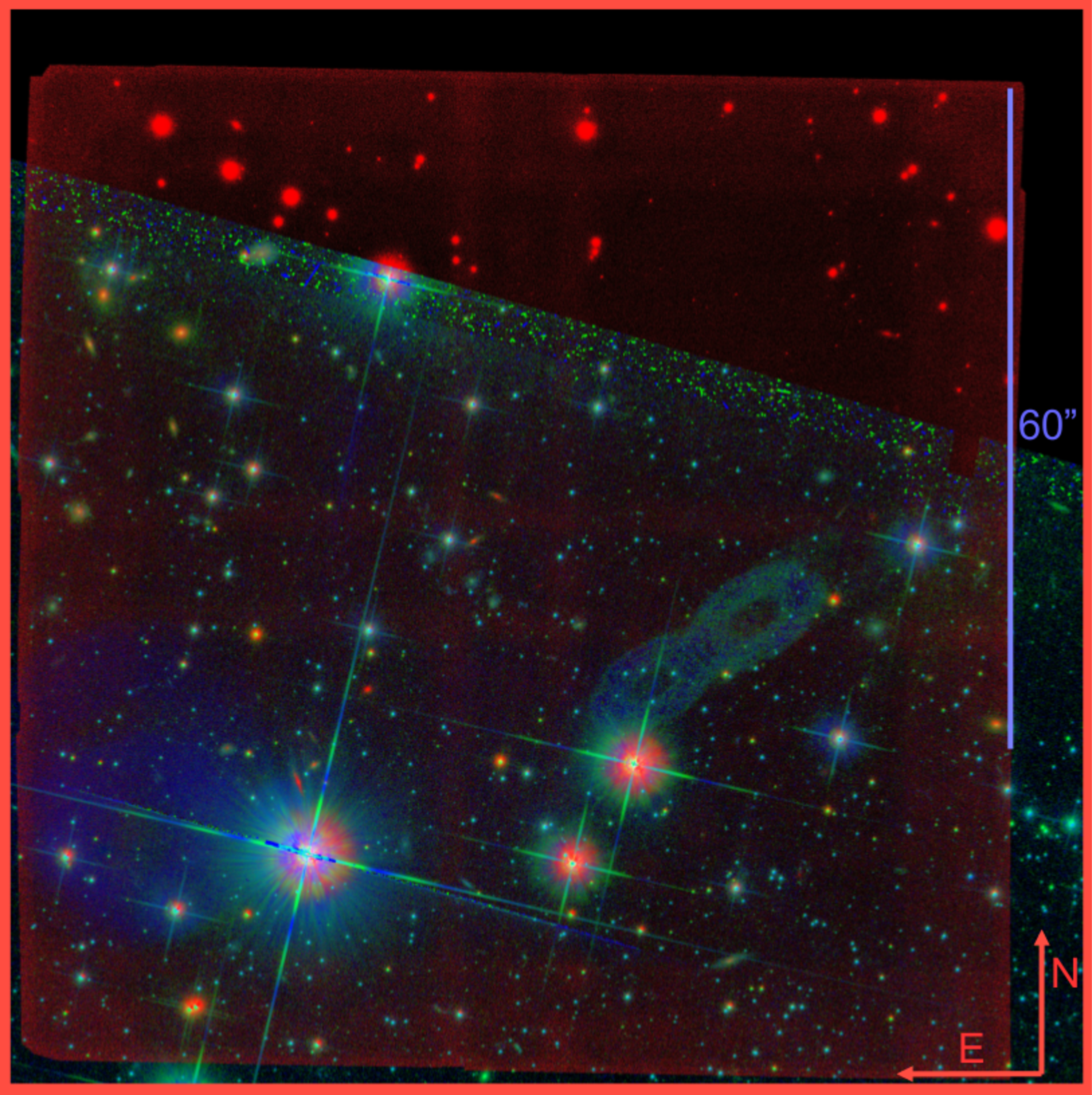}
\includegraphics[width=1.134 \columnwidth]{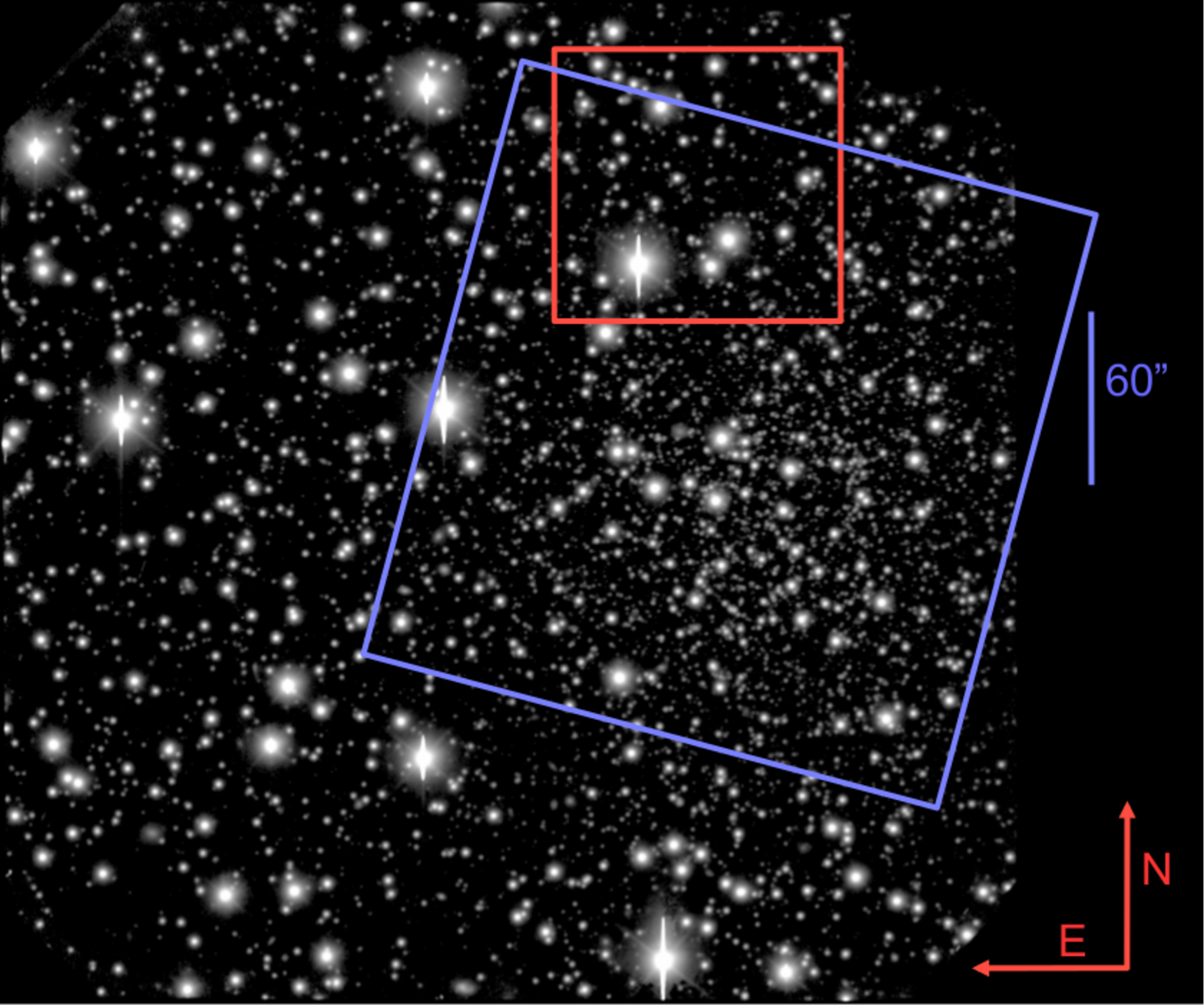}
\caption{ 
Pyxis images: (\textit{left}) 3-color image of the three high-resolution images used in this study (blue $F606W$, green $F814W$, red $K'$); 
  (\textit{right}) GMOS-S $i$-band image, the GSAOI field is indicated in red and the HST field in blue. The GMOS-S image was also obtained as part of our Long Term Gemini program, but is not used for science in this paper.}
\label{fig:datdis}
\end{center}
\end{figure*}

\subsection{HST Imaging} \label{sec:HST}

The first epoch consists of archival HST ACS/WFC $F606W$ and $F814W$ data\footnote{Proposal ID GO11586
PI-Aaron Dotter.}.
These images were obtained in 2009 (\textit{MJD}$=55115.7$) and provide a sufficient baseline in combination with our AO imaging.
In each band, there are 4 `long' images of about 540 seconds exposure time, which we use for the main astrometry. To recover photometry for bright stars that were saturated in the long exposure images, 
 we use single short exposure ($\approx$53 s) images obtained in both filters.
The individual frames are dithered by up to 17$\arcsec$, such that not all sources are contained in each individual exposure. 

For astrometry and photometry, we use the $\tt{flc}$ data products produced by the MAST pipeline.
The $\tt{flc}$ data products are bias-subtracted, dark-subtracted, flat-fielded $\tt{flt}$ images with the pixel-based charge transfer efficiency (CTE) correction applied \citep{Avila_16}.  
The $\tt{flc}$ images have a pixel scale of 0.05$\arcsec$ pixel$^{-1}$.
The pipeline produced $\tt{drz}$ images for each filter are also used for visual inspection. For some purposes, we create our own chip-merged frames using a chip-separation of 50 pixels between the two ACS chips.

Generally, cosmic rays are flagged by the $\tt{CALACS}$ pipeline in individual $\tt{flt}$/$\tt{flc}$ exposures in a given filter by comparison against other images of the same filter. The dithering between these exposures, however, leaves some area of the chip without a CR comparison image. In these cases, we flag cosmic rays manually by comparing this area to its counterpart in the other filter. The weight of pixels affected by cosmic rays is set to zero so that they do not affect our analysis.

\subsection{Gemini Imaging}\label{sec:lbt}
For the second epoch, we use the wide field AO system GeMS/GSAOI at Gemini South \citep{Rigaut_14,Carrasco_12}.  
The Gemini Multi-Conjugate Adaptive Optics System (GeMS) is a multi-conjugate adaptive optics system (MCAO) designed
 for the Gemini-S telescope.
For technical details on the AO system, we refer the reader to \citet{Rigaut_14} and for on-sky performance to \citet{Neichel_14}.
The Gemini South Adaptive Optics Imager (GSAOI) instrument is designed to work with GeMS and provides diffraction limited images over the 0.9 to 2.5 $\mu$m wavelength range.
The image plane is filled with a 2$\times$ 2 mosaic of Rockwell HAWAII-2RG 2048$\times$2048 pixel arrays producing
 a $85 \arcsec \times 85 \arcsec$ field-of-view at 0.02$\arcsec$ pixel$^{-1}$ resolution.
For additional technical details on GSAOI, we refer the reader to \citet{McGregor_04} for instrument design 
 and to \citet{Carrasco_12} for commissioning performance.

The Pyxis field was chosen to both meet the technical requirements of the AO system (three nearby bright, R$<$ 15.5 mag, stars for tip-tilt and
plate-scale mode variations) and to provide maximal overlap with the HST frame for data validation and comparison purposes.
We avoided the central part of the cluster where the density is high and background galaxies are more difficult to identify and fully characterize. 
The science data were obtained on 2015 January 7 (\textit{MJD}$=57030.3$) under good conditions,  
giving a 5-year baseline between the first (HST) and second (AO) epoch of imaging.
The data set consists of 30 science $K'$-band images of 120 seconds each. 
In addition, off-field images were observed for sky subtraction. The observations of the science frames were divided
in five groups of six images each. Within each group, the science images were observed using a small dither-steps
($<1\arcsec$) with the tip-anisoplanatic loop closed (accurate astrometry of the NGS probes in Canopus is derived before the
loop is closed). Larger dithers of 5$\arcsec$ were used between each group to cover the gap regions between detectors and
to improve the derived distortion solution for the GeMS/GSAOI data. Offsets above 1$\arcsec$ requires to open tip-
anisoplanatic loop, apply the large offset and re-do the astrometry of the NGS Canopus probes, before the
observations can be continued. The sky images were obtained between the groups.

We reduce the individual GSAOI frames in the standard way using domeflats, dead pixel masks, and sky images. The sky is constructed from off-Pyxis images because the source density for on-Pyxis images is too high. The data is also corrected for non-linearity and craters caused by bright stars by setting the affected pixels to values above the saturation limits. 
We construct noise maps from the data and find these to be dominated by sky noise. 

The four chips of the GSAOI images are arranged in a 2$\times$2 grid with average separations of 120 pixels in both the $x$- and $y$- image dimensions.
For the derivation of the distortion corrections, we treat the four chips mostly independently, 
see Section~\ref{sec:distortion}. 
For photometry and pixel positions, however, there are insufficient stars in any individual chip for the derivation of a point-spread-function (PSF). 
Thus, we must combine the four chips into a single frame for each exposure using the average chip separation. 
Our procedure assumes that the chips are perfectly aligned, which is not exactly true, 
 and we will evaluate the impact of this misalignment in our PSF modeling (Section~\ref{sec:psf_est}).

\begin{figure*}
\begin{center}
\includegraphics[width=1.0 \columnwidth]{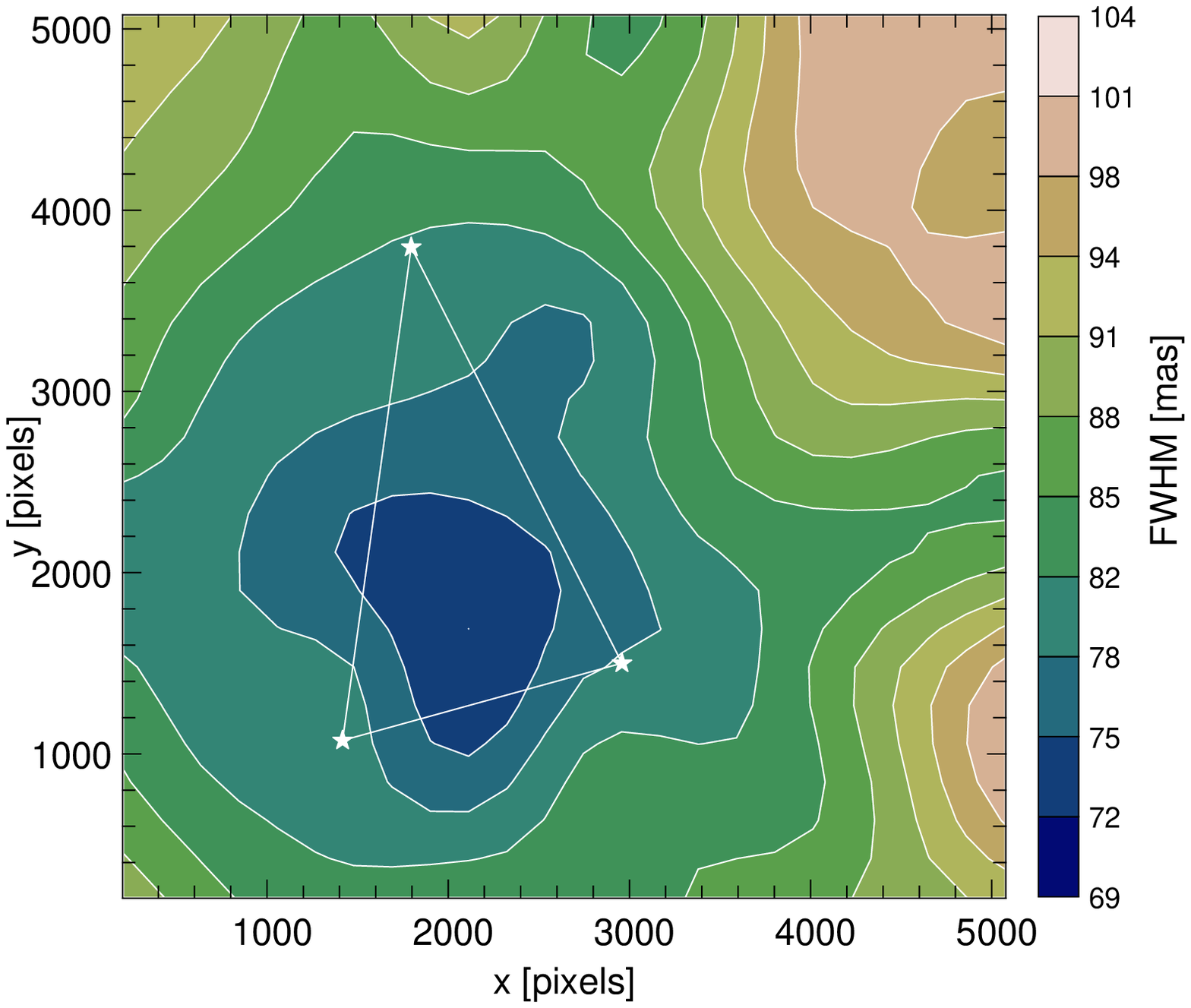}
\includegraphics[width=1.0 \columnwidth]{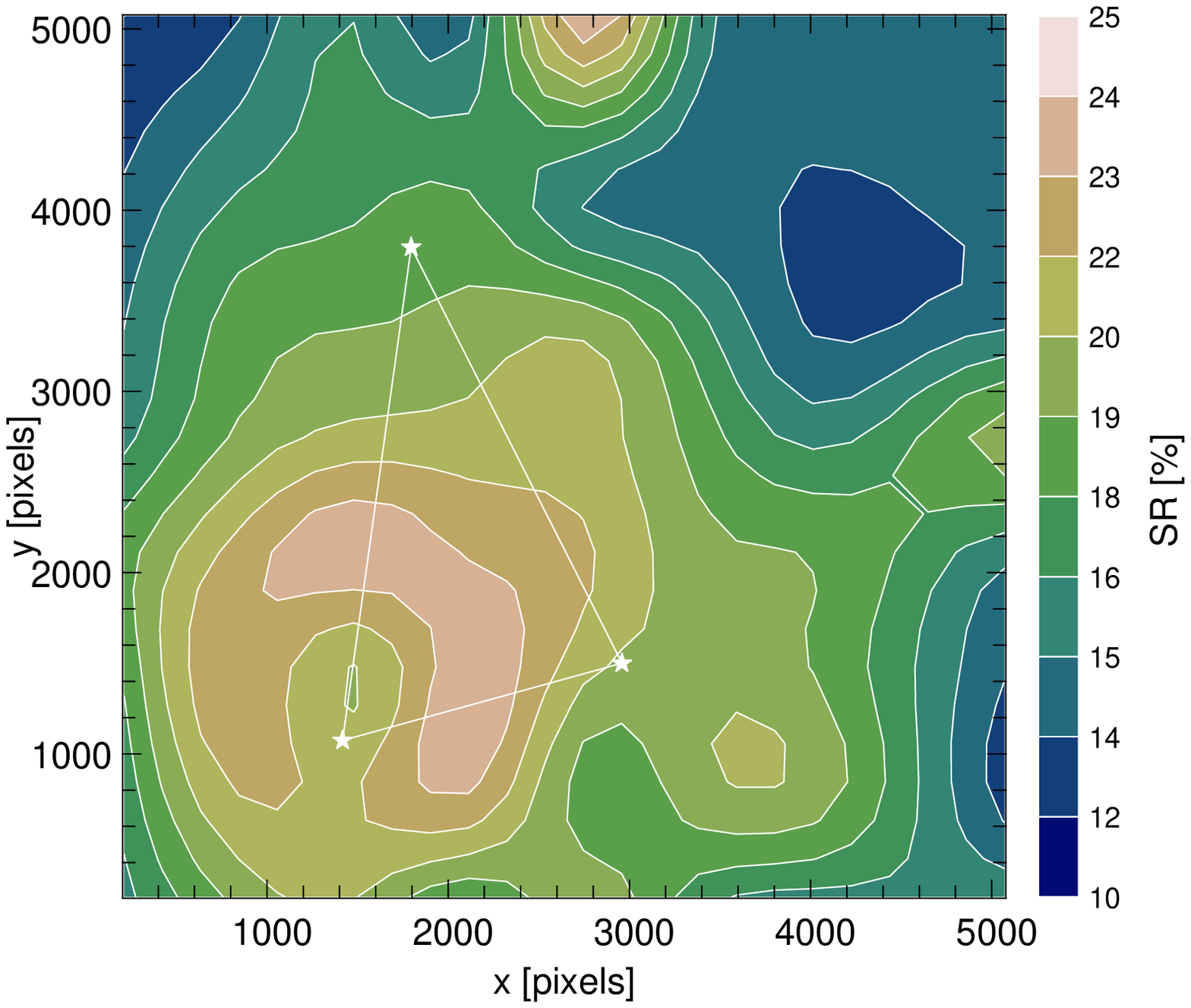}
\caption{Quality characteristics of the Pyxis data: On the left the FWHM is shown as a function of position, the unit is mas; on the right the Strehl Ratio is shown as a function of position, the unit is \%. Both are measured on the mosaic image. The white stars of the triangle mark the used tip tilt stars.}
\label{fig:ao_quality}
\end{center}
\end{figure*}

Most of our analysis is performed on the individual frames, but for some purposes, like source identification, we use the higher signal-to-noise combined mosaic image of the 30 individual science frames. The mosaic-ed combined image is constructed using the package THELI\footnote{Available: {\url https://www.astro.uni-bonn.de/theli/} \citep{Erben_05,Schirmer_13}}.
Astrometry and distortion correction are done using the program "SCAMP"
\citep{Bertin_06} called from THELI. The reference catalog is constructed from an 
archival image obtained in J-band with the Visible and Infrared Survey Telescope 
for Astronomy (VISTA) and the VISTA InfraRed Camera (VIRCAM) located at 
Paranal Observatory. After the astrometry and distortion correction are derived, 
the sky background subtraction is performed on individual images in order to achieve
a homogeneous  zero background level across the multi-array GSAOI frames. Finally 
the images are resampled to a common position, mosaic-ed and then combined using 
the software Swarp \citep{Bertin_10} called from inside THELI. The internal astrometric 
errors in the final co-added mosaic-ed image are less than 10 mas.
We did evaluate the use of this higher signal-to-noise image for our astrometric analysis, but found the distortion corrections applied to the individual images before co-addition were not sufficiently precise for our need of 1 mas precision.
The individual source shapes, however, were not strongly impacted by the residual distortion in the co-added image, and thus the image provides higher S/N for morphology assessment than the individual frames. In Figure~\ref{fig:ao_quality}, we demonstrate the quality of our mosaic with the FWHM and Strehl ratio computed across the field of view. As in \citet{Neichel_14} we use Yorick to measure these properties. The correction is relatively good with FWHM$<90$ mas SR$>15\%$ over most of the field.

\subsection{Image Overlap}

Figure \ref{fig:datdis} gives a summary of the imaging used for this project.
The left panel of Figure \ref{fig:datdis} is a color composite image of the overlap between
 the HST optical and GeMS/GSAOI NIR imaging.
Approximately 73\%  
of our GeMS/GSAOI field has HST imaging. 
The right panel of Figure \ref{fig:datdis} uses a GMOS-S image of Pyxis (located in the center towards the right)
 showing the locations of the imaging used for this paper.
While the HST pointing is on the center of the Pyxis cluster, our GSAOI field is offset to the North where the stellar density is lower. 
From visual inspection, the GSAOI images are of overall lower photometric depth than those of HST.

\section{Photometry and Initial Positions} \label{sec:catalogs}

In this section, we describe the creation of Pyxis astrometric and photometric catalogs from the optical HST and near-infrared AO imaging. 
We sub-divide this process into the procedures for stellar photometry and those for galaxy photometry. 
Stellar photometry for HST imaging is described in Section \ref{sec:hststars}. 
For the GSAOI images, we first develop appropriate PSF fitting techniques in Section \ref{sec:fullstars}, which are described in detail in those subsections. 
Our MCAO PSF is applied to stellar sources in Section \ref{sec:aostars}.
The estimation of galaxy positions is described in Section \ref{sec:galaxypositions}, which is done for HST and AO in concert. 
We determine corrections to the initial positions using a preliminary distortion correction 
 and evaluate the effect of differential chromatic refraction (DCR) in Section \ref{sec:dcr}.
This process is summarized in Section \ref{sec:catalogsummary}.

\subsection{Stellar Sources in HST} \label{sec:hststars}
There exist already well-vetted codes for extracting astrometry and photometry from HST images and we follow those outlined in \citet{Anderson_06}.
For the photometry and astrometry, the $\tt{flc}$ data products were used instead of the provided $\tt{drz}$ data products since the latter involve an image resampling that degrades the astrometry. 
Starting with the $\tt{flc}$ images, we perform PSF fitting, utilizing empirical `Anderson Core' PSFs constructed specifically for the $F606W$ and $F814W$ filters by \citet{Anderson_06}, to create a catalog of pixel positions and photometry of all point sources for all exposures. Both stellar and non-stellar sources are included in this catalog, but the positions for galaxies will be refined by 2-dimensional galaxy fitting in Section \ref{sec:galaxypositions}.

We calibrate the photometry using the STScI provided zero points \footnote{They are obtained from http://www.stsci.edu/hst/acs/analysis/ zeropoints/old\_page/localZeropoints.} of 26.406 and 25.520 mag in $F606W$ and $F814W$, respectively,
 and use the 0.5$\arcsec$ to infinity aperture-correct (0.091 mag in both bands from \citet{Sirianni_05}.

\subsection{Derivation of AO PSFs Using Stellar Sources} \label{sec:fullstars}

In MCAO imaging the PSF varies over the field of view. Thus,
  the PSF needs to be derived as a function of the coordinate
  position. Because Starfinder \citep{Diolaiti_00} derives only 1 PSF
  per attempt, it is a cumbersome tool to derive the PSF for the full field of view. Further, in our case of a low star density the advantage of
  Starfinder in the regime of a high star density is not valid. In
  contrast, the low star density, makes it more difficult to choose a
  local sample of stars, \citep[as in][]{Meyer_11,Neichel_14}, with
  sufficient number and spatial coverage.
DAOPHOT \citep{Stetson_87} was used as an option in other works \citep{Ortolani_11,Massari_16}. 
We here opt to test another option, PSF Extractor \citep{Bertin_11}.
First, we have to construct an initial high SNR stellar catalog from which the spatial and time-varying PSF can be derived (Section \ref{sec:sources}).
Second, we use PSF Extractor (PSFeX) to build model PSF grids from which the PSF at any given location can be derived (Section \ref{sec:psf_est}).
We evaluate the effectiveness of these models for astrometry in Section \ref{sec:psfastrom}.
Lastly, we develop model grids on the AO mosaic image for photometry in Section \ref{sec:photgrid}.

\subsubsection{Source Classification} \label{sec:sources}

Our source list is generated from the THELI-mosaicked GSAOI images as these images set the limiting photometric depth of our analysis.
For the initial selection of sources, we start with the function $\tt{find}$ in the $\tt{dpuser}$ image reduction\footnote{For more details on $\tt{dpuser}$ see {\url http://www.mpe.mpg.de/~ott/dpuser/index.html}}, which uses a globally-derived SNR criterion and FWHM to select stars. 
After an initial pass, additional SNR criteria based on the local noise around an individual source are applied; this step ensures that most selected stellar sources are real. 
Lastly, each source is then examined by eye in the images to remove spurious sources, including false detection caused by close saturated stars, image artifacts, or sky noise,
 or real sources that are compromised due to any of the previous spurious sources. 
Thus, our final source list contains visually verified objects that are either stars or galaxies.

Validation and addition of more galaxy candidates continues via a visual inspection of the $K'$-band mosaic. 
To ease detection of galaxies with lower surface brightness features, we first smooth the $K'$-band mosaic with a Gaussian of FWHM$=3$ pixels and refine the initial classification of star or galaxy. 
Galaxy candidates are then fit individually with a two-dimensional Gaussian and the fit parameters are compared to those fits of neighboring stars. 
Only candidates that are clearly different from confirmed stars are considered galaxies in this step. 
Not all objects are clearly classified with a Gaussian fit
 and for these borderline cases we also consider the fit parameters from the Galfit code \citep{Peng_02} (more detailed galaxy fitting will be provided in Section~\ref{sec:galaxypositions}). 
We compare the $\chi^2$ of a PSF fit and a Sersic fit and the obtained Sersic parameters. 
We exclude sources which are fit better by a Sersic, but for which the Sersic index is large, and the effective radius 
and the axis ratio are both small. These are unphysical parameters for galaxies and these objects are probably barely-resolved binary stars. Overall, only a few sources are reclassified in the second step. Due to the high spatial resolution there remain no borderline cases at the end. 

We confirm the classifications with visual inspection of the HST imaging 
 and we exclude a few red galaxies which are invisible in the HST images. 
Overall, 52 galaxies are confirmed in the GSAOI footprint.
The four saturated stars in the AO imaging are added by hand to our star list.
Overall, 450 stars are confirmed in the GSAOI footprint.

We note that both the star and galaxy samples were created conservatively owing to the needs of our analysis for clean samples of stars (for the proper motions) and galaxies (for the astrometric reference frame).
The faintest sources (galaxy or stellar) contribute only with a very small weight to the overall analysis 
 and the incompleteness of the sample does not affect our proper motion.

\subsubsection{PSF estimation for AO images} \label{sec:psf_est}

We use PSF Extractor \citep[PSFeX;][]{Bertin_11} to generate PSF models for the GSAOI imaging for use in both astrometry and photometry-based analyses. PSFeX is designed to automatically select point-like sources from SExtractor \citep{Bertin_96} catalogs and construct models of the point spread function (PSF) across an image. PSFeX models PSF variations as a user-defined polynomial function of position in an image. However, PSFex does not work directly on the images themselves. Instead, it operates on SExtractor catalogs that have a small image `vignette' recorded for each detection. We will first describe general considerations for using PSFeX before we discuss the detailed generation of our PSF grids for astrometry and photometry.

We start by generating SExtractor catalogs for each of our AO frames. PSFeX automatically pre-selects sources from this SExtractor input catalog that are likely to be stellar based on source characteristics, such as half-light radius and ellipticity, while also rejecting contaminated or saturated objects identified in SExtractor. Each iteration of the PSF modeling consists of computing the PSF, comparing the vignettes to the reconstructed model, and excluding detections that show too much departure between the data and the model. We use a pixel-based Principal Component Analysis to build a $\chi^2$-minimized image basis vector to represent the PSF. PSFeX uses these custom basis vectors to interpolate PSF models at specific locations for an output grid with the same resolution and size as each input GSAOI image. Thus, there are two critical inputs for PSFeX: (i) the SExtractor catalog from which sources are selected and (ii) the interpolation scheme used to generate the model output grid.

Our GSAOI images are atypical of the normal PSFeX applications due to a very high degree of PSF variability across the frame and a source density that is overall too low for high-fidelity measurements of that variability. Thus, we bypass the ``automatic'' selection and used our confirmed stellar sample (Section~\ref{sec:sources}) to optimize the basic PSFeX procedure for our GSAOI data in the following ways: (i) we use large image vignettes (VIGNET size and PSF\_SIZE) of 128x128 pixels, (ii) we require a minimum SNR of 20 in the central pixel, and (iii) we allow for large variations in both the maximum ellipticity (0-100\%) and FWHM (2.5 - 29.7 pixels) of the model sources. These modifications help to maximize the number of sources PSFeX uses to build the basis vectors and thus improve the accuracy of any higher-order interpolation to parts of the image where either the source identification is sparse or the background RMS is larger.

The size of the image vignettes we used to construct the model were chosen to fully encapsulate the spatial extent of the PSF for the used stars\footnote{The four brightest stars have also SNR outside that vignette, but we ignore these stars because they are saturated.}
The large range of acceptable source properties allowed to account for up to a 100\% variation in the typical PSF size as a function of both time and seeing conditions \citep{Dalessandro_16}. Typical values for the FWHM of accepted stars ranged from 5-7 pixels, with ellipticities that ranged from 5-10\%. By carefully removing close binaries (within 600 mas) from the sample of stars used by PSFeX, we have also minimized any effects of contamination from neighboring sources when using a large image vignette. The reliability of our PSF generation is thus insensitive to image-by-image changes in the shape of the PSF.

For our purposes, the output grids are configured to have 32 model PSFs in either dimension across the full span of the chip-merged image. Once grids are made, we determine the properly modeled PSF at the location of a given source by interpolating bi-linearly between the nearest four PSFs of the model grid. We use these model PSFs determined at the location for each star as the inputs for the astrometric and photometric
analyses to follow.

\subsubsection{PSF Model Grids for Astrometry} \label{sec:psfgrid}

Our astrometry is performed at the individual frame level and, thus, we must construct a model grid for each of the 30 GSAOI $K'$-band images of Pyxis independently.
We build two model grids for each frame. 
The first (PSF1) uses a set of `hand-selected' sources for the PSF modeling and employs a quadratic interpolation scheme to generate the output grid. The second (PSF2) uses a larger automated `SExtractor-expanded' source list and employs a cubic interpolation scheme. 
We discuss the details of each model separately.

The `hand-selected' (PSF1) source selection selects the highest SNR sources from our initial stellar list based on their properties in the individual frames.
We perform an additional verification of the image shapes as they appear in the individual frames to remove saturation and close neighbors.
Additionally, we exclude any stars within 35 pixels of the edges of each chip in an individual exposure to minimize the number of sources in the final list that would be greatly affected by the unreliable noise estimates there, and would thus be unreliable for building the PSF models.

Our `hand-selected' PSF grids were generated using an average of 20 sources per image, with 100\% of the sources passing the FWHM, SNR, and ellipticity selection criteria described above. A quadratic interpolation was preferred due to the small number of sources in each image, and, in particular, in some border regions of each GSAOI detector. We tested cubic interpolation, and while in many regions of the image it was reassuringly similar to quadratic, in sparser regions it showed too many artifacts.

Our `SExtractor-expanded' PSF (PSF2) grids were generated using an average of 40 sources per image, with $ \sim 98\%$ of the sources passing the automated PSFeX selection criteria for FWHM, SNR, and ellipticity. 
This catalog is effectively `blind' to our `hand-selected' catalog, but since our `hand-selected' catalog represents the `best' stars in the frame, many of those stars will be included.
Similar to the `hand-selected' case, we do additional source verification checks on individual frames. 
For this expanded list of sources, we then choose a cubic interpolation scheme (higher order than for `hand-selected') while all other parameters remained the same as in the generation of our `hand-selected' grids (PSF1). 
The lower limit on the acceptable FWHM for each source is important to filter out any cosmic rays detected when using the `SExtractor-expanded' (PSF2) source lists.

\begin{figure}
\begin{center}
\includegraphics[width=1.0 \columnwidth,angle=-90]{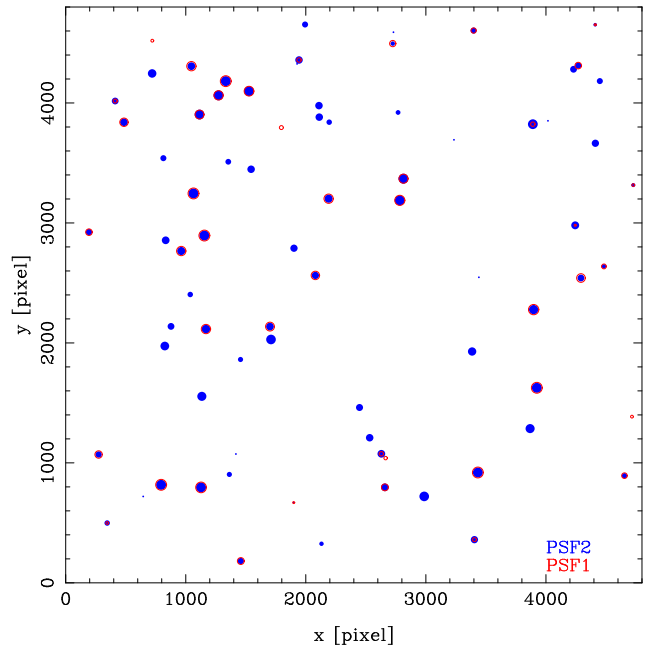}
\caption{The distribution of stars used by PSF1 (red open circles) and PSF2 (blue filled circles) to build the PSF model grids. 
The stars are shown on the mosaic.
Here, the area of the symbols is proportional to the number of images on which they are used (1-30), with the largest symbols corresponding to a star used to build the PSF model on all images.}
\label{fig:PSFstar_distr}
\end{center}
\end{figure}

For both PSF1 and PSF2, the actual sources used for any given image will vary due to several factors. The main factors are (i) the SNR of a star on an image which varies due to conditions, and (ii) whether the star is close to a border. (Stars within 35 pixels of a chip border on a particular image are excluded.)
  Figure~\ref{fig:PSFstar_distr} shows the distribution of stars used by PSF1 and PSF2.
 Overall it is clear that the stars used to build both PSF models do a good job at sampling the entire image, and are not systematically biased to any one location on the images.

The overall SNR of our `SExtractor-expanded' cubic PSF grid (PSF2) is slightly lower but comparable to our `hand-selected' quadratic PSF grid (PSF1). In general, if the number of detected sources is constant one would expect a quadratic interpolation scheme to produce a model grid with a higher integrated SNR because less parameters need to be fit compared to a cubic interpolation scheme. The fact that the two grids have similar integrated properties shows us that higher-order interpolation schemes are much more robust when nearly twice as many sources per image are used to build the model grid.

\subsubsection{Comparison of the PSF models for Astrometry} \label{sec:psfastrom}

Only relative differences induced by the two PSF grids creates motion uncertainty.
We compare velocities obtained with both PSFs to measure the
  contribution of the PSF modeling schemes adopted. For that we
 calculate our final velocity in Section~\ref{sec:absvel} with both PSFs and use the difference between the two motions to the get the additional error terms of
 0.07/0.09 mas yr$^{-1}$ in R.A./Dec. The relatively small error indicates that the effects of our two different PSF models on the motion are not the dominating error terms. This uncertainty is probably even an overestimate of PSF effects, since the complex distortion correction can add additional effects.
Using the two distortion corrections we also measure the astrometric scatter over the different detections.
The astrometric scatter for each stellar source is overall similar in both PSFs; for our galaxies, however, the scatter is slightly smaller for PSF1. 
We average the resulting positions from PSF1 and PSF2 to obtain our final measurement.

\subsubsection{Model Grids for Photometry} \label{sec:photgrid}
Our photometry is performed on the higher S/N K$'$-band mosaic image and, thus, we also build PSF models for the THELI mosaic image. 
Due to the higher S/N, we follow the general scheme of the `SExtractor-expanded' model to generate this PSF.
More specifically, we chose a pixel-based PCA, cubic interpolation scheme, and a model accuracy threshold of 10\% as with the individual frames. 
We modify the selection criteria to account for the higher S/N in the mosaic image and, thus, the range of allowed parameters for FWHM, MINSN, and MAXELLIP are tightened. 
We increased the minimum S/N to 400 and changed the FWHM range to $0-19.7$ pixels.
Cosmic rays were removed in the creation of the mosaic.
Our mosaic PSF grid was generated using 40 input sources and we require that 100\% of the sources pass the modified PSFeX selection criteria.

 \begin{figure}
 \begin{center}
   \includegraphics[width=0.70 \columnwidth,angle=-90]{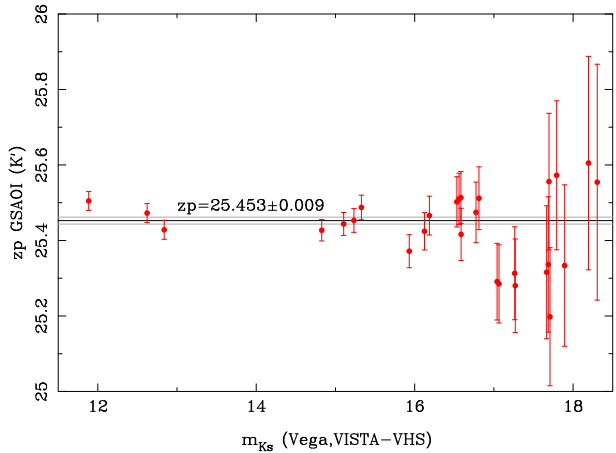}
 \caption{Photometric calibration of our GSAOI $K'$ data to VISTA-VHS $K_s$-band \citep{McMahon13}.
 Twenty-eight stars are used. The lines show the zero point and its formal uncertainty.  
 } 
 \label{fig:zp_gsaoi}
 \end{center}
 \end{figure}

\subsection{Application of AO PSFs to Stars} \label{sec:aostars}
For the AO data we use Galfit to fit for the positions of the stars, due to the difficulty in adapting the HST-specific codes (e.g., Section \ref{sec:hststars}) for images obtained with a substantially different instrument. 
The stars are fit with the PSFs developed for the AO mosaic in Section~\ref{sec:psf_est}. 
We first fit all objects on the mosaic using the mosaic PSF grid (Section \ref{sec:photgrid}), which provides good initial values for the objects. 
We use a fit window of 101 pixels for stars and of 151 pixels for galaxies, which again includes secondary stars and galaxies. In all cases we check whether the primary source is well fit and make some adjustments in details like the starting values, especially the magnitudes.

We then use a preliminary version of an inverse distortion correction (Section~\ref{sec:distortion}) to predict the positions of our sources in the individual frames and bi-linearly interpolate the PSF grid to the location of the source. 
We combine these positions and PSF models with the preliminary fit results from the mosaic (as a starting guess) to fit all sources in each of the individual frames. We do this for both PSF grids.
This method fails in a few cases for which the source has a much brighter neighbor. 
We evaluate sources that are flagged by Galfit as being overall less reliable 
with a $*$ and find that the positions residuals are not significantly different from more reliable sources. Thus, we use them. 

\begin{figure}
 \begin{center}
   \includegraphics[width=0.70 \columnwidth,angle=-90]{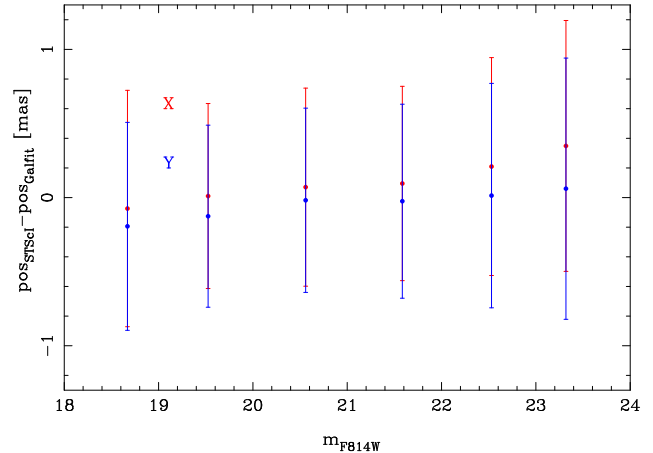}
 \caption{Comparison of star positions measured on $F814W$ images with the different codes. We show the average and standard deviation in magnitude bins.
 } 
 \label{fig:meth_pos_comp}
 \end{center}
 \end{figure}

We exclude detections of individual stars whose derived magnitudes are more than 1 magnitude different from the magnitude of that star in the mosaic image. Further, we exclude those objects for which the centroid is outside a chip or the position error is larger than 20 pixels. 
The position errors of stars depend on the flux measured by Galfit
because $\sigma_x\propto 1/\mathrm{SNR}$ \citep{Lindegren_78}. While
noise is always well-measured the signal is not well-measured when the
SNR is low. In this case, the position errors are too low or too high
when the measured flux is too high or too low. Therefore, we multiply
the position-errors measured by Galfit by the flux ratio between
between its measured flux in the single image and its flux in the
mosaic. Also, after adjustment, the position errors depend on
  the Strehl ratio and the noise of the single image.
In this process, we also compare the Galfit positions for stars on HST images with the well-established \citet{Anderson_06} techniques (Section \ref{sec:hststars}), see Figure~\ref{fig:meth_pos_comp}. We obtain for stars with 20$<m_\mathrm{F814W}<$23, that the difference in position is on average 0.1 mas per image. Since more than one image is used the bias on the final positions is even smaller, and the scatter in position is 0.65 mas. Thus, the two codes agree sufficiently for our purposes.

To calibrate our photometry, we cross match our stellar photometric catalog to the VISTA Hemisphere Survey\footnote{Publically available {\url http://www.vista-vhs.org/}} \citep[VISTA-VHS;][]{McMahon13} which is in turn tied to the 2MASS photometric system. 
There are 28 stars with reliable magnitudes measured in both catalogs.
Figure~\ref{fig:zp_gsaoi} compares the zero-point derived for each of the 28 stars to its VISTA-VHS magnitude down to the limiting depth of the VISTA-VHS catalog of $K_s$ = 18.1 mag.
To obtain a reduced $\chi^2$ of 1, we must add 0.025 mag in quadrature to the reported photometric uncertainties in the VISTA-VHS catalog.
The added value likely accounts for the GSAOI uncertainties, which are probably caused by imperfect PSF knowledge. 
We obtain a zero point of $25.453\pm0.009$ mag. 
The uncertainty is likely underestimated as several terms are not taken into account explicitly; for example, we do not consider the filter-curve differences between GSAOI and VISTA-VHS which will introduce color-dependent effects. Since we do not use K'-band photometry for source selection, the error underestimate does not affect our results.

\subsection{Positions for Galaxies} \label{sec:galaxypositions}

For galaxies we require a method that can both account for the spatially varying PSFs across an individual image and provide consistent results with different PSFs on the different instruments. Specifically, the temporal variability of the AO corrections makes it important to decompose the effects of the PSF from intrinsic shape effects \citep[see][]{Fritz_16}.

Our approach uses Galfit for the astrometry and photometry of galaxies in both the HST and AO imaging. 
Galfit fits source models to the data by minimizing the $\chi^2$ with a Levenberg-Marquardt algorithm,
 where the $\chi^2$ is determined between the image and model under consideration of the associated uncertainties on a per pixel basis.
The uncertainties for the output model parameters are based on diagonalizing and projecting the covariance matrix. 
We use Sersic profiles \citep{Sersic_68} as models for the galaxies. Even though this is a simplistic model, it is advantageous over more complicated models that are not point-symmetric and have less well-defined centers. In more complicated galaxy models, the galaxy center variations with wavelength might be even more problematic.
We use a preliminary version of the inverse distortion correction (Section~\ref{sec:distortion}) to obtain starting centroids for Galfit and mask out bad pixels. 

In a few cases, 
a multi-component (bulge and disc) Sersic improves the fit in the $K'$-band from visual inspection of the residual image and we fit also two Sersic components to the HST images. 
We force the two components to have the same center, but all other parameters are free in the fit.
 
Of the 52 galaxies in our initial list, we exclude one galaxy (number 27) due to its very peculiar shape (to be discussed further in Section~\ref{sec:absvel}).  
We then fit the other galaxies for all images and for all bands where they are present. 
Stars (as PSFs) or secondary galaxies (as 1 or 2 component Sersics) within the fitting window (a box 61 pixels on a side for HST) are also fit to take into account their influence on the primary galaxy. 
Diffraction spikes are fit as Sersics, albeit the resulting fit is very elongated.
The fit fails in a few cases to obtain a reasonable result; 
reasons for that are (i) the galaxy is invisible in the image, (ii) the galaxy is too faint compared to other neighboring sources or (iii) image artifacts (usually diffraction spikes). 
In general, fainter galaxies have more problems and fits fail more often; this is particularly problematic in $F606W$ since our $K'$-band selected galaxies are typically fainter in that band. Since faint sources are not constraining astrometrically, that is not a problem.

\subsection{Differential Chromatic Refraction} \label{sec:dcr}

The atmosphere of the earth refracts light: \begin{equation} \alpha=\alpha' \tan(\zeta) 
\label{eq:zeta}
\end{equation} Therein $\zeta$ is the angle from zenith and $\alpha'$ is the deflection at $45^\circ$. Most of the refraction is corrected for automatically in any linear transformation, see Section~\ref{sec:distortion} and \citet{Fritz_09}, such that it does not impact relative astrometry. It is not possible this way to correct refraction which depends on the color of the source, differential chromatic refraction (DCR). The refraction depends in the following way on $\lambda$:
 \begin{equation}
 \alpha'=\frac{n(\lambda)^2-1}{2\,n(\lambda)^2}.
 \end{equation}
Specifically, the effective wavelength of the source within the used band sets the refraction. While DCR is large in the optical \citep{Kaczmarczik_09,Fritz_15}, it is much smaller in the infrared \citep{Fritz_09}. Of the near infrared bands it is smallest in the K-band \citep{Trippe_10,Fritz_16}. This is one of the reasons why we use K'-band observations. Since essentially all stars (especially the relatively blue Pyxis stars) are in the Rayleigh-Jeans tail in the K'-band the DCR effect between the stars in our observations is less than 0.2 mas even at $\zeta=45\degree$ \citep{Fritz_09,Fritz_16}. 

In contrast, galaxies have a longer effective wavelength because they are redshifted. We use the catalog of \citet{Galametz_13} to estimate the observed $H-K_s$ color of our reference galaxies from the $K'$ magnitudes.
We obtain that the mean
color is $H-K_s=0.79$ (Vega) mag, and that the variation with magnitude is weak.  From photometric redshift catalogs, like that of \citet{Ilbert_09}, it follows from their magnitudes that our galaxies have a redshift of about z$=0.9$.
Using redshifted spectra of galaxies as in \citet{Fritz_16}, we obtain that the $H-K'$ 
 color of such redshifted galaxies is consistent with the colors in \citet{Galametz_13}. We then use redshifted galaxies like in \citet{Fritz_16} to obtain that the typical DCR shift between blue stars and galaxies is about 0.5 mas at $\zeta=45\degree$. 
This shift is probably overestimated, because the brighter, astrometrically more important galaxies, are slightly bluer and thus more similar to stars. 
Pyxis was observed with $\zeta$ between 7.2$\degree$ and 20.6$\degree$. Since DCR scales as total atmospheric refraction, it follows from Equation~\ref{eq:zeta} that the the DCR difference is at most 0.17 mas or 0.03 mas yr$^{-1}$. On average over the 30 images DCR is even smaller, it is 0.01 mas/yr in Dec. and even smaller in R.A.
This is negligible compared to our other errors, especially the reference frame (to be discussed in Section~\ref{sec:absvel}), 
so we did not attempt to correct it.

\subsection{Summary} \label{sec:catalogsummary}

We derive stellar photometry and astrometry from HST and AO imaging; for the former we use standard routines and for the latter we develop custom PSFs. 
We evaluate individual sources via visual inspection to classify spurious sources (e.g., cosmic ray hits), stellar sources, and non-point sources; the latter are quantitatively evaluated using Galfit and determined to be either unresolved binaries or galaxies.
We use 2-d Sersic models to determine the true photocenters of galaxies in both the HST and AO imaging. 
We use a preliminary distortion correction to refine the locations of
the sources for matching across catalogs and evaluate the impact of
DCR. At the conclusion of this process, we have velocity
  measurements for 349 stars and 32 galaxies.  That are less objects than detected in K'-band (Section~\ref{sec:sources}) because some are outside of the HST field of view or could not be fit 
  on the HST images due to faintness, complex source shape, or a too bright neighboring source or diffraction spike.

\section{Deriving the proper motion of the Pyxis cluster}
\label{sec:deriv_pm}

We describe here how we derive the proper motion of Pyxis using HST and GSAOI data.
For good proper motions a good distortion correction is necessary. Derivation of the distortion correction begins with classifying the stars as Pyxis members and non-members. The classification of stars in this way is an iterative process using both photometry and astrometry (Section~\ref{sec:selection}). 
This analysis starts with an isochrone analysis. That classification is then used in the preliminary determination of the distortion and preliminary relative proper motions (Section~\ref{sec:selection}). This then leads to a refinement of the membership classification using the preliminary proper motions. This cycle is then repeated until the membership uncertainty found is not an important source of proper motion error. The distortion correction is explained in detail in Section~\ref{sec:distortion}. 
This Section concludes with deriving position uncertainties and some additional checks for proper motion systematics.
Then, we determine the final relative proper motions for the stellar sources, including a full evaluation of the astrometric reference frame (Section \ref{sec:relvel}).
We summarize our error budget in Section \ref{sec:errorbudget} before deriving the final absolute proper motion in Section \ref{sec:motionfinal}.

\subsection{Pyxis Membership Determination} \label{sec:selection}

Our observations contain both Pyxis stars and unassociated field stars. The latter are usually foreground stars in the Galactic disk, because there are few stars around the distance of Pyxis in the halo. The target star selection is important; firstly, because only Pyxis stars should be used for calculating the motion, and secondly, whether stars are members or not is relevant for how they are used in the distortion derivation (Section~\ref{sec:distortion}). 
Our selection is an iterative process using photometry and astrometry. 

\begin{figure}
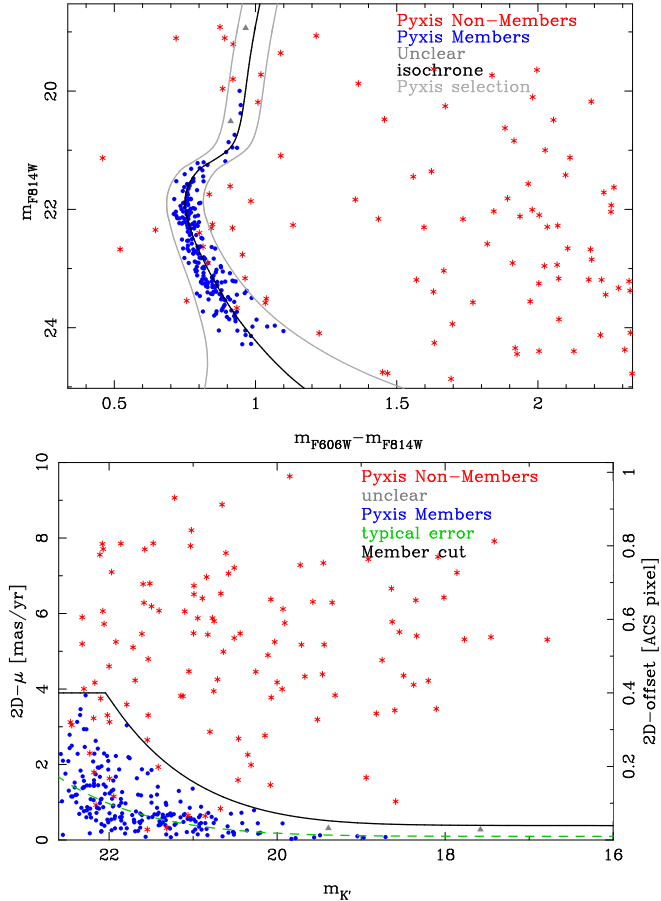

 \begin{center}
   \includegraphics[width=0.70 \columnwidth,angle=-90]{Figure6a.eps}
  \includegraphics[width=0.70 \columnwidth,angle=-90]{Figure6b.eps}
 \caption{Final selection of Pyxis stars, using photometry and astrometry; stars need to fulfill both criteria to be identified as Pyxis Members (blue dots), otherwise they are non-members (red stars). Two are unclear (gray triangles). Top: Color magnitude diagram. The range of color is restricted on the red side, to make the plot in the Pyxis region clearer. The Pyxis isochrone (black) is from the Dartmouth stellar evolution database using the determination by \citet{Dotter_11}. The gray lines show the selection box. Bottom: 2D proper motion/2D offset compared to the mean Pyxis motion. The dashed green line shows the typical error as a function of magnitude. The solid black line shows our selection criterion, stars above it are excluded from the sample. 
 } 
 \label{fig:sel1}
 \end{center}
 \end{figure}

We start with photometry. We use the optical HST photometry, because it has higher SNR for the rather blue Pyxis stars. 
To select members we use the best fitting isochrone from \citet{Dotter_11}.
 We obtain this isochrone from the Dartmouth stellar evolution database \citep{Dotter_08} which was also used by \citet{Dotter_11}. 
 We determine by hand which offset needs to be added to the isochrone so that it matches the observed Pyxis star sequence. Since the majority of the blue stars are Pyxis members (see Figure~\ref{fig:sel1}), the details do not matter much for Pyxis star selection.
 We obtain offsets of 18.859 magnitudes in $F606W$ and 18.525 magnitudes in $F814W$.
This procedure corrects for distance, extinction  and imperfect zeropoints. 
To select Pyxis stars we shift the isochrone slightly. 
The shift (0.062 mag at bright magnitudes and more at the faint end, see Figure~\ref{fig:sel1}) is chosen such that the box contains nearly all stars in the Pyxis sequence. 

We then use this first sample for the first run of the distortion correction, which uses only Pyxis stars, see Section~\ref{sec:distortion}. 
However, we do not use stars brighter than m$_\mathrm{F814W}=20.7$ in the first iteration because bright stars, which are not Pyxis members, can bias the distortion correction severely. We use then this preliminary distortion correction to calculate the preliminary relative proper motions, using the median position over all detections for the K'-band positions, and for HST positions, the average of all detections. The error of the proper motions is dominated by the scatter over the K'-detections. The error contribution from HST is relatively minor. In two iterations we then exclude stars whose motions diverge by more then $4\sigma$ or 0.4 pixels $=$ 3.9 mas yr$^{-1}$ from the Pyxis proper motion. 
That error is dominated by $K'$-band SNR, therefore it is a function of $K$-band magnitude, see Figure~\ref{fig:sel1}.
Four stars fainter than the limit are excluded with this cut. Of the stars brighter than this limit, three are clearly not members, two are clearly members, and two are borderline cases. We include them in our primary sample but we also check how the proper motion changes when we exclude them. The primary sample consists of 220 Pyxis stars.

As another test of whether our motion is sensitive to the Pyxis selection, we widen the selection box by a factor of three. That adds 18 stars, but 7 of them are astrometrically not Pyxis members. Thus, this variant only includes 11 additional stars, all of which are faint. The impact of including these stars is smaller than of using the two bright stars, because these stars are of lower weight in the motion due to their SNR.

Finally we calculate whether the selection of Pyxis stars impacts the velocity of Pyxis. 
Therefore we repeat the calculation in Section~\ref{sec:absvel} for the different Pyxis star samples. 
We obtain that the uncertainty in the selection of Pyxis stars adds an error of 0.05 mas yr$^{-1}$.

\begin{figure}
 \begin{center}
   \includegraphics[width=0.70 \columnwidth,angle=-90]{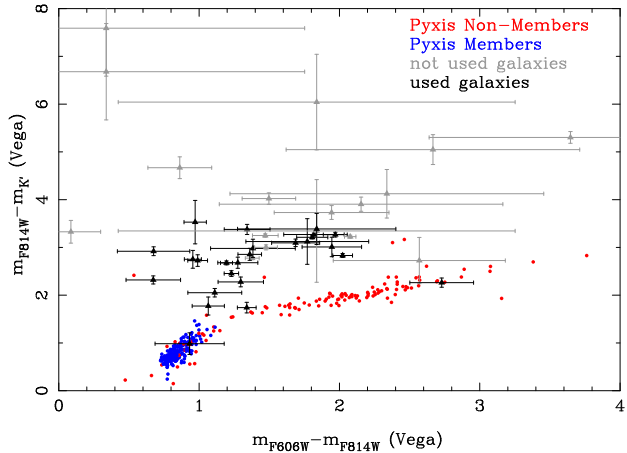}
 \caption{Color-color diagram of the observed sources. It is clearly visible that essentially all morphologically-selected stars (dots, error bars are omitted for clarity) share the same sequence, the stellar locus. Galaxies  (triangles with error bars) form mostly a cloud above this sequence.  The `used galaxies'  show the reference sample against which the proper motion of Pyxis is obtained.
 } 
 \label{fig:sel2}
 \end{center}
 \end{figure}

As an additional test, we show here the color-color diagram of all our sources, not just the stars, see Figure~\ref{fig:sel2}.  We use galaxies as reference objects, because due to tip-tilt star constraints, it is not possible to select a field with a quasar. Also galaxies are less affected by residual distortion, because since we can use several of them, these systematics roughly average out. The galaxies are selected morphologically, see Section~\ref{sec:sources}. For a given $F606W-F814W$ color in Figure~\ref{fig:sel2}, most galaxies are redder than stars in $F814W-K'$. This is expected because galaxies are redshifted compared to the stars. Some galaxies overlap with the stellar sequence, as it is expected for these and similar colors, see e.g. \citet{Galametz_13}. 
 Galaxies that are blue in $F814W-K'$ are all obviously extended in our images, such that it is clear from visual inspection that they are galaxies. In contrast, the more compact galaxies are all outside of the color-color stellar locus, which confirms our visual morphological selection.
 
We check also the locations of our stellar sources in the color-color diagram. The vast majority lie on the stellar locus and are thus obviously stars. There are two sources which are clearly redder in $F814W-K'$ as is expected for QSOs. We check these sources: one of them lies on a strong diffraction spike in the HST images, making its properties unreliable. The second source has a $K'$ = 21.6 mag and an absolute proper motion of 4.7 mas yr$^{-1}$ and we conclude that it is probably a star.  
Since the different optical and NIR imaging were not obtained simultaneously, it is also possibly that it is a variable star. Regardless, with a position uncertainty of about 0.5 mas yr$^{-1}$, the object has a low weight in the proper motion fitting and is not providing any useful constraints.
 
\subsection{Distortion correction, image registration and position errors} \label{sec:distortion}

For distortion correction we start with a similar approach as in \citet{Fritz_15},
 which assumes that one data set is clearly easier to correct for distortion.
For this analysis, HST data is better than the GSAOI data because the distortion of ACS/WFC is 
 well-characterized and stable in time apart from the linear terms \citep{Anderson_06}. 
An STScI-developed code \citep{Anderson_06} was used to perform the geometric distortion correction of the source positions.
This leaves offsets induced by the time dependence to be removed by a separate linear transformations. 
After applying this distortion correction we correct for all linear terms, including skew terms, by
means of fully general 6-parameter transformations into the $F814W$ image, which has the largest overlap with the GSAOI imaging.

The distortion of GSAOI is not known for this study, because it is not stable on long time scales \citep{Neichel_14}, thus we must derive the distortion solution empirically from our images.
Our full distortion correction for GSAOI is a combination of using an external true reference (HST in our case; Section~\ref{sec:hstdistortion}), as well as internal referencing within the GSAOI data itself (Section~\ref{sec:distortion_derivation}).

\subsubsection{External Distortion Correction for Pyxis using HST}
\label{sec:hstdistortion}
Since the HST data was observed at a different epoch, we can only use objects which have not moved relevantly as an external reference. Unfortunately, there are too few bright galaxies for the distortion correction. The non-Pyxis members of Section~\ref{sec:selection} are nearly all Galactic disk stars, which are likely to have larger proper motions than our distortion correction can tolerate. In other work,s often the members of the target satellite are used for distortion correction. The member stars in Pyxis have internal motion and thus move relative to each other. This motion could correspond to an additional error term which needs to be added to the other errors terms when a distortion solution is derived  \citep[for example as in][]{Dalessandro_16,Massari_16}. 
However, the relative importance of the internal motions can be ascertained from the internal velocity dispersion of Pyxis.

Since no measurement of the dispersion of Pyxis exists, we instead use data from other globular clusters to determine scaling relations for this quantity, using data from \citet{Harris_96}. We start with the luminosity, finding that faint globular clusters like Pyxis have  smaller dispersions. In addition, the dispersion depends on the mass concentration. We use as a measure for the size of the globular cluster the half light radius. We obtain that, as expected, $\sigma \propto \sqrt(L/r_\mathrm{half})$ is valid. Using for Pyxis M$_V=-7.0$ and $r_\mathrm{half}=17.7$ pc (see Section~\ref{sec:motionfinal}) we obtain for Pyxis $\sigma=0.97\pm0.16$ km/s, where the error is from the scatter in the relation over the globular clusters with both measurements. The properties of Pyxis are somewhat uncertain, but it is clear that the dispersion of Pyxis is less than 2 km/s, which translates at its distance into a proper motion dispersion of less than 0.01 mas yr$^{-1}$. This is negligible compared to other terms (Section~\ref{sec:distortion_derivation}~and~\ref{sec:errorbudget}.), and thus we ignore it.

Similar to the approach in \citet{Fritz_15}, we combine the determination of the distortion with the determination of the linear terms, which register the image and correct the image scale. 
For the fitting, we use the ${\tt mpfit}$ package \citep{Markwardt_09}.

The linear terms can be described as follows,
\begin{equation}
 \begin{split}
 \mathrm{R.A.} & = c_1+c_2\,x_\mathrm{cor}+c_3\,y_\mathrm{cor} \\
 \mathrm{Dec.} & = d_1+d_2\,x_\mathrm{cor}+d_3\,y_\mathrm{cor}
 \end{split}
  \label{eq:lin_corr}
 \end{equation}
  and are as expected not stable. This is primarily due to airmass variation, but other effects, such as atmospheric turbulence and corrections made by the the AO system, contribute also to modulations of the image scale.

For the distortion correction, we fit both cubic and quadratic order polynomials finding that the $\chi^2/\mathrm{d.o.f.}$ of both are quasi-identical. This suggests that the cubic approach overfits the data and we therefore adopt a quadratic correction of the form:

  \begin{equation}
  \begin {split}
 x_\mathrm{cor} & =a_1+a_2\,x'+a_3\,y'+a_4\,x'^2+a_5\,x'y'+a_6\,y'^2 \\
 y_\mathrm{cor} & =b_1+b_2\,x'+b_3\,y'+b_4\,x'^2+b_5\,x'y'+b_6\,y'^2 \\
 \end{split}
 \label{eq:dist_corr}
 \end{equation}
 
Here the constant and linear terms are used to account for the offsets and scale differences between the four chips. Chip 1 is the reference, thus $a_2$ and $b_3$ are for chip 1 defined to be 1, and $a_1,a_3,b_1$ and $b_2$ are defined to be 0. As usual all distortion parameters are different for the four distinct chips. Equation~\ref{eq:lin_corr} and~\ref{eq:dist_corr} are also used in Section~\ref{sec:distortion_derivation}.
In practice, we apply first the distortion correction (Equation \ref{eq:dist_corr}) to the data and then apply the linear terms (Equation \ref{eq:lin_corr}). 

 We test whether the four detectors should be treated independent of each other for the linear correction (Equation~\ref{eq:lin_corr}), but we find that in this case the errors are larger than when we fit these time-dependent linear terms to all four detectors simultaneously.  The preference for coupled chips is also supported by fact that the $\chi^2$ values are smaller when they are coupled. There are two reasons why fixing these two types of parameters is the better strategy: firstly, because the relative chip orientation is effectively stable. Secondly, because there are relatively few bright sources in the images, it is preferable to fit fewer terms. 

\subsubsection{Tests for Stability over Full Observation Window}
We test whether the distortion is stable over the 2.5 hours over which we observed Pyxis \citep{Fritz_16}. Thereby, we solve for different distortion coefficients when the astrometric loop is closed (i.e., within each six-image set), versus considering all the 30 images, i.e., even when the astrometric loop is not closed.
 This test is done relative to our mosaic since it is then possible to treat all stars, also not Pyxis members, the same way. We find a slight decrease in the error floor in the $x$-dimension from 0.40 mas (when only one distortion solution is derived for all the 30 images) to 0.29 mas (when one distortion solution is derived for only the images in the closed loop) to 0.24 mas (when different distortion solutions are derived for each of the 30 images), and in the $y$-dimension from 0.26 mas to 0.24 mas to 0.22 mas, for the three cases respectively.
Stated differently, we find a relatively small decrease in the error floor of 0.16 mas in $x$ and 0.02 mas in $y$.
While some improvement is expected when allowing these terms to vary
\citep[because the GeMS/GSAOI system is not stable; see discussions in][]{Neichel_14} and/or \citep[because of atmospheric variability; see discussions in][]{Massari_16b}, it is surprising that the error floor decreases when all images are used versus when the astrometric loop is closed since during this time the distortion should be stable. Therefore, the improvement is probably caused by more degrees of freedom. Further, the decrease of the error floor is small when compared to other errors. We therefore assume the distortion solution is stable over the total observation window for the analyses to follow.

\subsubsection{Internal Distortion Correction}
\label{sec:distortion_derivation}

It turns out that the Pyxis member-star sample does not obtain a sufficiently good distortion correction solution, due to the fact that the system is relatively sparse, and we find that when using only this sparse sample, the resulting offset between chips has a non-negligible uncertainty. 
The bulk of the Pyxis member-star sample has lower SNR (i.e., the bulk of the Pyxis stars are faint), while non-member stars are typically brighter. Thus we devise a method to leverage these brighter stars in the distortion solution. Since the non-members move, we cannot use their position on the HST image as reference. We must instead use their positions only within the GSAOI data which was all taken within the same epoch. 
We use the first GSAOI image as reference, because i) it is in the center of all images and thus has the biggest overlap with the other images, and ii) this image has also a relatively high Strehl ratio compared to the other images in the program.
 Since, the first GSAOI image is distorted as well, we also have to apply the same distortion correction to the positions in the first image, as described further below.
 This procedure adds 170 stars that we can use to help bootstrap the distortion terms. Overall, we end up using 390 of the 450 stars for deriving the distortion correction (5 stars are omitted because they are saturated or close to a saturated star in the first image and an additional 55 stars are excluded because they are not on the first image).
 
 We can no longer use {\tt mpfit} for this expanded sample, because for non-Pyxis members there are no true positions which remain unchanged between frames (i.e., the non-Pyxis stars have no known position, since they depend on the distortion which is yet to be determined). We instead develop our own Monte Carlo fitting method. As a starting point we use the distortion solution derived using only Pyxis stars. Because the fitted function contains 222 free parameters, the minimum cannot be reached in a sufficient time if all are allowed to remain free. After some initial iterations, we fix all parameters that affect only one image in the solution; these include all linear parameters to transform each individual GSAOI image to HST, with the exclusion of the parameters of the first GSAOI image, which is used as the `reference' image for all non-Pyxis stars. It is expected that the influence of these parameters is reduced by $\sqrt{N}$ since $N$ (up to 29) different images are used to calculate the final positions. After fixing these frame-specific parameters, the 48 parameters are left to vary in the fit, which increases the speed of the procedure. 
 
As a result of our custom Monte Carlo fitting procedure, the uncertainties for the fitted parameters in the distortion solution (e.g., the chip separations) are now small compared to the positional errors of the galaxies used to define the absolute frame. 
 The impact of imperfect linear terms in the distortion solution (those held partly fixed in the fitting process) is included de facto for galaxy positions, because the total uncertainty in the position of an individual source is derived from its frame-to-frame scatter.
The two PSF options (PSF1 and PSF2) obtain consistent proper motions.

\subsubsection{Position Errors}
Here we describe more fully the initial set-up and rejection criteria used for the Monte Carlo method above, and then give an estimate of the typical position errors that we achieve after distortion correction. For an initial set of input errors for the stars, we use the Galfit positional uncertainties.
In the first iteration, we identify the Pyxis members (Section~\ref{sec:selection}).
 Since the Pyxis membership of a few stars is uncertain, we use two different samples, which include the uncertain stars or not.
We also test positions obtained with both of our model PSFs. 
We then proceed with each possible combination of PSF and Pyxis membership sample.

We first clean each sample of outliers based on a comparison to its median position, defining a cutoff criterion as follows:
\begin{equation}
R_i=\sqrt{\Delta R_x,i^2+\Delta R_y,i^2} <\eta\,\sigma_i \label{eq:rej_criteria},
\end{equation} 
where $\Delta R_x,i$ and $\Delta R_y,i$ are the differences between the median position of the source and its position in the $i$-th image, $\sigma_i$ is the positional scatter of this star 
 and $\eta$ is a scaling factor that is changed iteratively. 
The behavior of the solution is monitored with stricter cuts on $\eta$ and, in our final iteration, we use $\eta=5$. Then we measure the scatter over all detections of a star (i.e., in all the images for which it has a good detection). 
We also calculate $1.483\times$ the median deviation of the stellar positions as a robust measure of the positional scatter, and adopt a ratio between the scatter and this median to adjust the final uncertainty for each source. However, sometimes low number statistics can cause non-physically small scatter values, and to safeguard against this, we also keep track of the typical error expected for stars of a given magnitude (as coded basically by their SNR and quality of fit to the PSF), and use this typical value if it is higher than the scatter.

\begin{figure}
 \begin{center}
\includegraphics[width=0.70 \columnwidth,angle=-90]{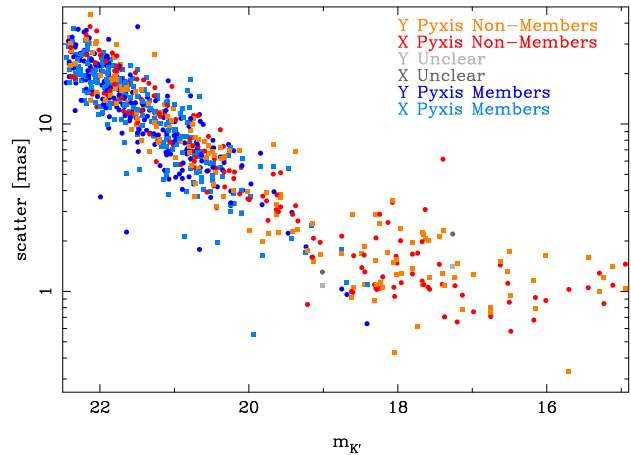} 
 \caption{GSAOI position scatter of the stars used in the transformation. Dots stand for the scatter in X (the X of the HST image), boxes for the scatter in Y. We use $1.483\cdot$ the median deviation as a robust scatter measure. The scatter is measured after transformation into the HST reference frame, thus it contains also a contribution from residual distortion. Only the scatter for PSF1 is shown, but the difference between PSF1 and PSF2 is small.
 } 
 \label{fig:residuals1}
\end{center}
\end{figure}

We modify the process slightly for non-Pyxis stars. Already in our first iteration we include an error floor estimate, because for some bright stars the uncertainty scaled by the SNR is much smaller than realistic errors.
In a second iteration, we then exclude outliers with $R_{i} > 5\sigma_i$ and again adjust uncertainty using the scatter as before. 
To avoid any single star having a very large weight in the final fit, we set a minimum uncertainty of 0.5 mas. This is somewhat smaller than our error floor ($\approx$ 1 mas) measured after the final iterations, see Figure~\ref{fig:residuals1}. This floor is mainly caused by residual distortion, and its level is consistent with other measurements for GSAOI, see \citet{Neichel_14}.

We now present the typical positional errors we obtain after the distortion correction process is done. In subsequent analysis, we use the median position of a star in the GSAOI images, after application of the distortion correction, as the star's final position for the 
GSAOI epoch. 
The uncertainty in this position is defined as the robust scatter of the set of frame positions divided by the square root of the number of images in which the source appears. 
Because the distortion residuals probably do not average out over the small (1$\arcsec$) dither between individual frames, but do average out over the large (7$\arcsec$) dither between image sets, we add an additional error term of 1 mas (the error floor) divided by the square root of the number of image sets in which the source appears (maximum of 5). 
This estimate is conservative since there are fewer large dither steps (5) than there are images (30).
In the best case, bright stars detected in 5 images sets, we obtain a precision of 0.4 mas in the star positions and of 0.08 mas yr$^{-1}$ in the proper motions. 
The final uncertainties for star positions only affect the relative motions;
for the absolute motions, the error contribution of the reference galaxies is the dominant term.

\subsubsection{Comparison to Other GSAOI Distortion Solutions}
\begin{figure}
 \begin{center}
\includegraphics[width=0.99 \columnwidth,angle=-90]{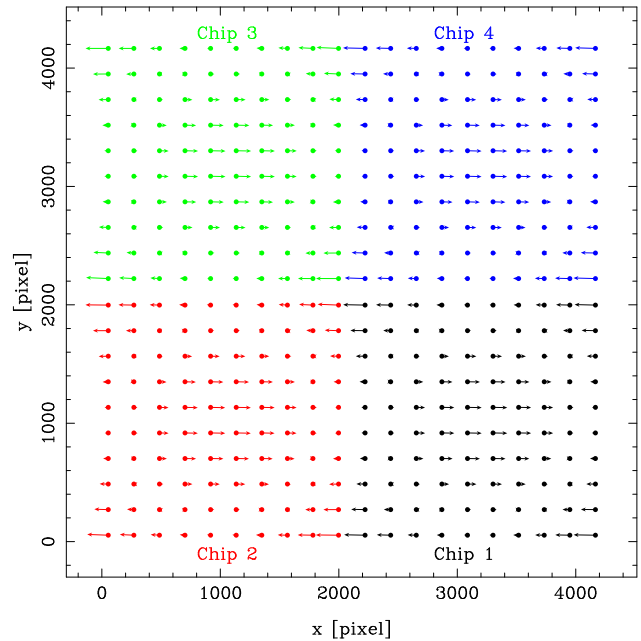} 
\caption{Distortion map of GSAOI derived from PSF1 data (but the difference is very small between the two PSFs). The vectors showing the distortion field are magnified by a factor of 25. Not shown are linear scale differences and tilts between the different chips. Also the chips do not exactly have the shown positions. Further, a linear transformation which sets the average shift of each chip to zero is applied for easy comparison with the literature.
 } 
 \label{fig:dis_map}
\end{center}
\end{figure}

We now analyze and compare our distortion field with those previously found in the literature. The maximum distortion vector of our distortion field is 12.9 pixel, which is 259 mas. To ease comparison with most literature, we set the average shift in our distortion map to zero by a linear transformation in the same way as \citet{Ammons_16}.
We show the derived distortion map in Figure~\ref{fig:dis_map}. The standard deviation over the distortion field is 63 mas in x and 1.8 mas in y. A stronger distortion in x and a similar distortion field was also observed by \citet{Ammons_16} and \citet{Massari_16} for GSAOI.

We compare our distortion field quantitatively with \citet{Ammons_16}\footnote{For quantitative use of the \citet{Ammons_16} field  
we use their distortion parameters directly, which were kindly provided to us.} and \citet{Dalessandro_16} (we applied on their field a linear transformation in the same way as to our data). After the transformation, the distortion field of \citet{Dalessandro_16} is similar to the others and shows much larger shifts in x.
We calculate the difference in the distortion field between us and the Ammons/Dalessandro field. This difference has similar distortion strengths in x and y, the scatter is in both about 4 mas for \citet{Ammons_16} and 3.5 mas for \citet{Dalessandro_16}.
This difference is much smaller than the distortion in x, but it is clearly bigger than the errors in our and the other distortion determinations. The data used for the \citet{Ammons_16} distortion determination was obtained at the end of 2012, and the data of the \citet{Dalessandro_16} distortion determination was obtained around May 2013, both well before our observations. Thus, it is not surprising that the distortion fields differ, as it is known that the distortion of GSAOI is not stable on long time scales \citep{Neichel_14}. The distortion also depends to some extent on the guide-star asterism used.

\begin{figure*}
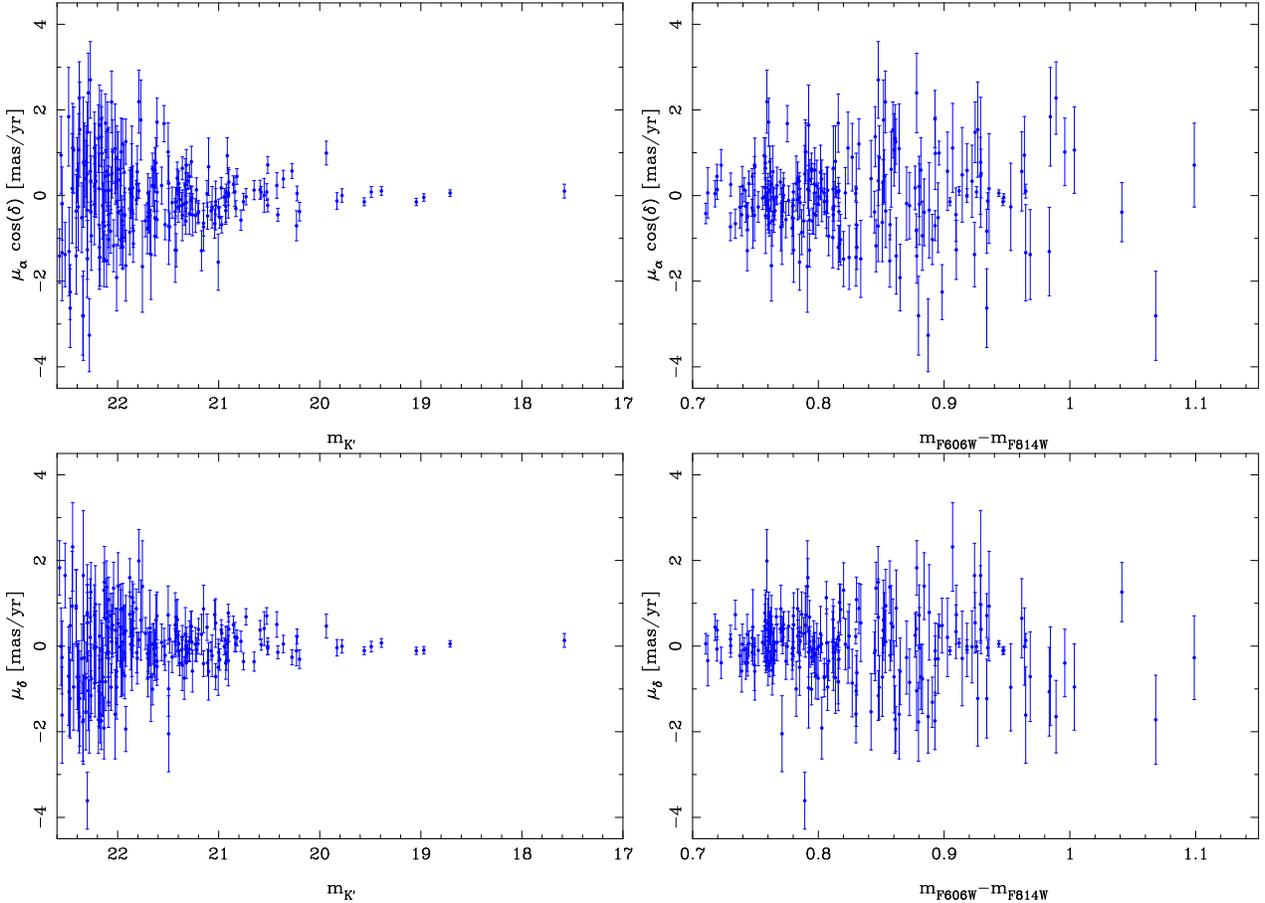

\begin{center}
\includegraphics[width=0.70 \columnwidth,angle=-90]{Figure10a.eps}
\includegraphics[width=0.70 \columnwidth,angle=-90]{Figure10b.eps}
\includegraphics[width=0.70 \columnwidth,angle=-90]{Figure10c.eps}
\includegraphics[width=0.70 \columnwidth,angle=-90]{Figure10d.eps}
\caption{Relatively proper motion of the Pyxis stars as function of K'-magnitude (left) and $F606W$-$F814W$ color (right).}
 \label{fig:col_trend}
\end{center}
\end{figure*}

As additional checks for systematics we test, similarly to \citet{Bellini_14,Dinescu_16}, whether the proper motion varies as a function of magnitude and color, see Figure~\ref{fig:col_trend}. We use only Pyxis stars since the other stars have large peculiar motions such that this would hide any trend. It is clear that there is no trend as a function of magnitude or color.

\subsection{Relative proper motions of stars} \label{sec:relvel}

The positions and motions are first measured in pixel space. To transform them into the world coordinate system (WCS), we again utilize the 
VISTA-VHS data \citep{McMahon13} obtained in April 2012, which is astrometrically tied to 2MASS. As before, we only use stars that are isolated in the VHS data. A total of 23 stars are common to the VISTA-VHS observations and our HST footprint. 
We perform a fit to determine a linear transformation between final pixel positions and the VISTA-VHS WCS.
The scatter between transformed HST positions and the VHS positions is 26/25 mas in R.A./Dec\footnote{Here, as always, the position/velocity in R.A. is multiplied by $\cos(\mathrm{Dec.})$}.  
The positional uncertainty in the VISTA-VHS WCS must also be considered and this is about 100 mas\footnote{For details see: {\url https://www.eso.org/sci/observing/phase3/ data\_releases/vhs\_dr2.pdf}}.
The precise error in the WCS is not important for us since we are not interested in absolute astrometry, only in absolute motions.

\begin{figure}
\begin{center}
\includegraphics[width=0.70 \columnwidth,angle=-90]{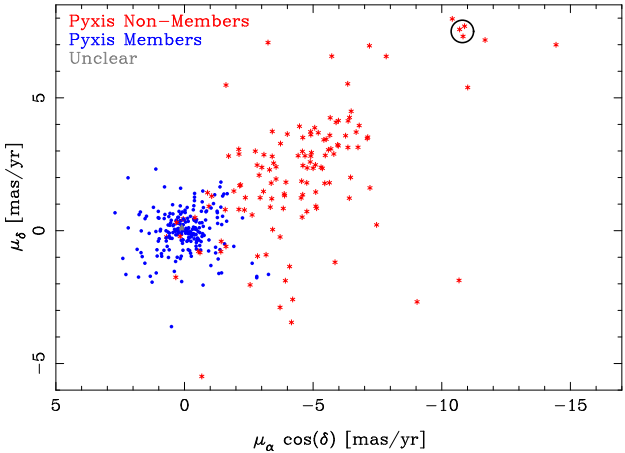} 
\caption{Motion of the stars relative to Pyxis. Pyxis members are shown as blue dots, non-members are shown as red stars, the two with unclear membership as gray triangles. For clarity the errors are not shown. The size of the per-star error is between 0.074 and 1.50 mas yr$^{-1}$, with a median of 0.44 mas yr$^{-1}$. The potential group of three stars is within the black circle. The result shows PSF1, but the differences between the two PSFs are small.
} 
\label{fig:mot_stars}
\end{center}
\end{figure}

We compute the average motion of the 23 VISTA-VHS stars relative to Pyxis member stars finding it to be $\sim$16 mas.
This difference does not impact our proper motion because we use the same WCS transformation on both the HST and GSAOI epochs.
The position uncertainty in the VISTA-VHS frame causes an uncertainty in the overall image scale, which is a systematic uncertainty on the final proper motion. 
Since the sources used for the WCS transformation extend over 96/56$\arcsec$ in R.A./Dec., the image scale uncertainty is at most 0.11/0.17\% using the 100 mas error in the VISTA-VHS WCS. 
Without accounting for this error term, the motion uncertainty is 0.86\% of the motion
 for the stars with the smallest fractional motion uncertainty. 
Thus, even in the most extreme case, the image scale error is smaller than the other error components (even more so for the absolute proper motion, see Section~\ref{sec:absvel}). 

As test we calculate the error weighted average motion of the Pyxis members. We obtain \\ $\mu_{\alpha} \cos{\delta}=-0.04\pm0.02$ mas yr$^{-1}$ and $\mu_{\delta}=0.01\pm0.02$ mas yr$^{-1}$. Given that Pyxis is the reference the motion should be exactly zero. It is not exactly zero because different stars have different weights in the distortion fit and here.
However, the motion is much smaller than the measured absolute proper motion of Pyxis (Section~\ref{sec:motionfinal}).
In Figure~\ref{fig:mot_stars} we show the relative motions of the stars. It is visible that most non-member stars show offsets in the same direction. That is expected, because the main foreground population is in the galactic disk. All velocities are shown in Appendix~\ref{sec:all_prom_mot}.

\subsubsection{Potential Moving Group}

\begin{table*}
\centering
\caption{Potential group members} \label{tab:stargroup}
\begin{tabular}{c c c c c c c c c}
 \hline \hline
Id & R.A.$^{a}$ & Dec.$^{a}$ & $\delta_X$ $^{b}$  & $\delta_Y$ $^{b}$ & $\mu_\alpha$  & $\mu_\delta$ &  m$_\mathrm{K'}$ & $F814W-K'$ \\
         & [$^{\circ}$]  & [$^{\circ}$]       & [arcsec]& [arcsec] & [mas/yr]         &  [mas/yr]          &         &    \\
 \hline 
33    & 137.0073024 &    -37.2081552 & 0 & 0      &  -10.69$\pm0.11$        &  7.51$\pm$0.11  & 15.49 &  2.03  \\
37    & 137.0071044 &   -37.2080924  & -0.568 & 0.226   &  -10.81$\pm0.10$        &  7.62$\pm$0.10  & 15.70 &  2.11  \\
84    & 136.9924541 &   -37.2060942  & -42.574 & 7.419   &  -10.85$\pm0.09$        &  7.28$\pm$0.09  & 18.01 &  1.70  \\
397    & 137.0055077 & -37.1883926   &-5.145 & 71.145     &  -10.41$\pm0.08$        &  7.97$\pm$0.08  & 17.37 &  1.71  \\
405    & 137.0104143  &  -37.1877676  & 8.925 & 73.395     &  -11.68$\pm0.12$        &  7.17$\pm$0.12  & 17.44 &  1.75  \\
\hline \hline
\multicolumn{9}{l}{$^{a}$ The uncertainty of the positions are about 7$\times10^{-6}$ degree compared to the absolute reference frame.} \\
\multicolumn{9}{l}{This does not affect the relative positions.} \\
\multicolumn{9}{l}{$^{b}$ Distance from star 33.} \\
\end{tabular}
\end{table*}

 There is a cluster of three to five stars with proper motions of $\sim -11$ mas yr$^{-1}$ in R.A. and $\sim7$ mas yr$^{-1}$ in Dec., visible in Figure~\ref{fig:mot_stars}. The properties of the stars are listed in Table~\ref{tab:stargroup}. Two of the stars (397 and 405) have motions which differ from the other three by more than the measured errors, and therefore are likely only a chance association, while the other three are consistent within 1$\sigma$/2$\sigma$ in R.A./Dec. Two of the remaining stars (33 and 37) are close neighbors (distance 0.66$\arcsec$), have consistent colors, and are bright; there are only 12 stars brighter over the full 9050 square arcsecond GSAOI footprint of our observations. Thus, these two stars are very likely associated and probably form a binary. If these stars are giants, i.e., luminous and distant, their high proper motions would imply high velocities, and thus that they are unbound from the Milky Way. Therefore, they are more likely nearby dwarfs. Using a solar metallicity isochrone from \citet{Dotter_08} and assuming that the stars have the same extinction as Pyxis, we obtain a distance of about 1000 pc and mass of about 0.5 M$_\odot$ each. That would imply a projected separation of 660 AU within the binary. The remaining star, 84, is fainter and slightly bluer. However, a co-evolved fainter dwarf star should be redder, instead of having about the same color as the other two bright stars. It is important to note that as a giant, this star would be also unbound. Photometric parallax predicts a distance of about 4000 pc and a similar mass compared to the binary stars. Thus, star 84 cannot belong to the same physical association as stars 33 and 37 and probably only has a nearly identical proper motion by chance, similar to stars 397 and 405.

\subsection{Reference frame systematics and random errors}
\label{sec:absvel}

\begin{figure}
\begin{center}
\includegraphics[width=0.99 \columnwidth]{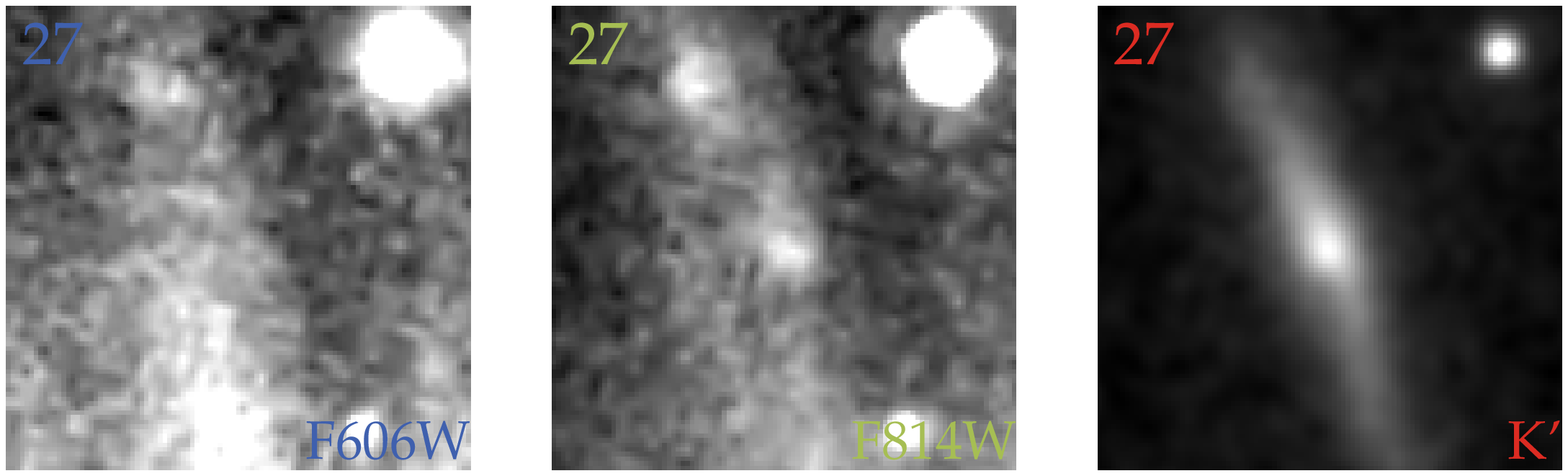} 
\includegraphics[width=0.99 \columnwidth]{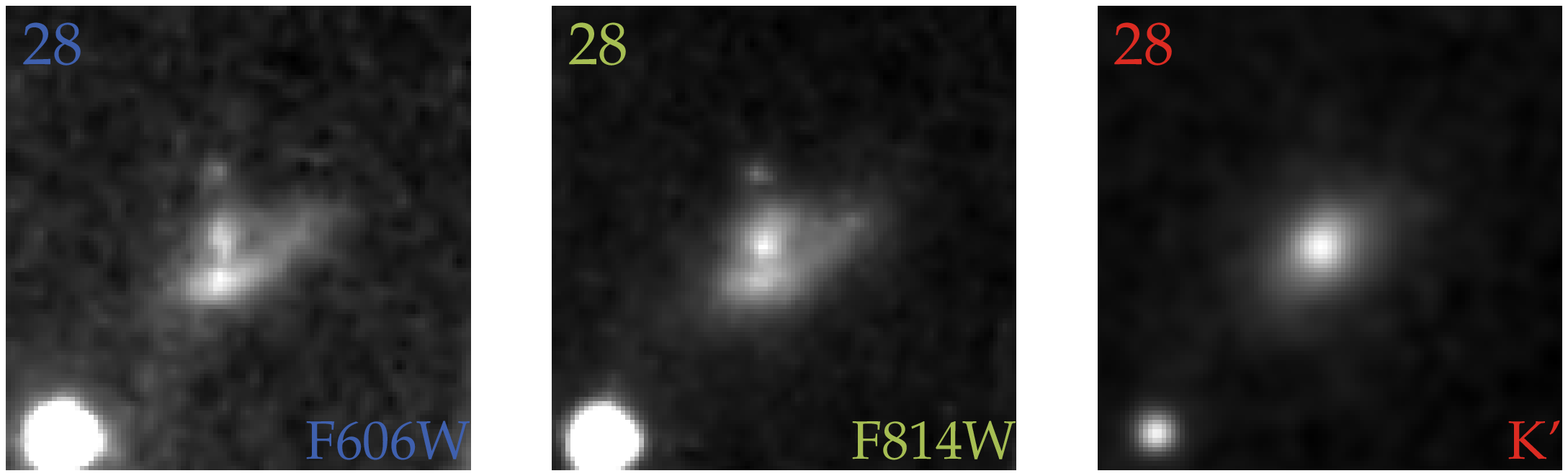} 
\includegraphics[width=0.99 \columnwidth]{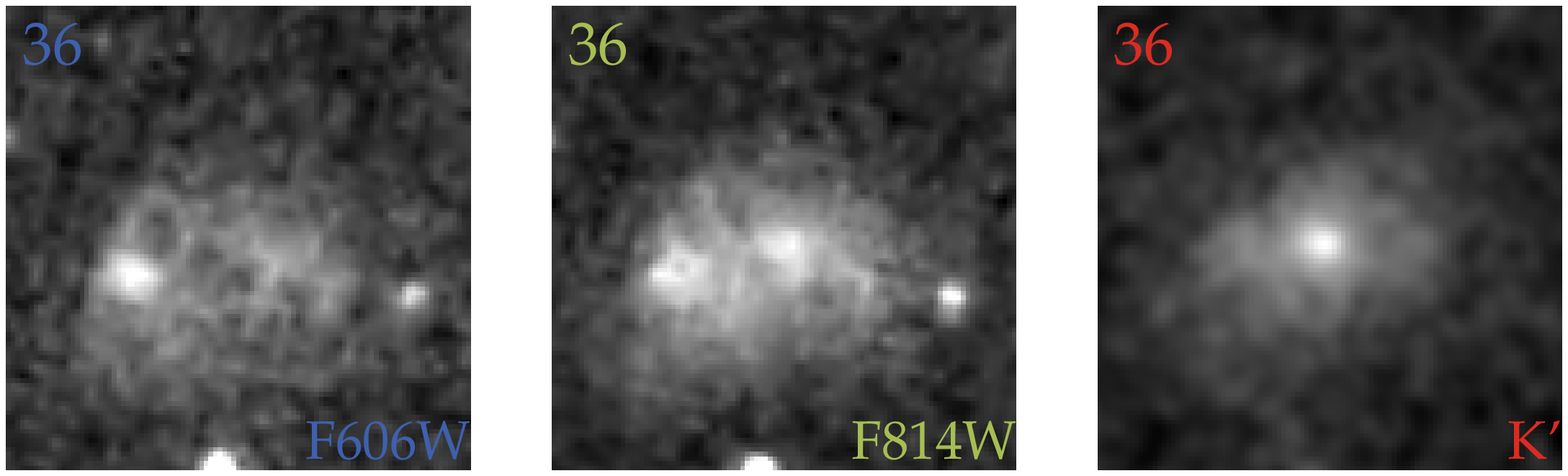} 
\includegraphics[width=0.99 \columnwidth]{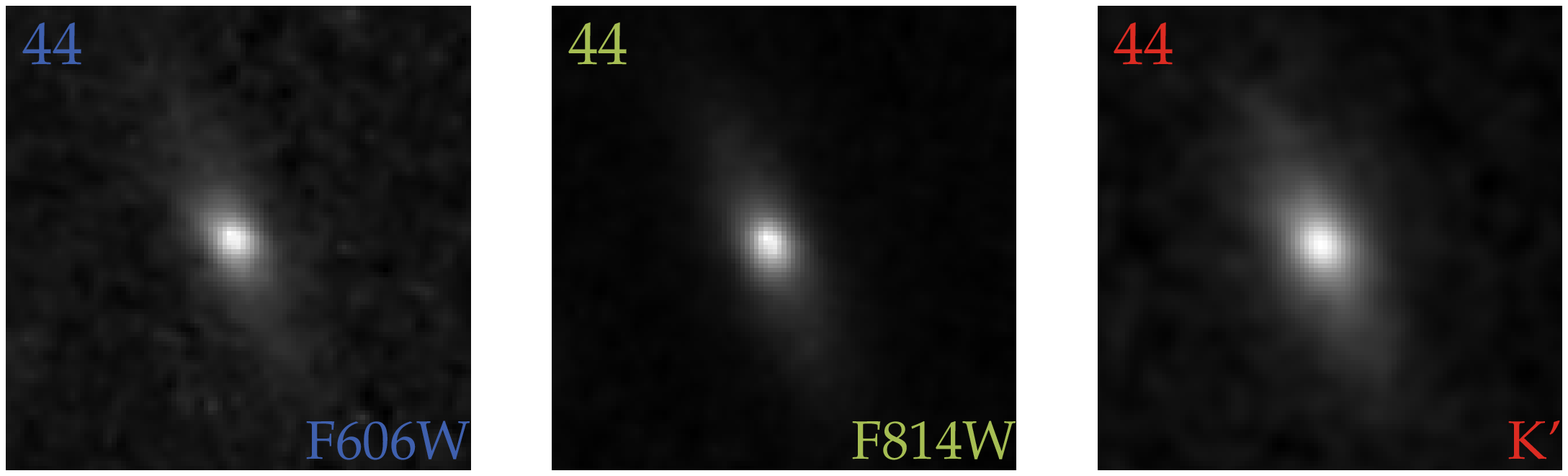} 
\caption{Example of galaxies. The top three show three astrometrically-bad galaxies. Relatively few galaxies are astrometrically bad. The bottom-most galaxy 44 is astrometrically good as the majority of the galaxies.
} 
\label{fig:gal_images}
\end{center}
\end{figure}

For each galaxy we calculate its position in the transformed reference frame. 
We therefore use the average of the positions. 
Following Equation \ref{eq:rej_criteria}, we exclude those positions ($i$) which are more than 5$\sigma$ offset  from the median position (as for stars we use the median deviation here).
The uncertainty is obtained by calculating the standard deviation and dividing it by $\sqrt{N}$ or the number of used positions. All individual positions are given the same weight.

In contrast to most high-resolution studies which use galaxies as references \citep{Sohn_12}, our images are not observed at the same wavelength in the two epochs. 
This is a potential problem, as galaxies are extended, and colors can vary intra-galaxy due to extinction and stellar-population effects. Therefore, the position of the photocenter of a galaxy can also vary across different band-passes.  Here we describe our methods of both characterizing and minimizing the impact of the wavelength-dependent galaxy photocenter. 

As a first check, 
we compare the shapes of the galaxy in the drizzled $F606W$ and $F814W$ images and the $K'$-band mosaic. 
One galaxy, number 27 (see Figure~\ref{fig:gal_images}), an edge-on disk, looks very different in the HST images: it is nearly invisible in $F606W$ and consists of two clearly-separated subclumps in $F814W$. Thus, given that its profile is already significantly different in the relatively closely-separated HST bandpasses, we exclude it from the galaxy sample.

There are other sources that despite having successful fits across all filters, have a photocenter that is clearly drifting as a function of wavelength. 
This is obvious for galaxies 28 and 36 (see Figure~\ref{fig:gal_images}), which seem to consist of two components whose relative contributions vary with wavelength. 
To further refine the sample of galaxies we look at the offset in the center of each galaxy in $K'$-band relative to the median offset 
for the full sample in both ($K'$-$F606W$) and ($K'$-$F814W$). 
We find that when $F606W$ is used as the reference for this median the random errors are larger and there are more outliers. 
That is not surprising, because due to $F606W$'s larger difference in wavelength from $K'$-band, it is to be expected that the center varies more than compared to $F814W$. Therefore, we use only $F814W$ positions from HST. 

 \begin{figure}
 \begin{center}
   \includegraphics[width=0.70 \columnwidth,angle=-90]{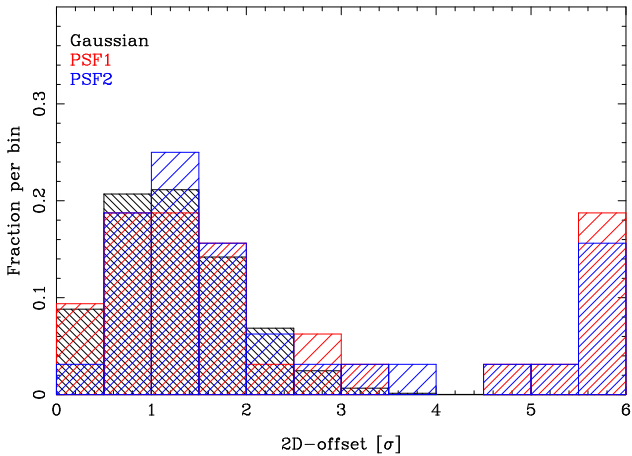}
 \caption{Histograms of 2D-position offsets for galaxies with the two PSFs and Gaussian error. The different histograms are scaled in x and y to be roughly matching for 2D-offset$<2$. It is visible that outliers with more than 3.5 $\sigma$ are unlikely for a Gaussian distribution. We therefore exclude them as outliers.}
 \label{fig:offset_2d}
 \end{center}
 \end{figure}

\begin{figure}
\begin{center}
\includegraphics[width=0.70 \columnwidth,angle=-90]{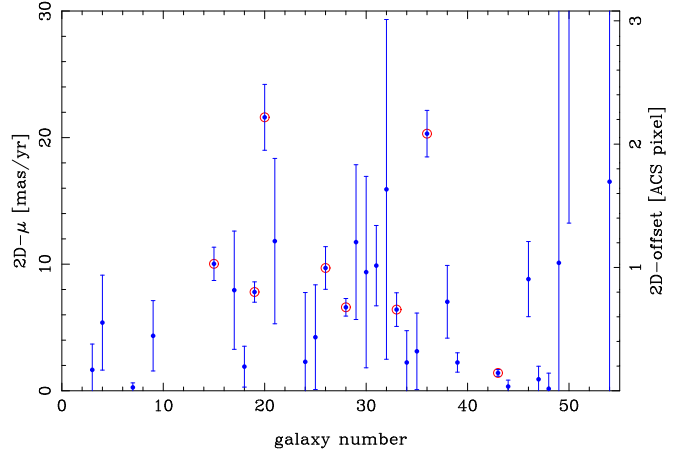} 
\caption{Velocities of all $K'$-band galaxies with a measurable $F814W$ HST counterpart. Error bars are 1-$\sigma$.
Red encircled galaxies are excluded from the sample because they are outliers by more than 3.5-$\sigma$.
The velocity shown on the $Y$-axis is measured relative to the mean error-weighted velocity of all good galaxies.
} 
\label{fig:offsets_r}
\end{center}
\end{figure}

To determine which galaxies are affected by color effects, we make histograms using again Equation \ref{eq:rej_criteria}, to calculate offsets in $\sigma$. First, the offsets are calculated relative to the median offset, which is later refined to the error-weighted median which is similar. We also multiply the errors by a factor to obtain that the histogram has a similar shape to a Gaussian for offsets $R<2$, see Figure~\ref{fig:offset_2d}. That is reached by using a factor of 1.2. It is visible that there are outliers. We define all with $R>3.5$ as outliers. For our sample size a larger offset occurs only with 5\% probability. That has also the advantage that the same galaxies are outliers for PSF1 and PSF2. We exclude that way 8 of 32 galaxies. Among these are also the previously discussed galaxies 28 and 36. 
In Figure~\ref{fig:offsets_r}, we show proper motion relative to the mean error-weighted proper motion for all good galaxies.

The fact that we need to scale up the errors to match the main histograms shows that all errors are slightly underestimated. Likely, that is due to a weak influence of color effects on all galaxies.
To correct for it we calculate the reduced $\chi^2$ over the good galaxies. We present statistics for PSF1 only, but using PSF2 obtains similar numbers. The reduced $\chi^2$ is 1.98/1.80 in R.A./Dec. 
 We then scale our errors such that the reduced $\chi^2$ is 1 in both $x$ and $y$. 
After rescaling of the reduced $\chi^2$ we obtain errors of 0.296/0.267 mas yr$^{-1}$ in R.A./Dec.

 \begin{figure}
\begin{center}
\includegraphics[width=0.70 \columnwidth,angle=-90]{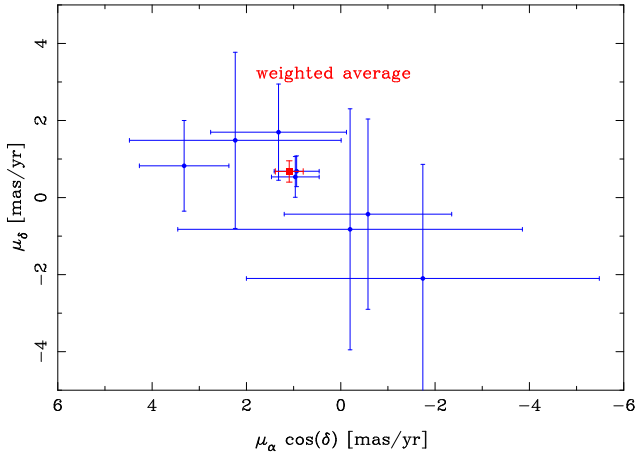} 
\caption{Absolute velocity of Pyxis. Each blue dot stands for a velocity derived from a single background galaxy. The red box shows the weighted average velocity derived from all galaxies. Only galaxies with small errors are shown for clarity.
} 
\label{fig:abs_m}
\end{center}
\end{figure}

\subsection{Error Budget for Proper Motion} \label{sec:errorbudget}

There are several components that contribute to the uncertainty on our proper motion. We now summarize them. 
First, we have uncertainty from our PSF modeling of the GSAOI images. In Section~\ref{sec:selection},
 this was estimated to be 0.07/0.09 mas yr$^{-1}$ in R.A./Dec.
Second, we have the uncertainty in the selection of Pyxis member stars, which was derived to be 0.05 mas yr$^{-1}$ in both R.A. and Dec. from Section~\ref{sec:selection}.
Lastly, we have the uncertainty in the absolute reference frame due to scatter in the individual measurements,
 which includes individual fitting uncertainties and distortion uncertainties.
In Section~\ref{sec:absvel}, we derive 0.296/0.267 mas yr$^{-1}$ in R.A./Dec and this is the dominant uncertainty in our measurement.

The final uncertainty is the quadrature sum of these individual components which is 0.31/0.29 mas yr$^{-1}$ in R.A./Dec.

\subsection{Final Proper Motion and Galactocentric Velocity} \label{sec:motionfinal}

The calculation of the average velocity of Pyxis uses the good galaxies with the discussed weights. 
The velocity of Pyxis is the error-weighted average of the galaxy velocities with the sign reversed plus the relative velocity of Pyxis (derived in Section~\ref{sec:relvel}).
In total we obtain a motion of $\mu_{\alpha} \cos{\delta}=$1.09$\pm$0.31 mas yr$^{-1}$ and $\mu_{\delta}=$0.68$\pm$0.29 mas yr$^{-1}$, see Figure~\ref{fig:abs_m}.

To convert the proper motion to physical units we additionally require both the distance to Pyxis and the Solar position and velocity relative to the galactic center.
 For Pyxis we adopt a distance of
 $39.4\pm4$ kpc \citep{Sarajedini_96}. 
 The distance modulus uncertainty ($\sigma_{\mu} = 0.22$ mag) is determined based on the (i) uncertainty of the absolute magnitude of the red horizontal branch (RHB)​, (ii) uncertainties based on the sparseness of the RHB for Pyxis (and concerns of statistical decontamination), and (iii) the large reddening along this line of sight (and the potential for differential reddening; E(B-V)$\approx0.25$).
 We repeated our analysis by doubling our uncertainty, and it did not influence the conclusions.

From the reflex motion of Sgr~A* \citep{Reid_04}, it is possible to infer the solar motion in the direction of the Galactic rotation (V') when the distance to the Galactic Center is known ($R_{GC}$). 
The circular velocity at the position of the sun is not necessary for the conversion. 
Recent determinations of $R_{GC}$ are around 8 kpc, some slightly larger \citep{Reid_14,Chatzopoulos_14} and slightly smaller \citep{Boehle_16,Hunt_16}, while recent reviews \citep{Bland_16,DeGrijs_16} obtain slightly larger values. 
We adopt $R_{GC}$=8 kpc as in \citet{Bovy_14b}. 
The solar velocity relative to the local standard of the rest is well-determined in the radial (U) and vertical (W) direction \citep{Reid_04,Schoenrich_10,Bovy_12b}, and thus we use these directly.
The resulting solar motion with respect to the Galactic center is U/V'/W$=$11.0/241.9/7.3 km/s. 
To obtain the full Galactocentric velocity of Pyxis, it is necessary to use the line-of-sight velocity, for which we average \citet{Palma_00} and \citet{Saviane_12} to  $35.7\pm3$ km/s heliocentric.

The uncertainties in parameters for the Sun are small compared to the uncertainties for Pyxis. We ignore the former in the following.
To estimate uncertainties including correlations between them in Galactocentric positions and velocities\footnote{We use the same coordinate-system conventions as \citet{Kallivayalil_13}.} we add to the properties with the largest errors (proper motions, distance) Gaussian random numbers with the width of the error. We repeat this 100,000 times and then calculate the median and 1-$\sigma$ median range for each parameter.  
We obtain for the position: X/Y/Z: $-13.9\pm0.6$/$-38.7\pm3.9$/$4.8\pm0.5$ kpc and for the velocities V$_\mathrm{X}$/V$_\mathrm{Y}$/V$_\mathrm{Z}$: 
52$\pm$55/227$\pm$11/245$\pm$61 km/s. 
The velocity is equivalent to Galactocentric V$_\mathrm{rad}=$-203$\pm$11 km/s and V$_\mathrm{tan}=$278$\pm$60 km/s in spherical coordinates. These and other properties are summarized in Table~\ref{tab:pyxsummary}.

\begin{table*}
\centering
\caption{Summary of Pyxis Properties. \label{tab:pyxsummary}}   
\begin{tabular}{ l l l}
\hline \hline 
Property & Measurement  & Source \\
\hline 
 R.A./Dec. 			& 09:07:57.8/-37:13:17 						& \citet{DaCosta_95,Irwin_95}\\
 l/b 				& 261.32$^\circ$/7.00$^\circ$ 				& \citet{DaCosta_95,Irwin_95}\\
 r$_\mathrm{core}$ 	& 15.8 pc$^{a}$								& \citet{DaCosta_95}\\
 concentration (c) 	& 0.65 										& \citet{DaCosta_95}\\
 r$_\mathrm{half}$ 	& 17.7 pc$^{b}$								& \citet{DaCosta_95}\\
 Luminosity 		& M$_V=-6.0$ $^{a}$ 						& \citet{DaCosta_95} \\
 Age     			& 11.5$\pm$1 Gyrs 							& \citet{Dotter_11} \\
 Metallicity 		& [Fe/H]$=-1.45\pm0.1$ 						& \citet{Palma_00,Dotter_11,Saviane_12}\\ 
 Distance 			& $39.4\pm4$ kpc$^{c}$						&  \citet{Sarajedini_96} and this work \\
 X/Y/Z 				& -13.9$\pm0.6$/-38.7$\pm3.9$/4.8$\pm0.5$ 	& this work\\
 $v_{los}$ 			&  $35.7\pm3$ km/s 							& \citet{Palma_00,Saviane_12} \\ 
 proper motion 		& 1.09$\pm$0.31/0.68$\pm$0.29 mas yr$^{-1}$ & this work\\
 V$_\mathrm{X}$/V$_\mathrm{Y}$/V$_\mathrm{Z}$ & 52$\pm$55/229$\pm$11/245$\pm$61 km/s & this work \\ 
 V$_\mathrm{rad}$/V$_\mathrm{tan}$/V$_\mathrm{tot}$ & -203$\pm$11/278$\pm$60/344$\pm$49 km/s  & this work \\ 
\hline \hline

\multicolumn{3}{l}{$^{a}$ Original measurements by \citet{DaCosta_95} are adjusted for the updated distance.} \\
\multicolumn{3}{l}{$^{b}$ Obtained from r$_\mathrm{core}$ and concentration with interpolation using a \citet{King_62} model} \\
\multicolumn{3}{l}{$^{c}$ The distance value is from \citet{Sarajedini_96}, and the error is from this work.} \\
\multicolumn{3}{l}{$^{d}$ For the metallicity we do not use the photometric metallicity of \citet{Sarajedini_96}, because it is different} \\
\multicolumn{3}{l}{~~~ from the newer photometric metallicity of \citet{Dotter_11}.} \\
\end{tabular}
\end{table*}

\section{Orbit, Origin, and Implications} \label{sec:origin}

Having determined the three-dimensional Galactocentric position and velocity of Pyxis, we can proceed to determine the orbital parameters most consistent with these measurements (Section \ref{sec:orbit}).
The orbit, in turn, permits detailed exploration of the relationship between Pyxis 
 and the ATLAS stellar stream proposed by \citet{Koposov_14} (Section \ref{sec:connection}). 
 We then use the orbit in conjunction with other properties of Pyxis to constrain its origin (Section~\ref{sec:origin2}). 
Lastly, we use our measured motions and the assumption that Pyxis is bound to the Milky Way 
 to estimate the mass of the Milky Way (Section \ref{sec:mw_mass}).

\subsection{Orbit of Pyxis} \label{sec:orbit}

To estimate the orbit of Pyxis, we use an orbit-integrator approach.
We integrate orbits with starting positions sampled from the uncertainty range for the current phase--space location using \texttt{galpy} \citep{Bovy_14b} in the Milky-Way-like potential \texttt{MWPotential2014}, which is fit to the most recent dynamical constraints on the Milky Way. We generate 1000 different orbits by drawing all of the measured parameters with their uncertainties.  The total velocity of Pyxis is 344$\pm$49$\,\mathrm{km\,s}^{-1}$.
In the $\tt{MWPotential2014}$, the motion of Pyxis is near the escape velocity at its position
 ($\approx360$ km s$^{-1}$). That causes that about 41\% of the possible orbits are consistent with Pyxis being unbound, but the halo of \texttt{MWPotential2014} is relatively light. It also has the consequence that the outer orbit parameters are unrealistic, because the calculation assumes that the Milky Way is isolated. Formally, we obtain median values of r$_\mathrm{peri}=$30.7 kpc,  r$_\mathrm{apo}=$1560 kpc, e$=$0.96, and a period larger than the age of the universe.

The halo mass in \texttt{MWPotential2014} is $8\times10^{11}$ M$_\odot$, which is smaller than the estimates of \citet{Boylan_13} and \citet{Marel_12b}. They obtain a total MW mass of $ 1.6\times10^{12}$ M$_\odot$. 
Therefore, we repeat the orbit integration process for a potential like $\tt{MWPotential2014}$, but with the halo mass doubled to  $ 1.6\times10^{12}$ M$_\odot$. 
For this we obtain a most likely orbit of r$_\mathrm{peri}=$28.7 kpc, r$_\mathrm{apo}=$123 kpc, e$=$0.64 and a period of about 1.6 Gyr. Only the determination of perigalacticon is consistent with that of the unaltered \texttt{MWPotential2014}. 

From this experiment, it is fair to conclude that the uncertainty in our knowledge of the Milky Way potential (mainly its total mass) creates large uncertainty in our ability to model the orbit of Pyxis. 
Indeed, only for the estimate of perigalacticon is the uncertainty in our measured parameters the dominant factor, producing an uncertainty of $\sigma_{peri}=$6 kpc. 
Regardless, it is clear that Pyxis is on a quite eccentric orbit $e\gtrsim$0.59 that does not come closer than $R_{GC}\approx$22 kpc. 
 For the median orbit of Pyxis there was no past perigalacticon passage for the less massive halo, since the most likely period is larger than the age of the universe. There is however enough uncertainty in the orbit, that it is possible then that Pyxis's first perigalacticon happened already in the past.
It had about 7 passages for the more massive halo.

\subsection{Connection of Pyxis with ATLAS Stream} \label{sec:connection}

In an evaluation of possible progenitors of the ATLAS stream, \citet{Koposov_14} determined Pyxis to be the most likely candidate 
based on orbit integrations consistent with the ATLAS stream
located at $(\mathrm{RA,Dec,distance}) \approx (20^\circ,-27^\circ,20\,\mathrm{kpc})$.
Plausible ATLAS orbits pass near the observed spatial locations of several globular clusters, but only Pyxis could not be ruled out based on kinematics. 
We convert our proper motion into the celestial coordinate system aligned with the ATLAS stream defined by \citet[$(\phi_1,\phi_2)$][]{Koposov_14} and determine $(\mu_{\phi_1},\mu_{\phi_2}) =(0.38,0.04)\,\mathrm{mas\,yr}^{-1}$. 
Comparing this value to \citet{Koposov_14}'s Figure 3 demonstrates that this proper motion is not aligned with orbits that go through both the ATLAS stream and Pyxis.

\begin{figure}
 \begin{center}
  \includegraphics[width=1.00 \columnwidth,angle=0]{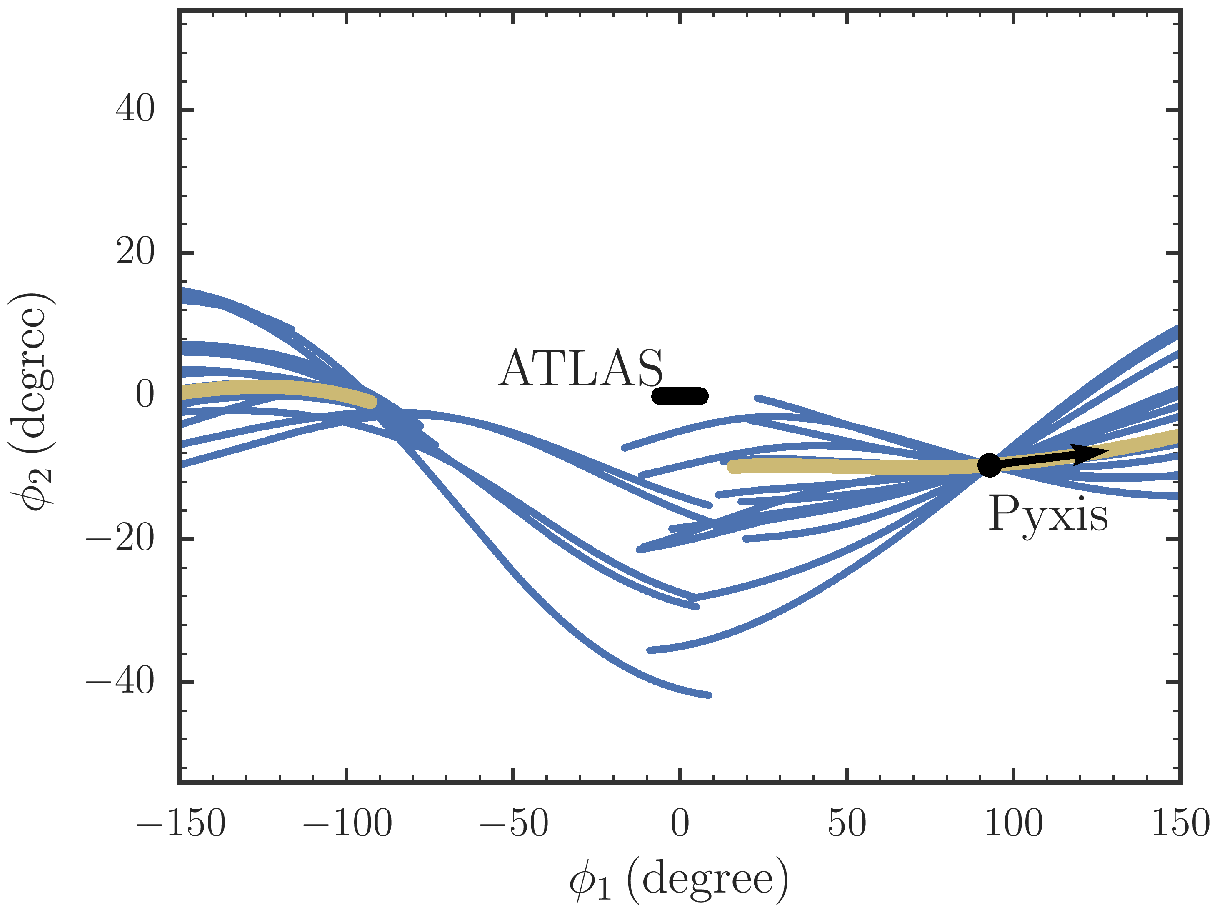}
  \includegraphics[width=1.00 \columnwidth,angle=0]{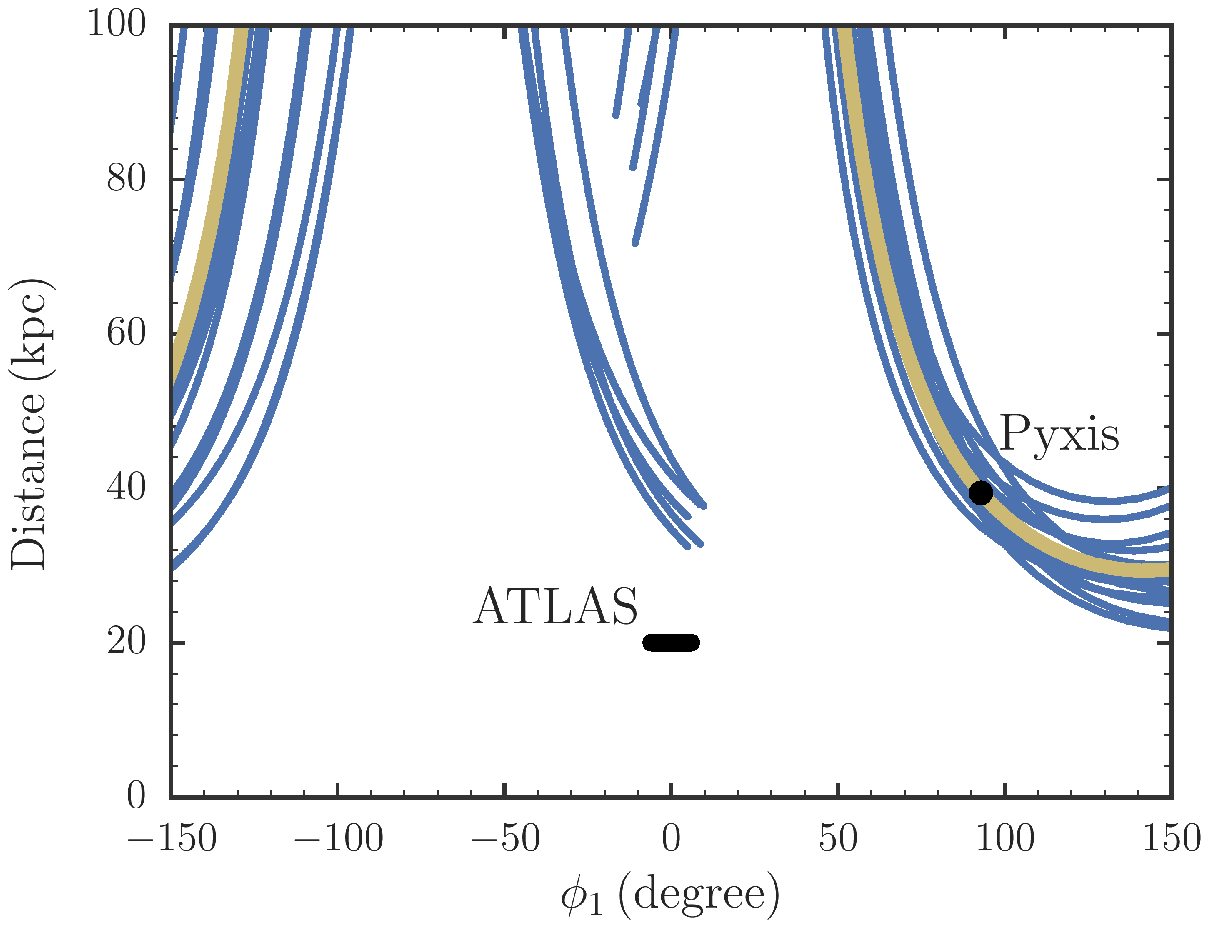}
 \caption{
 Comparison of possible orbits of Pyxis with the
location of the ATLAS stream \citep{Koposov_14}. In the top panel, we
show the stream on the sky in a custom system of angular coordinates
that is aligned with the ATLAS stream. Our measurement of the
direction of Pyxis' proper motion in this coordinate system is
indicated by the arrow. In the bottom panel, we display $\phi_1$
versus distance. In both plots, we show an ensemble of possible tracks exploring the errors with thin blue lines and the track of the measured properties with a thick yellow line. It is clear that Pyxis cannot be the progenitor of
the ATLAS stream.
} 
 \label{fig:atlas_stream}
 \end{center}
 \end{figure}

To investigate the ATLAS and other possible associations further, 
 we have integrated the Pyxis orbits from the previous sub-section both backwards and forwards in \texttt{MWPotential2014}. 
A random subset (1.5\%) of these orbits are shown in Figure~\ref{fig:atlas_stream} in the \citet{Koposov_14} celestial coordinate system $(\phi_1,\phi_2)$. Orbits are integrated until they reach $\phi_1 \approx 0$ with a maximum integration time of 10 Gyr.
It is clear that orbits originating or ending in Pyxis' current phase--space location do not go close to ATLAS's three-dimensional position. 
Repeating this comparison with the altered \texttt{MWPotential2014} with double the halo mass does not alter this conclusion. 
Thus, Pyxis is highly unlikely to be associated with the ATLAS stream. 
The progenitor of the ATLAS stream is either not yet identified, has already been fully tidally disrupted similar to conclusions drawn in regards to the Ophiuchus stream \citep{sesar_2015}, or the stream itself is the stretched out progenitor as is sometimes found in extra-galactic streams \citep[e.g.,][]{dmd_2015}.

\subsection{Origin of Pyxis} \label{sec:origin2}

The class of young halo globular clusters, first defined by \citet{Zinn_93}, does not follow the relatively strong correlation between age and metallicity of the inner globular clusters, and does not rotate with the Milky Way disk. 
These young halo clusters also appear to have an unusually low central concentration indicating that they never get close to the Milky Way  \citep{Vdbergh_94}. 
Two distinct populations of globular clusters are also seen in other galaxies
 \citep[see][for a review of work on external galaxies]{Brodie_06}. 
It is assumed that these young halo globular clusters did not form directly in the Milky Way, but in satellites which later merged with the Milky Way \citep{Zinn_93,Muratov_10,Renaud_16}. 
In the current epoch, the birth-satellites of the young globular clusters may be partially to fully disrupted. 
Sagittarius is an example which is in the process of being disrupted \citep{Majewski_03}, which likely donated some globular clusters to the Milky Way \citep{Siegel_11}. 

When it was discovered, Pyxis was classified
as a young halo globular cluster \citep{DaCosta_95,Irwin_95} due to its distance, concentration, and color magnitude diagram. 
That was further strengthened by analysis of better color magnitude diagrams and spectroscopy \citep{Sarajedini_96,Palma_00,Dotter_11}, which confirmed that the age (11.5 Gyrs) and metallicity combination is somewhat younger than that for the inner Milky Way clusters.  
The inner clusters are consistent with a wet merger origin \citep{Li_14,Weisz_16}. 
From our orbit analysis (Section~\ref{sec:orbit}) Pyxis spends most of its time at more than 60 kpc distance on average, and possibly at as much as 400 kpc. 
That implies that it is more distant than all but one of the old halo clusters \citep{Zinn_93},  
and more distant than any of the wet merger clusters in the simulation of \citet{Renaud_16}.
Thus, Pyxis probably formed in a dwarf galaxy that later fell into the Milky Way.

\subsubsection{Identifying Candidate Former Hosts}\label{sec:hostprop}

We now use the properties of Pyxis to infer the proprieties of its original host galaxy and determine candidate hosts from known satellite galaxies.
We consider three constraints: its luminosity, its metallicity, and its age.

\noindent {\sc Luminosity:}~~
It is very likely that the host galaxy was more luminous than Pyxis ($M_V=-6.0$) itself, 
which already excludes some of the ultrafaint spheroidal galaxies as the original hosts of Pyxis \citep{Mcconnachie_12}.

\noindent {\sc Metallicity:}~~ 
Globular clusters should be no more metal-rich than their parent galaxy and, because Globular clusters are typically quite old, are likely to be more metal poor than the peak metallicity for the parent.
That is the case for all globular clusters of the Fornax galaxy \citep{Buonanno_99,Strader_03,Kirby_11,Battaglia_06}.
This is also true for the known globular clusters of the Sagittarius dwarf galaxy, the core of which has a peak metallicity \textgreater-0.5 [Fe/H], but shows strong population gradients with radius \citep[][Hasselquist et al. in prep]{majewski_2013,Hyde_15,Gibbons_16}\footnote{There are several clusters in the halo, which are today not in Sagittarius, but probably were in the past \citep{Law_10b,Siegel_11}. One of them, Terzan 7 with [Fe/H]$\approx-0.6$ \citep{Sbordone_07}, has within the uncertainties the same metallicity as Sagittarius.}.

Using the assumption that Pyxis has a lower metallicity than its host galaxy we can use the mass-metallicity relation for Local Group satellite galaxies  measured by \citep{Kirby_13} to estimate a lower mass limit. 
Using [Fe/H] = -1.45 $\pm$ 0.1 dex, we infer the host of Pyxis had a stellar mass of \textgreater $\ 6.3\times10^6$ M$_\odot$ with a multiplicative uncertainty of 4.3. 
The uncertainty includes both the metallicity uncertainty of Pyxis and the scatter in the relation (the dominant term). 
The metallicity lower limit eliminates all satellites of the Milky Way besides the Magellanic Clouds, Sagittarius, Fornax and Leo I.
Two satellites of the Milky Way (Sculptor and Leo II) have a somewhat lower metallicity, but we nevertheless consider them for completeness in what follows.
Thus, only few surviving Milky Way satellites could have been the former host.

\noindent {\sc Age-Metallicity Relationship:}~~
We can compare the metallicity of Pyxis to the age-metallicity relationships
 for Fornax, SMC, LMC, and WLM compiled in \citet[][their figure 10]{Leaman_13}.
Pyxis matches very well to the relation for the LMC ($2.9\times10^9$ M$_\odot$), but is inconsistent with the SMC, WLM, and Fornax. 
There is however, some scatter between different galaxies, for example, Fornax and WLM have essentially the same relation, although WLM is two times as massive \citep[see discussion in][]{Weisz_16}.
Thus, also galaxies with about 50\% of the mass of the LMC are plausible former hosts.

We explore the range of galaxy size spanned by the constraints above, which means the former host galaxy mass is between between that of the LMC and Leo II. We now compare the dynamics for each of the candidate galaxies to that inferred for Pyxis, going down the mass scale.

\subsubsection{Connection of Pyxis with the Magellanic Clouds} \label{sec:lmc}

An association between Pyxis and the Magellanic Cloud system was proposed in the discovery \citep{Irwin_95} and early characterization \citep{Palma_00}. 
Also, in our analysis in the above Section~\ref{sec:hostprop}, the Magellanic Clouds overlap with the properties of Pyxis. Therefore, we now test whether they could be the former host of Pyxis. 
Because the Small Magellanic Cloud (SMC) follows closely the orbit of the Large Magellanic Cloud (LMC) \citep{Kallivayalil_13}, 
we concentrate our comparison on the LMC. 
Because the Magellanic system is rather complicated and the LMC is quite massive, a simple orbit-integration approach is not sufficient. 

We use cosmological simulations of a MW-like halo from the Aquarius project \citep{Springel2008b} to predict likely locations and orbits of material stripped from an LMC-like satellite. 
From the set of $6$ high-resolution dark matter halos in the Aquarius project, only one has an orbiting subhalo consistent with being an LMC.
The analog subhalo shows orbital properties consistent with those measured for the LMC and 
is sufficiently massive ($M_{\rm vir}=3.4 \times 10^{10} \; \rm M_\odot$). 
Furthermore, this subhalo has had two pericenter passages around its host. 
Thus, we can evaluate a Pyxis-LMC association for Pyxis having been stripped from the LMC either after a first or second pericenter passage.

Tidal disruption of orbiting satellites distributes the unbound
material not randomly, but rather 
along the orbit/energy of the progenitor system.    
The tidal disruption process imprints a relation between the area of the sky, the distance, and the velocity of all group members 
\citep[see also][in application to all dwarf satellites of the MW]{Sales2011,Sales2016}. 
Thus, if Pyxis was once part of the LMC system, 
we anticipate clear correlations between its phase--space and that occupied by the LMC over the course of its orbit.  

\begin{figure}
\begin{center}
   \includegraphics[width=1.00 \columnwidth,angle=0]{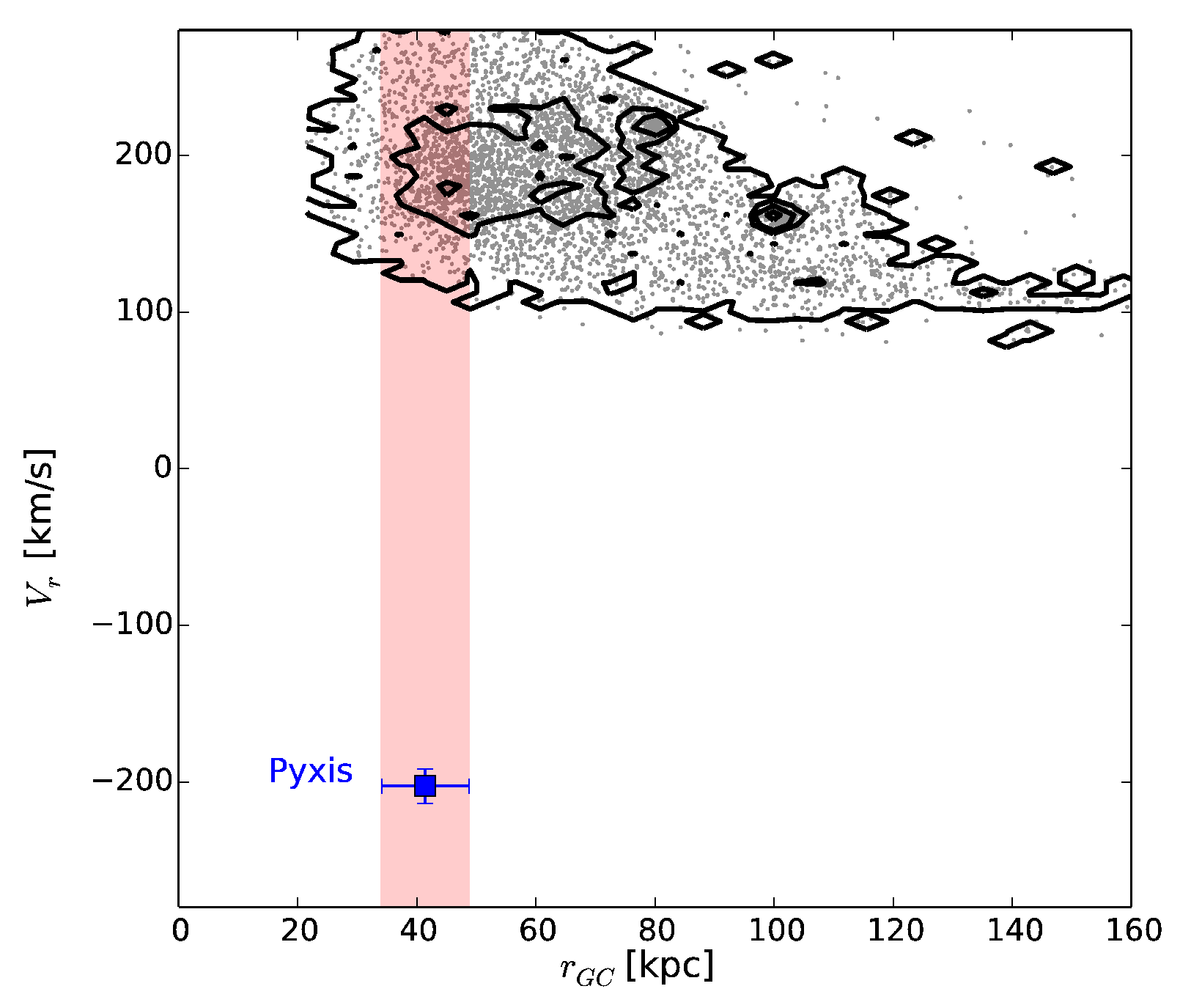}
   \includegraphics[width=1.00 \columnwidth,angle=0]{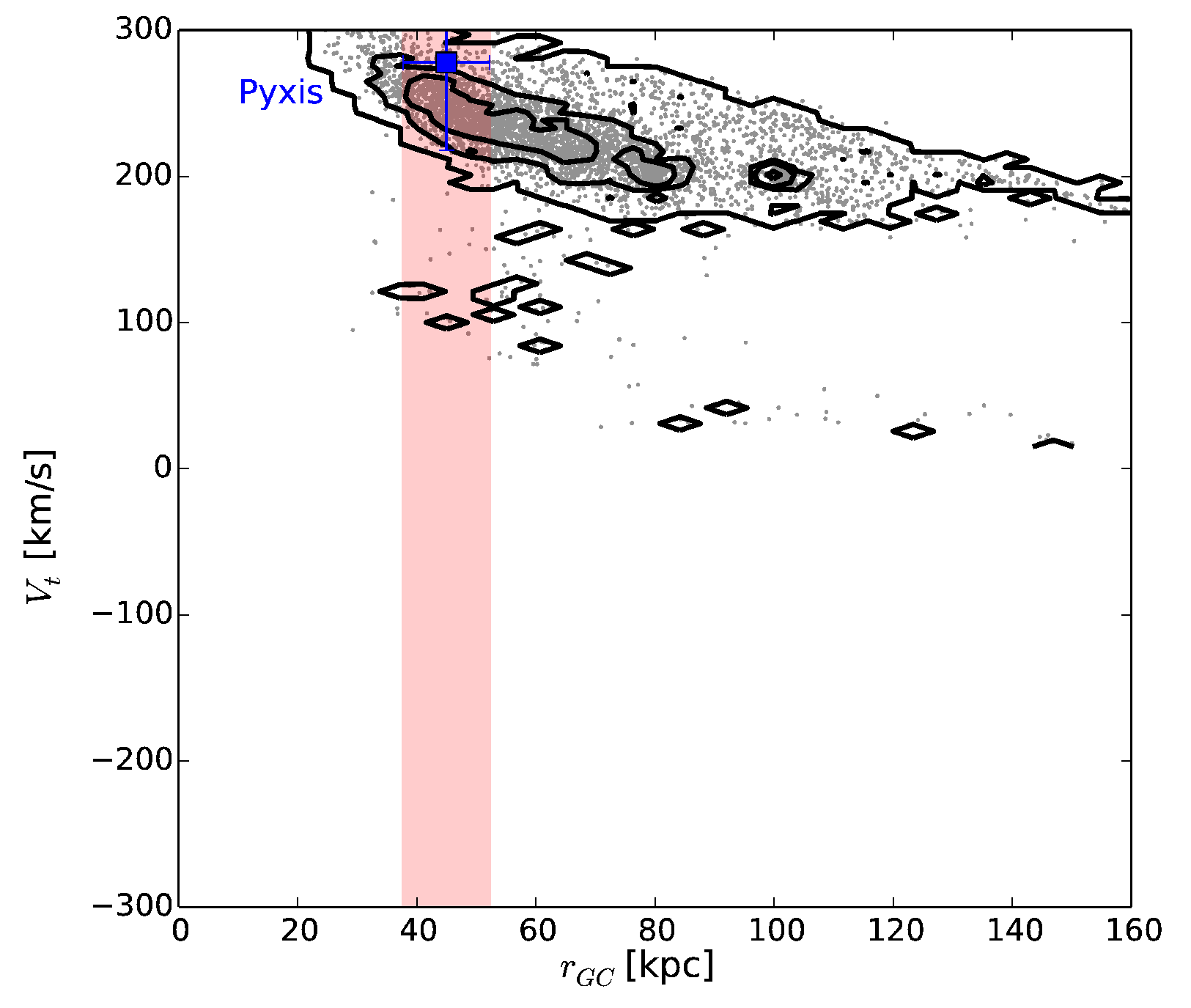}  
 \caption{Comparison of the Galactocentric motion of Pyxis with LMC analog debris at the coordinates of Pyxis assuming the LMC analog is on its first approach.
The radial velocity (top) and tangential velocity (bottom) of LMC debris particles are shown as gray points with contours drawn to show the concentration.
The motions of Pyxis are shown as the blue dot in both panels with the pink shading indicating the range of r$_{GC}$ for Pyxis. 
While the tangential motion is consistent with the LMC debris, the radial motion is inconsistent. 
This result is not demonstrably different assuming the LMC to be on a second passage.
 } 
 \label{fig:pyxis_lmc}
 \end{center}
 \end{figure}

We select all particles once part of the simulated LMC analog that lay on the sky within a circle of radius $5^\circ$ around the position of Pyxis, similar as in \citet{Sales2016}. 
The LMC is currently just passed its pericenter passage and is located in the Southern Galactic Hemisphere. 
Because Pyxis is in the Northern Hemisphere, 
it must fall along the {\it leading arm} of the system (i.e., the ``future'' for the LMC) and therefore corresponds to particles that show high positive radial velocities (moving away from the Milky Way).
The galactocentric distance ($r_{\rm GC}$) and velocities (radial or tangential velocity; $V_\mathrm{rad}$ or $V_\mathrm{tan}$, respectively) of these particles is shown in Figure \ref{fig:pyxis_lmc} as gray dots.   
We overplot in Figure~\ref{fig:pyxis_lmc} our measured values for Pyxis with
blue squares and error bars, which show a reasonable agreement in the
tangential velocity but a very different radial component. 
The situation is very similar if we study the second pericenter passage of the LMC analog instead. 
We conclude that Pyxis is unlikely to have been stripped from the LMC based on the LMC analog in the Aquarius simulation. The disagreement with $V_\mathrm{rad}$ is so great that the fact that the LMC analog in the simulation is lighter than the real LMC \citep{Marel_14} makes no difference to this conclusion.

The incompatibility of Pyxis with the LMC is due to increased knowledge of both Pyxis and the LMC in contrast to earlier works; for the former, we have orbital parameters for Pyxis, and for the latter, we know the LMC and SMC to have a large tangential velocity \citep{Kallivayalil_06a,Kallivayalil_13}.
The large tangential velocity of the Magellanic system implies the system has only recently fallen into the Galaxy \citep{Besla_07}.
The argument of the periapsis can only evolve strongly over a larger number of orbits and, with the limited number implied for the Magellanic System, this cannot be reconciled with the Pyxis orbit. 
 Therefore, neither the LMC nor the SMC can have been associated with Pyxis.
 
\subsubsection{Other Known Milky Way Satellites as Former Hosts} \label{sec:hosts}

We now explore whether one of the other satellites allowed by the constraints in Section~\ref{sec:hostprop} could be the former host. Since they are less massive than the LMC, a simple comparison of the orbits is sufficient. 

\noindent {\sc Sagittarius:}~~~~ The orbital plane of Sagittarius \citep{Law_10} is misaligned by nearly 90 degrees compared to the Magellanic stream \citep{Majewski_03, Pawlowski_15}. Pyxis shares the orbital plane of the Magellanic system. Sagittarius is thus excluded. 

\noindent {\sc Fornax:}~~ Several works measured the proper motion of Fornax, see \citet{Piatek_07,Mendez_11}. All agree that the most likely pericenter is at $\approx150$ kpc. \citet{Piatek_07} conclude that there is only a very small probability (2.5\%) that the pericenter of Fornax is smaller than 66 kpc.  That is clearly larger than the pericenter of Pyxis and would imply that the globular cluster Pyxis was at a distance of 40 kpc from Fornax when tides of the Milky Way separated them. That is larger than the distance of nearly all in-situ formed globular clusters \citet{Zinn_93,Renaud_16} in the case of the Milky Way. In a smaller galaxy the distance of globular clusters is very likely smaller. Thus, Fornax cannot be the host galaxy of Pyxis.  

\noindent{\sc Leo I:}~~ \citet{Sohn_13} measured the proper motion of Leo I. 
From the proper motion it follows that Leo I is likely currently on first approach, or at most on its second approach in its orbit. 
The pericenter is $91\pm36$ kpc, which is clearly larger than in the case of Pyxis, and again makes the distance between Pyxis and its host galaxy unrealistically large.
Further, Leo I shows signs of relatively recent star formation, which makes it unlikely that it is already on its second approach. 
In a first approach it is unlikely that a globular cluster is already separated by 270 kpc from its parent galaxy as in the case of Pyxis and Leo I. Finally, the orbital poles of their orbits deviate by 102$^\circ$, i.e., by about 4 $\sigma$, based on the measurement errors.

\noindent {\sc Sculptor:}~~
Sculptor's proper motion was best measured by \citet{Piatek_06}. It orbits with the wrong sign compared to Pyxis. Thus, it cannot have been the host of Pyxis.  

\noindent {\sc Leo II:}~~ Leo II orbital pole disagrees only by 1.5 $\sigma$ from the pole of Pyxis using the proper motion of \citet{Piatek_16}. However, according to that motion Leo II has a perigalacticon  of less than 36 kpc only with  2.5\% probability. That makes it unlikely that Leo II got close enough to the Milky Way for tidal disruption. Finally, Leo II is currently at 233 kpc \citep{Bellazzini_05}, while based on our orbital analysis Pyxis will be close to pericenter when it will be close to the sky coordinates of Leo II. Thus, as for the LMC, the argument of the periapsis disagrees. Further, Leo II's period is probably quite large, since it is so distant. Therefore, Leo II is not associated with Pyxis.

While dynamical friction can change the orbits of dwarf galaxies, it acts to decrease the distance of a galaxy. Thus, if we were to include dynamical friction in our analysis, the past distances of the discussed galaxies would increase further compared to the distances used here. Since these distances are already too large to match Pyxis, the problem would only be exacerbated.
We have exhausted the known Milky Way satellites with masses larger or slightly smaller than that implied assuming the metallicity of Pyxis places a lower limit on the mass of its host, and then using the mass-metallicity relationship for dwarf galaxies to constrain the mass.

\subsubsection{Pyxis as the Debris of an Accretion Event} \label{sec:detection}

Having found no suitable former hosts among the known satellite galaxies, we now explore whether or not either the intact host, or the debris from the disrupted host,  would be detectable.  

Pyxis is located only 7$^\circ$ off the Galactic plane. 
A galaxy which contains gas and on-going star formation, as would be the case if Pyxis were on first infall \citep{Wetzel_15}, would be detected also within a 7$^\circ$ distance from the Galactic plane. 
Thus, it is unlikely that Pyxis is on first infall towards the Milky Way, since no galaxy is detected there.  
On the other hand, a galaxy which contains only stars (i.e., no gas or active star formation markers) would be much more difficult to detect, even if it were not disrupted. 
However, that scenario is overall still unlikely since only in a small area of sky the galaxy would be undetectable.

More likely the parent galaxy of Pyxis has already been disrupted. It is in principle possible that a stream is left behind.  For the former host we assume that it is as small possible (Section~\ref{sec:hostprop}), similar to Leo II. Leo~II has  M$_V=-9.8$ of Leo II \citep{Mcconnachie_12} and about m$_\mathrm{tot}=4\times10^7M_{\odot}$. We now check whether a stream produced by a galaxy with these properties would be detectable. We assume that the stream would be on a similar orbit as Pyxis.

From the fact that a stream as faint as Orphan \footnote{Orphan itself cannot be associated with Pyxis as its orbit does not match that of Pyxis.} 
(34.6 magnitudes/arsec$^2$ \citet{Belokurov_07}) has been detected, we conclude that streams down to a surface brightness of 34.6 magnitudes/arsec$^2$ are detectable in the SDSS. 
We utilize the analytical relations describing stellar stream debris derived in \citet{Johnston_01}, with the detection limit of SDSS, to place additional constraints. 
A stream of a galaxy with m$_\mathrm{tot}=4\times10^7M_{\odot}$ should be 3$^\circ$ wide, when the orbit is similar to Pyxis (r$_\mathrm{peri}=$30 kpc), and  v$_\mathrm{circ}=$200 km/s \citep{Kuepper_15,Bovy_16}.
The pericenter of Pyxis's orbit is around R.A./Dec. of 170/0$^\circ$. If there were a stream belonging to Pyxis, the part of it that was at 35 kpc distance or less would be about 50$^\circ$ long within the footprint of SDSS. 
From that we derive that the brightest undetectable stream has a total integrated m$_r\approx11.4$ within the SDSS footprint. 

Correcting that for distance (30 kpc), and using a factor of 7.6 for not including stars that are outside of the matched filter \citep{Belokurov_07},
results in a limiting absolute magnitude of M$_r\approx-8.2$ for an undetected stream.
That is a factor of 4 less than the luminosity of the Leo II like galaxy (M$_r\approx-9.8$).
Since the full stream is not within the SDSS footprint, a stream produced by a Leo II like galaxy is probably at the detection limit.
However, Leo II was our analog for the lowest luminosity dwarf galaxy that could have been the host for Pyxis. Therefore, any dwarf galaxy brighter than Leo II would produce streams that are even easier to detect, and thus we conclude that it is unlikely that the former host of Pyxis has a still-detectable stream associated with it.

Pyxis is on eccentric orbit (0.59$ < e< 1$). Galaxies on eccentric orbits ($e\geq0.86$) usually produce shells after tidal disruption \citep{Johnston_08,Hendel_15}. Thus, possibly the remnant of the host galaxy is more shell-like. In that case, it would be more difficult to detect in the Milky Way. Possibly, the recently discovered clouds, such as the Virgo overdensity \citep{Vivas_01,Grillmair_16} are shells. Finally, the galaxy could have fallen in so early that its stellar distribution is now very diffuse \citep{Johnston_08}.

Since destruction of galaxies is easier closer to the Milky Way, comparing Sagittarius \citep{Law_10} with Hercules \citep{Kuepper_16}, a possible scenario is the following:
assuming a separation of 10 kpc between Pyxis and its galaxy and a perigalacticon of 22 kpc for Pyxis (within the error range), a perigalaction of 12 kpc for the galaxy is possible. 
There the tides of the MW could have separated the two, later dynamical friction would have further decreased the distance of the host, further easing full destruction of the progenitor. 
That hypothesis could be confirmed by detecting more globular clusters (former members of that galaxy) on similar orbits.

~~Since the number of young halo globulars (30; \citet{Mackey_05}) is smaller than the number of globular clusters for LMC-mass galaxies (about 50; \citep{Zaritsky_16a})
 it seems unlikely that an LMC-mass galaxy was accreted in the past, even taking into account uncertainties in both numbers. Also, the absence of a classical bulge \citep{Bland_16,Ness_16} in the Milky Way makes an accretion of an LMC mass galaxy unlikely. 
Finally, the orbital information of the member stars of this galaxy would not be fully lost, thus when enough stars have good proper motions measured by $\tt{Gaia}$, it should be possible to test this hypothesis further, and maybe also to determine the mass of the galaxy. 
A non-detection of any associated debris would be surprising, and would indicate that Pyxis formed in a different way.

\subsection{Mass of Milky Way} \label{sec:mw_mass}
Nearly all satellite-subhalos identified in simulations are bound to their parent \citep[see e.g.][]{DiCintio_12,Boylan_13}. 
While Pyxis itself does probably not contain dark matter, it still very likely originated in a galaxy which contains dark matter and is thus a subhalo (Section~\ref{sec:origin2}). Therefore, Pyxis shares the orbital properties of subhalos. 
We now estimate the mass of the MW assuming that Pyxis is bound. 
Similar to Section~\ref{sec:orbit}, we use $\tt{MWPotential2014}$ to fix both the disk and bulge, but allow the mass of the halo to vary.
 We obtain that the halo virial mass is with 68\% probability larger than 0.58$\times10^{12}$ M$_\odot$, which corresponds to a total mass of 0.65$\times10^{12}$ M$_\odot$ when including bulge and disk. The mass limit depends only very weakly on the concentration of the halo. Smaller concentrations like in \citet{Marel_12a} increase the mass of the limit, while larger would decrease it. Our concentrations of $c=15.3$ is relatively large compared to e.g. \citet{Xue_08}, it is similar to the value of $c=16$ obtained in the recent review of \citet{Bland_16}. 

From our analysis in Section~\ref{sec:detection}, it follows that Pyxis is likely not in its first approach after its formation 11.5 Gyrs ago. 
We repeat this analysis excluding cases when Pyxis is its first approach. 
This requires a more massive Milky Way. To calculate it we just exclude all cases which have not a recent enough first approach, that results in a halo mass lower limit of 0.88$\times10^{12}$ M$_\odot$ for the virial mass (Virial radius of 245 kpc) and total mass lower limit of 0.95$\times10^{12}$ M$_\odot$. 
This mass depends slightly stronger on the concentration. However, even for the unlikely case of $c=6$ the halo mass limit is still only
$1.03\times10^{12}$ M$_\odot$.
That is in some tension with the recent measurement by \citet{Gibbons_14} of M$_\mathrm{200\, kpc}=(0.56\pm0.12)\times10^{12}\,M_\odot$, but 
most measurements are larger. For example, the review of \citet{Bland_16} obtains M$_\mathrm{vir}=(1.3\pm0.3)\times10^{12}\,M_\odot$.

\section{Summary} \label{sec:summary}

\begin{enumerate}
\item We obtained near-infrared MCAO imaging of the Pyxis globular cluster with the GeMS/GSAOI imager on Gemini-S. These images provide a spatial resolution of 0.02$\arcsec$ pixel$^{-1}$ and sources have a FWHM of 0.08 $\arcsec$ over a 85$\arcsec \times$ 85$\arcsec$ field-of-view. We combine this imaging with archival HST imaging of the Pyxis cluster in the $F606W$ and $F814W$ filters at 0.05$\arcsec$ pixel$^{-1}$ (Section \ref{sec:dataset}). 
\item We developed PSF modeling techniques for our MCAO imaging (Section \ref{sec:psf_est}) to take into account the position and time dependent PSF in the AO imaging. With this approach we produce positions precise to 0.4 mas 
relatively over the full 85$\arcsec \times$ 85$\arcsec$ field-of-view. These efforts support future proper motion analyses with wide-field AO imaging planned for 30-meter class facilities.
\item We determined an absolute reference frame for MCAO-HST imaging using background galaxies for the first time. This method uses the PSFs and Sersic models to fit the galaxies.  
\item We determined the first proper motion measurement of the Pyxis star cluster over a 5 year baseline. We obtain a proper motion of $\mu_{\alpha}\,\cos(\delta) =$1.09$\pm$0.31 mas yr$^{-1}$ and $\mu_{\delta} = $0.68$ \pm $0.29 mas yr$^{-1}$.
\item This proper motion and the line-of-sight velocity implies in an eccentric orbit, with an apogalacticon between about 120 and 1600 kpc (That Pyxis is unbound is also possible.) and a perigalacticon of 30$\pm$6 kpc. The uncertainty in the apogalacticon is dominated by uncertainty of the mass of the Milky Way. 
\item Our orbit excludes Pyxis as the progenitor of the ATLAS stream, as was proposed by \citet{Koposov_14}.
\item The orbit also excludes that Pyxis was once a member of the LMC and of all Milky satellites down to at least the mass of Leo II. 
\item The orbit, metallicity, and age points to an extragalactic origin of Pyxis, in a rather massive dwarf galaxy (in mass between Leo II and LMC). Because all known satellite galaxies in that mass range are excluded, the host is probably already disrupted. It is unlikely that a satellite on its first passage would be totally destroyed and, thus, it is unlikely that Pyxis is on its first approach.
\item We use the velocity of Pyxis, the assumption that it is bound to the Milky Way, and the assumption that it is likely not on first approach to obtain a 68\% lower limit on the mass of the Milky Way of 0.95$\times10^{12}$ M$_\odot$. 
\end{enumerate}

\acknowledgements
This work was supported by the NSF CAREER award 1455260.\\

JB received support from the Natural Sciences and Engineering Reseach Council of Canada, an Alfred P. Sloan Fellowship, and from the Simons Foundation.\\

Based on observations obtained at the Gemini Observatory acquired through the Gemini Observatory Archiv acquired through the Gemini Observatory/Science Archive, which is operated by the Association of Universities for Research in Astronomy, Inc., under a cooperative agreement with the NSF on behalf of the Gemini partnership: the National Science Foundation (United States), the National Research Council (Canada), CONICYT (Chile), Ministerio de Ciencia, Tecnolog\'{i}a e Innovaci\'{o}n Productiva (Argentina), and Minist\'{e}rio da Ci\^{e}ncia, Tecnologia e Inova\c{c}\~{a}o (Brazil).

Some of the data presented in this paper were obtained from the Mikulski Archive for Space Telescopes (MAST). STScI is operated by the Association of Universities for Research in Astronomy, Inc., under NASA contract NAS5-26555. Support for MAST for non-HST data is provided by the NASA Office of Space Science via grant NNX09AF08G and by other grants and contracts.

Based on observations obtained as part of the VISTA Hemisphere Survey, ESO Progam, 179.A-2010 (PI: McMahon)

 \facility{HST (ACS/WFC), Gemini:South (GeMS/GSAOI)}

\software{ THELI  \citep{Erben_05,Schirmer_13}, "SCAMP" \citep{Bertin_06}, Swarp \citep{Bertin_10}, dpuser, Galfit \citep{Peng_02}, PSFeX \citep{Bertin_11},     \texttt{galpy} \citep{Bovy_14b} }

\bibliography{mspap}

\begin{thebibliography}{}
\expandafter\ifx\csname natexlab\endcsname\relax\def\natexlab#1{#1}\fi
\providecommand{\url}[1]{\href{#1}{#1}}

\bibitem[{{Ammons} {et~al.}(2016){Ammons}, {Garcia}, {Salama}, {Neichel}, {Lu},
  {Marois}, {Macintosh}, {Savransky}, {Bendek}, {Guyon}, {Marin}, {Garrel}, \&
  {Sivo}}]{Ammons_16}
{Ammons}, S.~M., {Garcia}, E.~V., {Salama}, M., {et~al.} 2016, in \procspie,
  Vol. 9909, Society of Photo-Optical Instrumentation Engineers (SPIE)
  Conference Series, 99095T

\bibitem[{{Anderson} \& {King}(2006)}]{Anderson_06}
{Anderson}, J., \& {King}, I.~R. 2006, {PSFs, Photometry, and Astronomy for the
  ACS/WFC}, Tech. rep., Space Telescope Science Institute

\bibitem[{{Avila} {et~al.}(2016){Avila}, {Grogin}, {Anderson}, {Bohlin},
  {Borncamp}, {Chiaberge}, {Coe}, {Golimowski}, {Kozhurina-Platais}, {Lucas},
  {Maybhate}, {McMaster}, \& {Ogaz}}]{Avila_16}
{Avila}, R., {Grogin}, N., {Anderson}, J., {et~al.} 2016, {Advanced Camera for
  Surveys Instrument Handbook for Cycle 24 v. 15.0} (Space Telescope Science
  Institute)

\bibitem[{{Battaglia} {et~al.}(2006){Battaglia}, {Tolstoy}, {Helmi}, {Irwin},
  {Letarte}, {Jablonka}, {Hill}, {Venn}, {Shetrone}, {Arimoto}, {Primas},
  {Kaufer}, {Francois}, {Szeifert}, {Abel}, \& {Sadakane}}]{Battaglia_06}
{Battaglia}, G., {Tolstoy}, E., {Helmi}, A., {et~al.} 2006, \aap, 459, 423

\bibitem[{{Bellazzini} {et~al.}(2005){Bellazzini}, {Gennari}, \&
  {Ferraro}}]{Bellazzini_05}
{Bellazzini}, M., {Gennari}, N., \& {Ferraro}, F.~R. 2005, \mnras, 360, 185

\bibitem[{{Bellini} {et~al.}(2014){Bellini}, {Anderson}, {van der Marel},
  {Watkins}, {King}, {Bianchini}, {Chanam{\'e}}, {Chandar}, {Cool}, {Ferraro},
  {Ford}, \& {Massari}}]{Bellini_14}
{Bellini}, A., {Anderson}, J., {van der Marel}, R.~P., {et~al.} 2014, \apj,
  797, 115

\bibitem[{{Belokurov} {et~al.}(2007){Belokurov}, {Evans}, {Irwin},
  {Lynden-Bell}, {Yanny}, {Vidrih}, {Gilmore}, {Seabroke}, {Zucker},
  {Wilkinson}, {Hewett}, {Bramich}, {Fellhauer}, {Newberg}, {Wyse}, {Beers},
  {Bell}, {Barentine}, {Brinkmann}, {Cole}, {Pan}, \& {York}}]{Belokurov_07}
{Belokurov}, V., {Evans}, N.~W., {Irwin}, M.~J., {et~al.} 2007, \apj, 658, 337

\bibitem[{{Bertin}(2006)}]{Bertin_06}
{Bertin}, E. 2006, in Astronomical Society of the Pacific Conference Series,
  Vol. 351, Astronomical Data Analysis Software and Systems XV, ed.
  C.~{Gabriel}, C.~{Arviset}, D.~{Ponz}, \& S.~{Enrique}, 112

\bibitem[{{Bertin}(2010)}]{Bertin_10}
{Bertin}, E. 2010, {SWarp: Resampling and Co-adding FITS Images Together},
  Astrophysics Source Code Library, , , ascl:1010.068

\bibitem[{{Bertin}(2011)}]{Bertin_11}
{Bertin}, E. 2011, in Astronomical Society of the Pacific Conference Series,
  Vol. 442, Astronomical Data Analysis Software and Systems XX, ed. I.~N.
  {Evans}, A.~{Accomazzi}, D.~J. {Mink}, \& A.~H. {Rots}, 435

\bibitem[{{Bertin} \& {Arnouts}(1996)}]{Bertin_96}
{Bertin}, E., \& {Arnouts}, S. 1996, \aaps, 117, 393

\bibitem[{{Besla} {et~al.}(2007){Besla}, {Kallivayalil}, {Hernquist},
  {Robertson}, {Cox}, {van der Marel}, \& {Alcock}}]{Besla_07}
{Besla}, G., {Kallivayalil}, N., {Hernquist}, L., {et~al.} 2007, \apj, 668, 949

\bibitem[{{Bland-Hawthorn} \& {Gerhard}(2016)}]{Bland_16}
{Bland-Hawthorn}, J., \& {Gerhard}, O. 2016, \araa, 54, 529

\bibitem[{{Boehle} {et~al.}(2016){Boehle}, {Ghez}, {Sch{\"o}del}, {Meyer},
  {Yelda}, {Albers}, {Martinez}, {Becklin}, {Do}, {Lu}, {Matthews}, {Morris},
  {Sitarski}, \& {Witzel}}]{Boehle_16}
{Boehle}, A., {Ghez}, A.~M., {Sch{\"o}del}, R., {et~al.} 2016, \apj, 830, 17

\bibitem[{{Bovy}(2015)}]{Bovy_14b}
{Bovy}, J. 2015, \apjs, 216, 29

\bibitem[{{Bovy} {et~al.}(2016{\natexlab{a}}){Bovy}, {Bahmanyar}, {Fritz}, \&
  {Kallivayalil}}]{Bovy_16b}
{Bovy}, J., {Bahmanyar}, A., {Fritz}, T.~K., \& {Kallivayalil}, N.
  2016{\natexlab{a}}, ApJ, in press, arXiv:1609.01298

\bibitem[{{Bovy} {et~al.}(2016{\natexlab{b}}){Bovy}, {Erkal}, \&
  {Sanders}}]{Bovy_16}
{Bovy}, J., {Erkal}, D., \& {Sanders}, J.~L. 2016{\natexlab{b}}, MNRAS,
  submitted, arXiv:1606.03470

\bibitem[{{Bovy} {et~al.}(2012){Bovy}, {Allende Prieto}, {Beers}, {Bizyaev},
  {da Costa}, {Cunha}, {Ebelke}, {Eisenstein}, {Frinchaboy}, {Garc{\'{\i}}a
  P{\'e}rez}, {Girardi}, {Hearty}, {Hogg}, {Holtzman}, {Maia}, {Majewski},
  {Malanushenko}, {Malanushenko}, {M{\'e}sz{\'a}ros}, {Nidever}, {O'Connell},
  {O'Donnell}, {Oravetz}, {Pan}, {Rocha-Pinto}, {Schiavon}, {Schneider},
  {Schultheis}, {Skrutskie}, {Smith}, {Weinberg}, {Wilson}, \&
  {Zasowski}}]{Bovy_12b}
{Bovy}, J., {Allende Prieto}, C., {Beers}, T.~C., {et~al.} 2012, \apj, 759, 131

\bibitem[{{Boylan-Kolchin} {et~al.}(2013){Boylan-Kolchin}, {Bullock}, {Sohn},
  {Besla}, \& {van der Marel}}]{Boylan_13}
{Boylan-Kolchin}, M., {Bullock}, J.~S., {Sohn}, S.~T., {Besla}, G., \& {van der
  Marel}, R.~P. 2013, \apj, 768, 140

\bibitem[{{Brodie} \& {Strader}(2006)}]{Brodie_06}
{Brodie}, J.~P., \& {Strader}, J. 2006, \araa, 44, 193

\bibitem[{{Buonanno} {et~al.}(1999){Buonanno}, {Corsi}, {Castellani},
  {Marconi}, {Fusi Pecci}, \& {Zinn}}]{Buonanno_99}
{Buonanno}, R., {Corsi}, C.~E., {Castellani}, M., {et~al.} 1999, \aj, 118, 1671

\bibitem[{{Carrasco} {et~al.}(2012){Carrasco}, {Edwards}, {McGregor}, {Winge},
  {Young}, {Doolan}, {van Harmelen}, {Rigaut}, {Neichel}, {Trancho}, {Artigau},
  {Pessev}, {Colazo}, {Tigner}, {Mauro}, {L{\"u}hrs}, \&
  {Rambold}}]{Carrasco_12}
{Carrasco}, E.~R., {Edwards}, M.~L., {McGregor}, P.~J., {et~al.} 2012, in
  \procspie, Vol. 8447, Adaptive Optics Systems III, 84470N

\bibitem[{{Casetti-Dinescu} \& {Girard}(2016)}]{Dinescu_16}
{Casetti-Dinescu}, D.~I., \& {Girard}, T.~M. 2016, \mnras, 461, 271

\bibitem[{{Chatzopoulos} {et~al.}(2015){Chatzopoulos}, {Fritz}, {Gerhard},
  {Gillessen}, {Wegg}, {Genzel}, \& {Pfuhl}}]{Chatzopoulos_14}
{Chatzopoulos}, S., {Fritz}, T.~K., {Gerhard}, O., {et~al.} 2015, \mnras, 447,
  948

\bibitem[{{Da Costa}(1995)}]{DaCosta_95}
{Da Costa}, G.~S. 1995, \pasp, 107, 937

\bibitem[{{Dalessandro} {et~al.}(2016){Dalessandro}, {Saracino}, {Origlia},
  {Marchetti}, {Ferraro}, {Lanzoni}, {Geisler}, {Cohen}, {Mauro}, \&
  {Villanova}}]{Dalessandro_16}
{Dalessandro}, E., {Saracino}, S., {Origlia}, L., {et~al.} 2016, \apj, 833, 111

\bibitem[{{Davies} \& {Kasper}(2012)}]{Davies_12}
{Davies}, R., \& {Kasper}, M. 2012, \araa, 50, 305

\bibitem[{{Davies} {et~al.}(2016){Davies}, {Schubert}, {Hartl}, {Alves},
  {Cl{\'e}net}, {Lang-Bardl}, {Nicklas}, {Pott}, {Ragazzoni}, {Tolstoy},
  {Agocs}, {Anwand-Heerwart}, {Barboza}, {Baudoz}, {Bender}, {Bizenberger},
  {Boccaletti}, {Boland}, {Bonifacio}, {Briegel}, {Buey}, {Chapron}, {Cohen},
  {Czoske}, {Dreizler}, {Falomo}, {Feautrier}, {F{\"o}rster Schreiber},
  {Gendron}, {Genzel}, {Gl{\"u}ck}, {Gratadour}, {Greimel}, {Grupp},
  {H{\"a}user}, {Haug}, {Hennawi}, {Hess}, {H{\"o}rmann}, {Hofferbert}, {Hopp},
  {Hubert}, {Ives}, {Kausch}, {Kerber}, {Kravcar}, {Kuijken}, {Lang-Bardl},
  {Leitzinger}, {Leschinski}, {Massari}, {Mei}, {Merlin}, {Mohr}, {Monna},
  {M{\"u}ller}, {Navarro}, {Plattner}, {Przybilla}, {Ramlau}, {Ramsay},
  {Ratzka}, {Rhode}, {Richter}, {Rix}, {Rodeghiero}, {Rohloff}, {Rousset},
  {Ruddenklau}, {Schaffenroth}, {Schlichter}, {Sevin}, {Stuik}, {Sturm},
  {Thomas}, {Tromp}, {Turatto}, {Verdoes-Kleijn}, {Vidal}, {Wagner}, {Wegner},
  {Zeilinger}, {Ziegler}, \& {Zins}}]{Davies_16}
{Davies}, R., {Schubert}, J., {Hartl}, M., {et~al.} 2016, SPIE, in press,
  arXiv:1607.01954

\bibitem[{{de Grijs} \& {Bono}(2016)}]{DeGrijs_16}
{de Grijs}, R., \& {Bono}, G. 2016, \apjs, 227, 5

\bibitem[{{Di Cintio} {et~al.}(2012){Di Cintio}, {Knebe}, {Libeskind},
  {Hoffman}, {Yepes}, \& {Gottl{\"o}ber}}]{DiCintio_12}
{Di Cintio}, A., {Knebe}, A., {Libeskind}, N.~I., {et~al.} 2012, \mnras, 423,
  1883

\bibitem[{{Dinescu} {et~al.}(1997){Dinescu}, {Girard}, {van Altena}, {Mendez},
  \& {Lopez}}]{Dinescu_97}
{Dinescu}, D.~I., {Girard}, T.~M., {van Altena}, W.~F., {Mendez}, R.~A., \&
  {Lopez}, C.~E. 1997, \aj, 114, 1014

\bibitem[{{Dinescu} {et~al.}(1999){Dinescu}, {van Altena}, {Girard}, \&
  {L{\'o}pez}}]{Dinescu_99a}
{Dinescu}, D.~I., {van Altena}, W.~F., {Girard}, T.~M., \& {L{\'o}pez}, C.~E.
  1999, \aj, 117, 277

\bibitem[{{Diolaiti} {et~al.}(2000){Diolaiti}, {Bendinelli}, {Bonaccini},
  {Close}, {Currie}, \& {Parmeggiani}}]{Diolaiti_00}
{Diolaiti}, E., {Bendinelli}, O., {Bonaccini}, D., {et~al.} 2000, \aaps, 147,
  335

\bibitem[{{Dotter} {et~al.}(2008){Dotter}, {Chaboyer}, {Jevremovi{\'c}},
  {Kostov}, {Baron}, \& {Ferguson}}]{Dotter_08}
{Dotter}, A., {Chaboyer}, B., {Jevremovi{\'c}}, D., {et~al.} 2008, \apjs, 178,
  89

\bibitem[{{Dotter} {et~al.}(2011){Dotter}, {Sarajedini}, \&
  {Anderson}}]{Dotter_11}
{Dotter}, A., {Sarajedini}, A., \& {Anderson}, J. 2011, \apj, 738, 74

\bibitem[{{Erben} {et~al.}(2005){Erben}, {Schirmer}, {Dietrich}, {Cordes},
  {Haberzettl}, {Hetterscheidt}, {Hildebrandt}, {Schmithuesen}, {Schneider},
  {Simon}, {Deul}, {Hook}, {Kaiser}, {Radovich}, {Benoist}, {Nonino}, {Olsen},
  {Prandoni}, {Wichmann}, {Zaggia}, {Bomans}, {Dettmar}, \&
  {Miralles}}]{Erben_05}
{Erben}, T., {Schirmer}, M., {Dietrich}, J.~P., {et~al.} 2005, Astronomische
  Nachrichten, 326, 432

\bibitem[{{Erkal} {et~al.}(2016){Erkal}, {Belokurov}, {Bovy}, \&
  {Sanders}}]{Erkal_16}
{Erkal}, D., {Belokurov}, V., {Bovy}, J., \& {Sanders}, J.~L. 2016, \mnras,
  463, 102

\bibitem[{{Fritz} {et~al.}(2010){Fritz}, {Gillessen}, {Trippe}, {Ott},
  {Bartko}, {Pfuhl}, {Dodds-Eden}, {Davies}, {Eisenhauer}, \&
  {Genzel}}]{Fritz_09}
{Fritz}, T., {Gillessen}, S., {Trippe}, S., {et~al.} 2010, \mnras, 401, 1177

\bibitem[{{Fritz} \& {Kallivayalil}(2015)}]{Fritz_15}
{Fritz}, T.~K., \& {Kallivayalil}, N. 2015, \apj, 811, 123

\bibitem[{{Fritz} {et~al.}(2016){Fritz}, {Kallivayalil}, {Carrasco}, {Neichel},
  {Davies}, {Beaton}, {Angell}, {Linden}, {Zivick}, {Majewski}, {Damke},
  {Boylan-Kolchin}, {van der Marel}, \& {Sohn}}]{Fritz_16}
{Fritz}, T.~K., {Kallivayalil}, N., {Carrasco}, E.~R., {et~al.} 2016, ArXiv
  e-prints, arXiv:1601.00965

\bibitem[{{Galametz} {et~al.}(2013){Galametz}, {Grazian}, {Fontana},
  {Ferguson}, {Ashby}, {Barro}, {Castellano}, {Dahlen}, {Donley}, {Faber},
  {Grogin}, {Guo}, {Huang}, {Kocevski}, {Koekemoer}, {Lee}, {McGrath}, {Peth},
  {Willner}, {Almaini}, {Cooper}, {Cooray}, {Conselice}, {Dickinson}, {Dunlop},
  {Fazio}, {Foucaud}, {Gardner}, {Giavalisco}, {Hathi}, {Hartley}, {Koo},
  {Lai}, {de Mello}, {McLure}, {Lucas}, {Paris}, {Pentericci}, {Santini},
  {Simpson}, {Sommariva}, {Targett}, {Weiner}, {Wuyts}, \& {the CANDELS
  Team}}]{Galametz_13}
{Galametz}, A., {Grazian}, A., {Fontana}, A., {et~al.} 2013, \apjs, 206, 10

\bibitem[{{Gibbons} {et~al.}(2014){Gibbons}, {Belokurov}, \&
  {Evans}}]{Gibbons_14}
{Gibbons}, S.~L.~J., {Belokurov}, V., \& {Evans}, N.~W. 2014, \mnras, 445, 3788

\bibitem[{{Gibbons} {et~al.}(2017){Gibbons}, {Belokurov}, \&
  {Evans}}]{Gibbons_16}
---. 2017, \mnras, 464, 794

\bibitem[{{Gillessen} {et~al.}(2009){Gillessen}, {Eisenhauer}, {Trippe},
  {Alexander}, {Genzel}, {Martins}, \& {Ott}}]{Gillessen_09}
{Gillessen}, S., {Eisenhauer}, F., {Trippe}, S., {et~al.} 2009, \apj, 692, 1075

\bibitem[{{Grillmair} \& {Carlin}(2016)}]{Grillmair_16}
{Grillmair}, C.~J., \& {Carlin}, J.~L. 2016, in Astrophysics and Space Science
  Library, Vol. 420, Astrophysics and Space Science Library, ed. H.~J.
  {Newberg} \& J.~L. {Carlin}, 87

\bibitem[{{Harris}(1996)}]{Harris_96}
{Harris}, W.~E. 1996, \aj, 112, 1487

\bibitem[{{Hendel} \& {Johnston}(2015)}]{Hendel_15}
{Hendel}, D., \& {Johnston}, K.~V. 2015, \mnras, 454, 2472

\bibitem[{{Hunt} {et~al.}(2016){Hunt}, {Bovy}, \& {Carlberg}}]{Hunt_16}
{Hunt}, J.~A.~S., {Bovy}, J., \& {Carlberg}, R.~G. 2016, ApJL, submitted,
  arXiv:1610.02030

\bibitem[{{Hyde} {et~al.}(2015){Hyde}, {Keller}, {Zucker}, {Ibata}, {Siebert},
  {Lewis}, {Penarrubia}, {Irwin}, {Gilmore}, {Lane}, {Koch}, {Conn},
  {Diakogiannis}, \& {Martell}}]{Hyde_15}
{Hyde}, E.~A., {Keller}, S., {Zucker}, D.~B., {et~al.} 2015, \apj, 805, 189

\bibitem[{{Ilbert} {et~al.}(2009){Ilbert}, {Capak}, {Salvato}, {Aussel},
  {McCracken}, {Sanders}, {Scoville}, {Kartaltepe}, {Arnouts}, {Le Floc'h},
  {Mobasher}, {Taniguchi}, {Lamareille}, {Leauthaud}, {Sasaki}, {Thompson},
  {Zamojski}, {Zamorani}, {Bardelli}, {Bolzonella}, {Bongiorno}, {Brusa},
  {Caputi}, {Carollo}, {Contini}, {Cook}, {Coppa}, {Cucciati}, {de la Torre},
  {de Ravel}, {Franzetti}, {Garilli}, {Hasinger}, {Iovino}, {Kampczyk},
  {Kneib}, {Knobel}, {Kovac}, {Le Borgne}, {Le Brun}, {F{\`e}vre}, {Lilly},
  {Looper}, {Maier}, {Mainieri}, {Mellier}, {Mignoli}, {Murayama}, {Pell{\`o}},
  {Peng}, {P{\'e}rez-Montero}, {Renzini}, {Ricciardelli}, {Schiminovich},
  {Scodeggio}, {Shioya}, {Silverman}, {Surace}, {Tanaka}, {Tasca}, {Tresse},
  {Vergani}, \& {Zucca}}]{Ilbert_09}
{Ilbert}, O., {Capak}, P., {Salvato}, M., {et~al.} 2009, \apj, 690, 1236

\bibitem[{{Irwin} {et~al.}(1995){Irwin}, {Demers}, \& {Kunkel}}]{Irwin_95}
{Irwin}, M.~J., {Demers}, S., \& {Kunkel}, W.~E. 1995, \apjl, 453, L21

\bibitem[{{Johnston} {et~al.}(2008){Johnston}, {Bullock}, {Sharma}, {Font},
  {Robertson}, \& {Leitner}}]{Johnston_08}
{Johnston}, K.~V., {Bullock}, J.~S., {Sharma}, S., {et~al.} 2008, \apj, 689,
  936

\bibitem[{{Johnston} {et~al.}(2001){Johnston}, {Sackett}, \&
  {Bullock}}]{Johnston_01}
{Johnston}, K.~V., {Sackett}, P.~D., \& {Bullock}, J.~S. 2001, \apj, 557, 137

\bibitem[{{Kaczmarczik} {et~al.}(2009){Kaczmarczik}, {Richards}, {Mehta}, \&
  {Schlegel}}]{Kaczmarczik_09}
{Kaczmarczik}, M.~C., {Richards}, G.~T., {Mehta}, S.~S., \& {Schlegel}, D.~J.
  2009, \aj, 138, 19

\bibitem[{{Kalirai} {et~al.}(2007){Kalirai}, {Anderson}, {Richer}, {King},
  {Brewer}, {Carraro}, {Davis}, {Fahlman}, {Hansen}, {Hurley}, {L{\'e}pine},
  {Reitzel}, {Rich}, {Shara}, \& {Stetson}}]{Kalirai_07}
{Kalirai}, J.~S., {Anderson}, J., {Richer}, H.~B., {et~al.} 2007, \apjl, 657,
  L93

\bibitem[{{Kallivayalil} {et~al.}(2006){Kallivayalil}, {van der Marel},
  {Alcock}, {Axelrod}, {Cook}, {Drake}, \& {Geha}}]{Kallivayalil_06a}
{Kallivayalil}, N., {van der Marel}, R.~P., {Alcock}, C., {et~al.} 2006, \apj,
  638, 772

\bibitem[{{Kallivayalil} {et~al.}(2013){Kallivayalil}, {van der Marel},
  {Besla}, {Anderson}, \& {Alcock}}]{Kallivayalil_13}
{Kallivayalil}, N., {van der Marel}, R.~P., {Besla}, G., {Anderson}, J., \&
  {Alcock}, C. 2013, \apj, 764, 161

\bibitem[{{King}(1962)}]{King_62}
{King}, I. 1962, \aj, 67, 471

\bibitem[{{Kirby} {et~al.}(2013){Kirby}, {Cohen}, {Guhathakurta}, {Cheng},
  {Bullock}, \& {Gallazzi}}]{Kirby_13}
{Kirby}, E.~N., {Cohen}, J.~G., {Guhathakurta}, P., {et~al.} 2013, \apj, 779,
  102

\bibitem[{{Kirby} {et~al.}(2011){Kirby}, {Lanfranchi}, {Simon}, {Cohen}, \&
  {Guhathakurta}}]{Kirby_11}
{Kirby}, E.~N., {Lanfranchi}, G.~A., {Simon}, J.~D., {Cohen}, J.~G., \&
  {Guhathakurta}, P. 2011, \apj, 727, 78

\bibitem[{{Koposov} {et~al.}(2014){Koposov}, {Irwin}, {Belokurov},
  {Gonzalez-Solares}, {Yoldas}, {Lewis}, {Metcalfe}, \& {Shanks}}]{Koposov_14}
{Koposov}, S.~E., {Irwin}, M., {Belokurov}, V., {et~al.} 2014, \mnras, 442, L85

\bibitem[{{Koposov} {et~al.}(2010){Koposov}, {Rix}, \& {Hogg}}]{Koposov_10}
{Koposov}, S.~E., {Rix}, H.-W., \& {Hogg}, D.~W. 2010, \apj, 712, 260

\bibitem[{{K{\"u}pper} {et~al.}(2015){K{\"u}pper}, {Balbinot}, {Bonaca},
  {Johnston}, {Hogg}, {Kroupa}, \& {Santiago}}]{Kuepper_15}
{K{\"u}pper}, A.~H.~W., {Balbinot}, E., {Bonaca}, A., {et~al.} 2015, \apj, 803,
  80

\bibitem[{{K{\"u}pper} {et~al.}(2016){K{\"u}pper}, {Johnston}, {Mieske},
  {Collins}, \& {Tollerud}}]{Kuepper_16}
{K{\"u}pper}, A.~H.~W., {Johnston}, K.~V., {Mieske}, S., {Collins}, M.~L.~M.,
  \& {Tollerud}, E.~J. 2016, AAS, submitted, arXiv:1608.05085

\bibitem[{{Law} \& {Majewski}(2010{\natexlab{a}})}]{Law_10b}
{Law}, D.~R., \& {Majewski}, S.~R. 2010{\natexlab{a}}, \apj, 718, 1128

\bibitem[{{Law} \& {Majewski}(2010{\natexlab{b}})}]{Law_10}
---. 2010{\natexlab{b}}, \apj, 714, 229

\bibitem[{{Leaman} {et~al.}(2013){Leaman}, {Venn}, {Brooks}, {Battaglia},
  {Cole}, {Ibata}, {Irwin}, {McConnachie}, {Mendel}, {Starkenburg}, \&
  {Tolstoy}}]{Leaman_13}
{Leaman}, R., {Venn}, K.~A., {Brooks}, A.~M., {et~al.} 2013, \apj, 767, 131

\bibitem[{{Li} \& {Gnedin}(2014)}]{Li_14}
{Li}, H., \& {Gnedin}, O.~Y. 2014, \apj, 796, 10

\bibitem[{{Lindegren}(1978)}]{Lindegren_78}
{Lindegren}, L. 1978, in IAU Colloq. 48: Modern Astrometry, ed. F.~V.
  {Prochazka} \& R.~H. {Tucker}, 197--217

\bibitem[{{Mackey} \& {van den Bergh}(2005)}]{Mackey_05}
{Mackey}, A.~D., \& {van den Bergh}, S. 2005, \mnras, 360, 631

\bibitem[{{Majewski} {et~al.}(2003){Majewski}, {Skrutskie}, {Weinberg}, \&
  {Ostheimer}}]{Majewski_03}
{Majewski}, S.~R., {Skrutskie}, M.~F., {Weinberg}, M.~D., \& {Ostheimer}, J.~C.
  2003, \apj, 599, 1082

\bibitem[{{Majewski} {et~al.}(2013){Majewski}, {Hasselquist}, {{\L}okas},
  {Nidever}, {Frinchaboy}, {Garc{\'{\i}}a P{\'e}rez}, {Johnston},
  {M{\'e}sz{\'a}ros}, {Shetrone}, {Allende Prieto}, {Beaton}, {Beers},
  {Bizyaev}, {Cunha}, {Damke}, {Ebelke}, {Eisenstein}, {Hearty}, {Holtzman},
  {Johnson}, {Law}, {Malanushenko}, {Malanushenko}, {O'Connell}, {Oravetz},
  {Pan}, {Schiavon}, {Schneider}, {Simmons}, {Skrutskie}, {Smith}, {Wilson}, \&
  {Zasowski}}]{majewski_2013}
{Majewski}, S.~R., {Hasselquist}, S., {{\L}okas}, E.~L., {et~al.} 2013, \apjl,
  777, L13

\bibitem[{{Markwardt}(2009)}]{Markwardt_09}
{Markwardt}, C.~B. 2009, in Astronomical Society of the Pacific Conference
  Series, Vol. 411, Astronomical Data Analysis Software and Systems XVIII, ed.
  D.~A. {Bohlender}, D.~{Durand}, \& P.~{Dowler}, 251

\bibitem[{{Mart{\'{\i}}nez-Delgado} {et~al.}(2015){Mart{\'{\i}}nez-Delgado},
  {D'Onghia}, {Chonis}, {Beaton}, {Teuwen}, {GaBany}, {Grebel}, \&
  {Morales}}]{dmd_2015}
{Mart{\'{\i}}nez-Delgado}, D., {D'Onghia}, E., {Chonis}, T.~S., {et~al.} 2015,
  \aj, 150, 116

\bibitem[{{Massari} {et~al.}(2016{\natexlab{a}}){Massari}, {Fiorentino},
  {McConnachie}, {Bellini}, {Tolstoy}, {Turri}, {Andersen}, {Bono}, {Stetson},
  \& {Veran}}]{Massari_16}
{Massari}, D., {Fiorentino}, G., {McConnachie}, A., {et~al.}
  2016{\natexlab{a}}, \aap, 595, L2

\bibitem[{{Massari} {et~al.}(2016{\natexlab{b}}){Massari}, {Fiorentino},
  {Tolstoy}, {McConnachie}, {Stuik}, {Schreiber}, {Andersen}, {Cl{\'e}net},
  {Davies}, {Gratadour}, {Kuijken}, {Navarro}, {Pott}, {Rodeghiero}, {Turri},
  \& {Verdoes Kleijn}}]{Massari_16b}
{Massari}, D., {Fiorentino}, G., {Tolstoy}, E., {et~al.} 2016{\natexlab{b}}, in
  \procspie, Vol. 9909, Society of Photo-Optical Instrumentation Engineers
  (SPIE) Conference Series, 99091G

\bibitem[{{McConnachie}(2012)}]{Mcconnachie_12}
{McConnachie}, A.~W. 2012, \aj, 144, 4

\bibitem[{{McGregor} {et~al.}(2004){McGregor}, {Hart}, {Stevanovic}, {Bloxham},
  {Jones}, {Van Harmelen}, {Griesbach}, {Dawson}, {Young}, \&
  {Jarnyk}}]{McGregor_04}
{McGregor}, P., {Hart}, J., {Stevanovic}, D., {et~al.} 2004, in \procspie, Vol.
  5492, Ground-based Instrumentation for Astronomy, ed. A.~F.~M. {Moorwood} \&
  M.~{Iye}, 1033--1044

\bibitem[{{McMahon} {et~al.}(2013){McMahon}, {Banerji}, {Gonzalez}, {Koposov},
  {Bejar}, {Lodieu}, {Rebolo}, \& {VHS Collaboration}}]{McMahon13}
{McMahon}, R.~G., {Banerji}, M., {Gonzalez}, E., {et~al.} 2013, The Messenger,
  154, 35

\bibitem[{{M{\'e}ndez} {et~al.}(2011){M{\'e}ndez}, {Costa}, {Gallart},
  {Pedreros}, {Moyano}, \& {Altmann}}]{Mendez_11}
{M{\'e}ndez}, R.~A., {Costa}, E., {Gallart}, C., {et~al.} 2011, \aj, 142, 93

\bibitem[{{Meyer} {et~al.}(2011){Meyer}, {K{\"u}rster}, {Arcidiacono},
  {Ragazzoni}, \& {Rix}}]{Meyer_11}
{Meyer}, E., {K{\"u}rster}, M., {Arcidiacono}, C., {Ragazzoni}, R., \& {Rix},
  H.-W. 2011, \aap, 532, A16

\bibitem[{{Muratov} \& {Gnedin}(2010)}]{Muratov_10}
{Muratov}, A.~L., \& {Gnedin}, O.~Y. 2010, \apj, 718, 1266

\bibitem[{{Neichel} {et~al.}(2014){Neichel}, {Lu}, {Rigaut}, {Ammons},
  {Carrasco}, \& {Lassalle}}]{Neichel_14}
{Neichel}, B., {Lu}, J.~R., {Rigaut}, F., {et~al.} 2014, \mnras, 445, 500

\bibitem[{{Ness} \& {Lang}(2016)}]{Ness_16}
{Ness}, M., \& {Lang}, D. 2016, \aj, 152, 14

\bibitem[{{Ortolani} {et~al.}(2011){Ortolani}, {Barbuy}, {Momany}, {Saviane},
  {Bica}, {Jilkova}, {Salerno}, \& {Jungwiert}}]{Ortolani_11}
{Ortolani}, S., {Barbuy}, B., {Momany}, Y., {et~al.} 2011, \apj, 737, 31

\bibitem[{{Palma} {et~al.}(2000){Palma}, {Kunkel}, \& {Majewski}}]{Palma_00}
{Palma}, C., {Kunkel}, W.~E., \& {Majewski}, S.~R. 2000, \pasp, 112, 1305

\bibitem[{{Pawlowski} {et~al.}(2015){Pawlowski}, {McGaugh}, \&
  {Jerjen}}]{Pawlowski_15}
{Pawlowski}, M.~S., {McGaugh}, S.~S., \& {Jerjen}, H. 2015, \mnras, 453, 1047

\bibitem[{{Peng} {et~al.}(2002){Peng}, {Ho}, {Impey}, \& {Rix}}]{Peng_02}
{Peng}, C.~Y., {Ho}, L.~C., {Impey}, C.~D., \& {Rix}, H.-W. 2002, \aj, 124, 266

\bibitem[{{Piatek} {et~al.}(2006){Piatek}, {Pryor}, {Bristow}, {Olszewski},
  {Harris}, {Mateo}, {Minniti}, \& {Tinney}}]{Piatek_06}
{Piatek}, S., {Pryor}, C., {Bristow}, P., {et~al.} 2006, \aj, 131, 1445

\bibitem[{{Piatek} {et~al.}(2007){Piatek}, {Pryor}, {Bristow}, {Olszewski},
  {Harris}, {Mateo}, {Minniti}, \& {Tinney}}]{Piatek_07}
---. 2007, \aj, 133, 818

\bibitem[{{Piatek} {et~al.}(2016){Piatek}, {Pryor}, \& {Olszewski}}]{Piatek_16}
{Piatek}, S., {Pryor}, C., \& {Olszewski}, E.~W. 2016, \aj, 152, 166

\bibitem[{{Pryor} {et~al.}(2015){Pryor}, {Piatek}, \& {Olszewski}}]{Pryor_15}
{Pryor}, C., {Piatek}, S., \& {Olszewski}, E.~W. 2015, \aj, 149, 42

\bibitem[{{Reid} \& {Brunthaler}(2004)}]{Reid_04}
{Reid}, M.~J., \& {Brunthaler}, A. 2004, \apj, 616, 872

\bibitem[{{Reid} {et~al.}(2014){Reid}, {Menten}, {Brunthaler}, {Zheng}, {Dame},
  {Xu}, {Wu}, {Zhang}, {Sanna}, {Sato}, {Hachisuka}, {Choi}, {Immer},
  {Moscadelli}, {Rygl}, \& {Bartkiewicz}}]{Reid_14}
{Reid}, M.~J., {Menten}, K.~M., {Brunthaler}, A., {et~al.} 2014, \apj, 783, 130

\bibitem[{{Renaud} {et~al.}(2016){Renaud}, {Agertz}, \& {Gieles}}]{Renaud_16}
{Renaud}, F., {Agertz}, O., \& {Gieles}, M. 2016, MNRAS, accepted,
  arXiv:1610.03101

\bibitem[{{Rigaut} {et~al.}(2014){Rigaut}, {Neichel}, {Boccas}, {d'Orgeville},
  {Vidal}, {van Dam}, {Arriagada}, {Fesquet}, {Galvez}, {Gausachs}, {Cavedoni},
  {Ebbers}, {Karewicz}, {James}, {L{\"u}hrs}, {Montes}, {Perez}, {Rambold},
  {Rojas}, {Walker}, {Bec}, {Trancho}, {Sheehan}, {Irarrazaval}, {Boyer},
  {Ellerbroek}, {Flicker}, {Gratadour}, {Garcia-Rissmann}, \&
  {Daruich}}]{Rigaut_14}
{Rigaut}, F., {Neichel}, B., {Boccas}, M., {et~al.} 2014, \mnras, 437, 2361

\bibitem[{{Sales} {et~al.}(2011){Sales}, {Navarro}, {Cooper}, {White}, {Frenk},
  \& {Helmi}}]{Sales2011}
{Sales}, L.~V., {Navarro}, J.~F., {Cooper}, A.~P., {et~al.} 2011, \mnras, 418,
  648

\bibitem[{{Sales} {et~al.}(2016){Sales}, {Navarro}, {Kallivayalil}, \&
  {Frenk}}]{Sales2016}
{Sales}, L.~V., {Navarro}, J.~F., {Kallivayalil}, N., \& {Frenk}, C.~S. 2016,
  MNRAS, submitted, arXiv:1605.03574

\bibitem[{{Sarajedini} \& {Geisler}(1996)}]{Sarajedini_96}
{Sarajedini}, A., \& {Geisler}, D. 1996, \aj, 112, 2013

\bibitem[{{Saviane} {et~al.}(2012){Saviane}, {da Costa}, {Held}, {Sommariva},
  {Gullieuszik}, {Barbuy}, \& {Ortolani}}]{Saviane_12}
{Saviane}, I., {da Costa}, G.~S., {Held}, E.~V., {et~al.} 2012, \aap, 540, A27

\bibitem[{{Sbordone} {et~al.}(2007){Sbordone}, {Bonifacio}, {Buonanno},
  {Marconi}, {Monaco}, \& {Zaggia}}]{Sbordone_07}
{Sbordone}, L., {Bonifacio}, P., {Buonanno}, R., {et~al.} 2007, \aap, 465, 815

\bibitem[{{Schirmer}(2013)}]{Schirmer_13}
{Schirmer}, M. 2013, \apjs, 209, 21

\bibitem[{{Sch{\"o}nrich} {et~al.}(2010){Sch{\"o}nrich}, {Binney}, \&
  {Dehnen}}]{Schoenrich_10}
{Sch{\"o}nrich}, R., {Binney}, J., \& {Dehnen}, W. 2010, \mnras, 403, 1829

\bibitem[{{Sersic}(1968)}]{Sersic_68}
{Sersic}, J.~L. 1968, {Atlas de galaxias australes} (Cordoba, Argentina:
  Observatorio Astronomico, 1968)

\bibitem[{{Sesar} {et~al.}(2015){Sesar}, {Bovy}, {Bernard}, {Caldwell},
  {Cohen}, {Fouesneau}, {Johnson}, {Ness}, {Ferguson}, {Martin},
  {Price-Whelan}, {Rix}, {Schlafly}, {Burgett}, {Chambers}, {Flewelling},
  {Hodapp}, {Kaiser}, {Magnier}, {Platais}, {Tonry}, {Waters}, \&
  {Wyse}}]{sesar_2015}
{Sesar}, B., {Bovy}, J., {Bernard}, E.~J., {et~al.} 2015, \apj, 809, 59

\bibitem[{{Siegel} {et~al.}(2011){Siegel}, {Majewski}, {Law}, {Sarajedini},
  {Dotter}, {Mar{\'{\i}}n-Franch}, {Chaboyer}, {Anderson}, {Aparicio}, {Bedin},
  {Hempel}, {Milone}, {Paust}, {Piotto}, {Reid}, \& {Rosenberg}}]{Siegel_11}
{Siegel}, M.~H., {Majewski}, S.~R., {Law}, D.~R., {et~al.} 2011, \apj, 743, 20

\bibitem[{{Sirianni} {et~al.}(2005){Sirianni}, {Jee}, {Ben{\'{\i}}tez},
  {Blakeslee}, {Martel}, {Meurer}, {Clampin}, {De Marchi}, {Ford}, {Gilliland},
  {Hartig}, {Illingworth}, {Mack}, \& {McCann}}]{Sirianni_05}
{Sirianni}, M., {Jee}, M.~J., {Ben{\'{\i}}tez}, N., {et~al.} 2005, \pasp, 117,
  1049

\bibitem[{{Sohn} {et~al.}(2012){Sohn}, {Anderson}, \& {van der
  Marel}}]{Sohn_12}
{Sohn}, S.~T., {Anderson}, J., \& {van der Marel}, R.~P. 2012, \apj, 753, 7

\bibitem[{{Sohn} {et~al.}(2013){Sohn}, {Besla}, {van der Marel},
  {Boylan-Kolchin}, {Majewski}, \& {Bullock}}]{Sohn_13}
{Sohn}, S.~T., {Besla}, G., {van der Marel}, R.~P., {et~al.} 2013, \apj, 768,
  139

\bibitem[{{Springel} {et~al.}(2008){Springel}, {Wang}, {Vogelsberger},
  {Ludlow}, {Jenkins}, {Helmi}, {Navarro}, {Frenk}, \& {White}}]{Springel2008b}
{Springel}, V., {Wang}, J., {Vogelsberger}, M., {et~al.} 2008, \mnras, 391,
  1685

\bibitem[{{Stetson}(1987)}]{Stetson_87}
{Stetson}, P.~B. 1987, \pasp, 99, 191

\bibitem[{{Strader} {et~al.}(2003){Strader}, {Brodie}, {Forbes}, {Beasley}, \&
  {Huchra}}]{Strader_03}
{Strader}, J., {Brodie}, J.~P., {Forbes}, D.~A., {Beasley}, M.~A., \& {Huchra},
  J.~P. 2003, \aj, 125, 1291

\bibitem[{{Trippe} {et~al.}(2010){Trippe}, {Davies}, {Eisenhauer}, {Schreiber},
  {Fritz}, \& {Genzel}}]{Trippe_10}
{Trippe}, S., {Davies}, R., {Eisenhauer}, F., {et~al.} 2010, \mnras, 402, 1126

\bibitem[{{van den Bergh}(1994)}]{Vdbergh_94}
{van den Bergh}, S. 1994, \aj, 108, 2145

\bibitem[{{van der Marel} {et~al.}(2012{\natexlab{a}}){van der Marel}, {Besla},
  {Cox}, {Sohn}, \& {Anderson}}]{Marel_12a}
{van der Marel}, R.~P., {Besla}, G., {Cox}, T.~J., {Sohn}, S.~T., \&
  {Anderson}, J. 2012{\natexlab{a}}, \apj, 753, 9

\bibitem[{{van der Marel} {et~al.}(2012{\natexlab{b}}){van der Marel},
  {Fardal}, {Besla}, {Beaton}, {Sohn}, {Anderson}, {Brown}, \&
  {Guhathakurta}}]{Marel_12b}
{van der Marel}, R.~P., {Fardal}, M., {Besla}, G., {et~al.} 2012{\natexlab{b}},
  \apj, 753, 8

\bibitem[{{van der Marel} \& {Kallivayalil}(2014)}]{Marel_14}
{van der Marel}, R.~P., \& {Kallivayalil}, N. 2014, \apj, 781, 121

\bibitem[{{Vivas} {et~al.}(2001){Vivas}, {Zinn}, {Andrews}, {Bailyn}, {Baltay},
  {Coppi}, {Ellman}, {Girard}, {Rabinowitz}, {Schaefer}, {Shin}, {Snyder},
  {Sofia}, {van Altena}, {Abad}, {Bongiovanni}, {Brice{\~n}o}, {Bruzual},
  {Della Prugna}, {Herrera}, {Magris}, {Mateu}, {Pacheco}, {S{\'a}nchez},
  {S{\'a}nchez}, {Schenner}, {Stock}, {Vicente}, {Vieira}, {Ferr{\'{\i}}n},
  {Hernandez}, {Gebhard}, {Honeycutt}, {Mufson}, {Musser}, \&
  {Rengstorf}}]{Vivas_01}
{Vivas}, A.~K., {Zinn}, R., {Andrews}, P., {et~al.} 2001, \apjl, 554, L33

\bibitem[{{Weisz} {et~al.}(2016){Weisz}, {Koposov}, {Dolphin}, {Belokurov},
  {Gieles}, {Mateo}, {Olszewski}, {Sills}, \& {Walker}}]{Weisz_16}
{Weisz}, D.~R., {Koposov}, S.~E., {Dolphin}, A.~E., {et~al.} 2016, \apj, 822,
  32

\bibitem[{{Wetzel} {et~al.}(2015){Wetzel}, {Tollerud}, \& {Weisz}}]{Wetzel_15}
{Wetzel}, A.~R., {Tollerud}, E.~J., \& {Weisz}, D.~R. 2015, \apjl, 808, L27

\bibitem[{{Xue} {et~al.}(2008){Xue}, {Rix}, {Zhao}, {Re Fiorentin}, {Naab},
  {Steinmetz}, {van den Bosch}, {Beers}, {Lee}, {Bell}, {Rockosi}, {Yanny},
  {Newberg}, {Wilhelm}, {Kang}, {Smith}, \& {Schneider}}]{Xue_08}
{Xue}, X.~X., {Rix}, H.~W., {Zhao}, G., {et~al.} 2008, \apj, 684, 1143

\bibitem[{{Yoon} {et~al.}(2011){Yoon}, {Johnston}, \& {Hogg}}]{Yoon_11}
{Yoon}, J.~H., {Johnston}, K.~V., \& {Hogg}, D.~W. 2011, \apj, 731, 58

\bibitem[{{Zaritsky} {et~al.}(2016){Zaritsky}, {McCabe}, {Aravena},
  {Athanassoula}, {Bosma}, {Comer{\'o}n}, {Courtois}, {Elmegreen}, {Elmegreen},
  {Erroz-Ferrer}, {Gadotti}, {Hinz}, {Ho}, {Holwerda}, {Kim}, {Knapen},
  {Laine}, {Laurikainen}, {Mu{\~n}oz-Mateos}, {Salo}, \&
  {Sheth}}]{Zaritsky_16a}
{Zaritsky}, D., {McCabe}, K., {Aravena}, M., {et~al.} 2016, \apj, 818, 99

\bibitem[{{Zinn}(1993)}]{Zinn_93}
{Zinn}, R. 1993, in Astronomical Society of the Pacific Conference Series,
  Vol.~48, The Globular Cluster-Galaxy Connection, ed. G.~H. {Smith} \& J.~P.
  {Brodie}, 38

\end{thebibliography}
\appendix

 \section{All Proper Motions}
\label{sec:all_prom_mot}

Here, we present all the stellar proper motion in Table~\ref{tab:all_mot} in electronic form.

\begin{table*}
\centering
\caption{Stellar proper motions} \label{tab:all_mot}
\begin{tabular}{c c c c c c c c c}
 \hline \hline
Id & Member & R.A.$^{a}$ & Dec.$^{a}$ &  $\mu_\alpha$  & $\mu_\delta$ &  m$_\mathrm{K'}$ & $F606W-F814W$ & $F814W-K'$ \\
         & & [$^{\circ}$]  & [$^{\circ}$]       & [mas/yr]         &  [mas/yr]          &         &    &  \\
 \hline 
 1 & Member & 136.9898206 & -37.2982895 & 0.58$\pm$0.17 & -0.26$\pm$0.17 & 19.93 & 0.79 & 1.29 \\
2 & Member & 136.9901661 & -37.2971284 & -0.66$\pm$0.34 & 0.73$\pm$0.34 & 20.74 & 0.73 & 1.2 \\
3 & Non-Member & 136.9880481 & -37.2971010 & -5.8$\pm$0.29 & 2.89$\pm$0.29 & 19.22 & 1.97 & 2.35 \\
4 & Member & 136.9825880 & -37.2967648 & -1.56$\pm$0.66 & -0.49$\pm$0.66 & 20.73 & 0.79 & 1.26 \\
5 & Non-Member & 136.9803312 & -37.2963152 & -2.11$\pm$0.6 & 3.06$\pm$0.6 & 18.91 & 2.1 & 2.51 \\
6 & Member & 136.9893551 & -37.2937942 & 1.1$\pm$1.04 & -1.59$\pm$1.04 & 21.77 & 0.86 & 1.45 \\
7 & Member & 136.9886615 & -37.2930079 & 0.09$\pm$0.34 & 0.21$\pm$0.34 & 20.38 & 0.82 & 1.14 \\
8 & Non-Member & 137.032578 & -37.2927165 & -4.2$\pm$0.85 & -2.59$\pm$0.85 & 21.08 & 1.57 & 2.11 \\
9 & Member & 136.9970668 & -37.2926264 & -0.1$\pm$0.46 & 0.87$\pm$0.46 & 21.49 & 0.78 & 1.22 \\
10 & Member & 136.9863243 & -37.2922259 & -0.7$\pm$0.35 & -0.11$\pm$0.35 & 19.84 & 0.89 & 1.34 \\
\hline \hline
\multicolumn{9}{l}{$^{a}$ The uncertainty of the positions are about 7$\times10^{-6}$ degree compared to the absolute reference frame.} \\
\multicolumn{9}{l}{This does not affect the relative positions.} \\

\end{tabular}
\end{table*}

\end{document}